\documentclass[letterpaper,fleqn,superscriptaddress,preprintnumbers,prd,onecolumn,longbibliography]{revtex4-2}

\usepackage{tikz}
\usepackage{mathtools}
\usetikzlibrary{shapes.geometric, arrows}
\tikzstyle{startstop} = [rectangle, rounded corners, minimum width=3cm, minimum height=1cm,text width=5cm,text centered, draw=black, fill=red!30]
\tikzstyle{io} = [trapezium, trapezium left angle=70, trapezium right angle=110, minimum width=3cm, minimum height=1cm,text width=4cm, text centered, draw=black, fill=blue!25]
\tikzstyle{cloud} = [draw, ellipse,fill=blue!40, node distance=2.5cm, minimum height=2em,text width=2.cm, text centered]
\tikzstyle{process} = [rectangle, minimum width=3cm, minimum height=1cm, text width=5.cm,text centered, draw=black, fill=orange!30]
\tikzstyle{decision} = [diamond, minimum width=3cm, minimum height=1cm, text width=3.5cm, text centered, draw=black, fill=green!30,node distance=4.5cm, inner sep=0pt]
\tikzstyle{arrow} = [thick,->,>=stealth]
 
\makeatletter
\g@addto@macro\bfseries{\boldmath}
\makeatother

\usepackage{graphicx,color}
\usepackage{amsmath,amssymb,amsfonts}
\usepackage{stackrel}
\usepackage{tabularx}
\usepackage{ulem}
\newcolumntype{Y}{>{\centering\arraybackslash}X}
\usepackage{booktabs}
\AtBeginDocument{
    \heavyrulewidth=.08em
        \lightrulewidth=.05em
        \cmidrulewidth=.03em
        \belowrulesep=.65ex
        \belowbottomsep=0pt
        \aboverulesep=.4ex
        \abovetopsep=0pt
        \cmidrulesep=\doublerulesep
        \cmidrulekern=.5em
        \defaultaddspace=.5em
}
\usepackage{xcolor}
\usepackage{hyperref}
\hypersetup{
    colorlinks=true,
        linkcolor={blue!80!black},
        citecolor={blue!80!black},
        urlcolor={blue!80!black} 
}

\providecommand{\openone}{\leavevmode\hbox{\large1\kern-7.3pt\normalsize1}}
\newcommand{\be}{\begin{equation}}
\newcommand{\ee}{\end{equation}}
\newcommand{\ba}{\begin{eqnarray}}
\newcommand{\ea}{\end{eqnarray}}

\newcommand{\fig}{Fig.~}

\renewcommand{\ln}{{\rm ln}}

\newcommand{\fKep}{f_{\rm Kep}}
\newcommand{\omegaKep}{\Omega_{\rm Kep}}
\newcommand{\mTOV}{M_{\rm TOV}}
\newcommand{\mNonrot}{M_{\rm nonrot}}
\newcommand{\rNonrot}{R_{\rm nonrot}}
\newcommand{\mKep}{M_{\rm Kep}}
\newcommand{\rKep}{R_{\rm Kep}}
\newcommand{\mTOVo}{M_{\rm TOV,B}}

\newcommand{\Mch}{\mathcal{M}_{\rm chirp}}

\newcommand{\JMs}{\left| \chi \right|}

\newcommand{\mSupra}{M_{\rm supra}}
\newcommand{\mSuprao}{M_{\rm supra,B}}
\newcommand{\mCrito}{M_{\rm crit,B}}
\newcommand{\mRemno}{M_{\rm remn,B}}
\newcommand{\mEjeo}{M_{\rm eje,B}}

\newcommand{\dint}[3]{\int_{#2}^{#3}\!\mathrm{d}#1 \,} 
\newcommand{\edit}[1]{\textcolor{black}{#1}}

\renewcommand{\emph}{\textit}

\graphicspath{{figures/}}

\begin{document}

\title{Multimessenger constraints for ultra-dense matter}

\preprint{HIP-2021-11/TH}
\author{Eemeli Annala}
\affiliation{Department of Physics and Helsinki Institute of Physics, P.O.~Box 64, FI-00014 University of Helsinki, Finland}
\author{Tyler Gorda}
\affiliation{Technische Universit\"{a}t Darmstadt, Department of Physics, D–64289 Darmstadt, Germany}
\affiliation{Helmholtz Research Academy for FAIR, D–64289 Darmstadt, Germany}
\author{Evangelia Katerini}
\affiliation{Department of Physics Aristotle University of Thessaloniki, University Campus, 54124, Thessaloniki, Greece}
\author{Aleksi Kurkela}
\affiliation{Faculty of Science and Technology, University of Stavanger, 4036 Stavanger, Norway}
\author{Joonas N\"attil\"a}
\affiliation{Physics Department and Columbia Astrophysics Laboratory, Columbia University, 538 West 120th Street, New York, NY 10027, USA and \\
Center for Computational Astrophysics, Flatiron Institute, 162 Fifth Avenue, New York, NY 10010, USA}
\author{Vasileios Paschalidis}
\affiliation{Departments of Astronomy and Physics, University of Arizona, Tucson, AZ, USA}
\author{Aleksi Vuorinen}
\affiliation{Department of Physics and Helsinki Institute of Physics, P.O.~Box 64, FI-00014 University of Helsinki, Finland}

\begin{abstract}
\noindent Recent rapid progress in neutron-star (NS) observations offers great potential to constrain the properties of strongly interacting matter under the most extreme conditions. In order to fully exploit the current observational inputs and to study the impact of future observations of NS masses, radii, and tidal deformabilities, we analyze a large ensemble of randomly generated viable NS-matter equations of state (EoSs) and the corresponding rotating stellar structures. We discuss the compatibility and impact of various hypotheses and measurements on the EoS, including those involving the merger product of the gravitational-wave (GW) event GW170817, the binary merger components in GW190814, and radius measurements of the pulsar PSR J0740+6620. We obtain an upper limit for the dimensionless spin of a rigidly rotating NS, $\JMs < 0.81$, an upper limit for the compactness of a NS, $GM/(Rc^2) < 0.33$, and find that the conservative hypothesis that the remnant in GW170817 ultimately collapsed to a black hole strongly constrains the EoS and the maximal mass of NSs, implying $\mTOV < 2.53 M_\odot$ (or $\mTOV < 2.19 M_\odot$ if we assume that a hypermassive NS was created). Additionally, we derive a novel lower limit for the tidal deformability as a function of NS mass and provide fitting formulae that can be used to set priors for parameter estimation and to discern whether neutron stars or other compact objects are involved in future low-mass GW events. Finally, we find that the recent NICER results for the radius of the massive NS PSR J0740+6620 place strong constraints for the behavior of the EoS, and that the indicated radius values $R(2M_\odot) \gtrsim 11$~km are compatible with moderate speeds of sound in NS matter and thus with the existence of quark matter cores in massive NSs. 
\end{abstract}

\maketitle

\section{Introduction}

The past ten years have firmly established neutron stars (NSs) as the leading laboratory for the study of ultra-dense strongly interacting matter. NS mass and radius determinations of increasing accuracy \cite{Demorest:2010bx,Antoniadis:2013pzd,Cromartie:2019kug,Fonseca:2021wxt,Fonseca:2016tux,Nattila:2017wtj,Riley:2019yda,Miller:2019cac,Miller:2021qha,Riley:2021pdl} and the breakthrough detection of gravitational waves (GWs) from a binary NS merger by LIGO and Virgo \cite{TheLIGOScientific:2017qsa,Abbott:2018exr} have together provided a large amount of robust observational data for NS properties. 
By means of general-relativistic, spherically-symmetric stellar-structure equations (so called Tolman-Oppenheimer-Volkov equations; TOV~\cite{1939PhRv...55..364T,Oppenheimer:1939ne}), these data have been repeatedly compared to predictions derived from microscopic calculations in nuclear and particle theory \cite{Tews:2012fj,Drischler:2017wtt,Lynn:2015jua,Holt:2014hma,Kurkela:2009gj,Gorda:2018gpy,Gorda:2021znl,Gorda:2021kme}. As a result, the uncertainties of the cold NS-matter equation of state (EoS), i.e.~the functional relationship between its energy density $\epsilon$ and pressure $p$, have been systematically reduced \cite{Annala:2017llu,Margalit:2017dij,Rezzolla:2017aly,Ruiz:2017due,Bauswein:2017vtn,Radice:2017lry,Most:2018hfd,Dietrich:2020efo,Capano:2019eae,Landry:2018prl,Raithel:2018ncd,Raithel:2019ejc,Raaijmakers:2019dks,Essick:2019ldf,Jokela:2020piw,Al-Mamun:2020vzu,Essick:2021kjb} (see also~\cite{Baym:2017whm,Gandolfi:2019zpj,Raithel:2019uzi,Horowitz:2019piw,Baiotti:2019sew,Chatziioannou:2020pqz,Radice:2020ddv} for reviews), which has even led to the first indications of deconfined quark matter (QM) residing inside massive NSs \cite{Paschalidis:2017qmb,Annala:2019puf,Ferreira:2020kvu,Minamikawa:2020jfj,Blacker:2020nlq}.

While steady progress towards deducing the true EoS realized in Nature is expected to continue with forthcoming improvements in theoretical calculations and NS observations, it is important to also ask whether we have exhausted current observational information in this task. Indeed, the observation of an electromagnetic signal from the binary-NS merger event GW170817 offers indirect information about the EoS that until now has not been fully explored. It is widely accepted that the GW170817 binary-merger remnant underwent gravitational collapse to a black hole (BH)~\cite{Margalit:2017dij,Rezzolla:2017aly,Ruiz:2017due,Shibata:2017xdx,Shibata:2019ctb}, which we shall refer to as the BH formation hypothesis in the following. Furthermore, there are indications that the collapse likely took place in a time frame of order $1\,\mathrm{s}$ by first forming a differentially~\footnote{A differentially rotating stellar configuration has an angular velocity that is a function of the distance from the axis of rotation.} (instead of uniformly) rotating hypermassive NS (HMNS) that later collapsed into a BH \cite{GBM:2017lvd}. However, given that neither the maximum mass of stable NSs nor the angular momentum at which the collapse took place are known \emph{a priori}~\cite{Shibata:2019ctb}, the remnant may also have been a supramassive NS that quickly lost its differential-rotation support and then spun down further due to, e.g., dipole radiation and collapsed to a BH.

To undergo eventual collapse to a BH, the remnant rest mass must have been above an (EoS-dependent) critical rest mass value we denote by $\mCrito$~\footnote{The definition of the rest or baryon mass of a star is $M_{\rm B}=\int {\rm d}^3x\sqrt{-g}\rho_{\rm B}u^0$, where $g$ is the determinant of the spacetime metric, $\rho_{\rm B}$ the rest-mass density, and $u^0$ the time component of the perfect fluid four-velocity. The rest-mass density is the product of the baryon number density $n_{\rm B}$ times a mean baryon mass. Strictly speaking, the mean baryon mass depends on the composition of matter, but for the calculations performed in this work we take it to be the neutron rest mass. This choice leads to rest mass values that are accurate to within roughly 1 part in $10^3$, but since we use this definition consistently throughout our work, our results are insensitive to the mean baryon mass value.}. The value of this critical rest mass depends also on the state of rotation and the remnant total angular momentum~\cite{Bozzola:2017qbu,Weih:2017mcw}. As we would like to separately study the more conservative BH formation hypothesis and the more speculative scenario involving a HMNS, there exist two limiting cases for the value of the critical mass that are of relevance to us. These include: a) the maximum rest mass that can be supported by a non-rotating NS, i.e.,~the TOV limit rest mass $\mTOVo$ (corresponding to the BH formation hypothesis); and b) the maximum rest mass supported by maximal uniform rotation, i.e.,~the supramassive limit rest mass $\mSuprao$ (corresponding to the formation of a HMNS)~\cite{1992ApJ...398..203C}. By separately adopting these two values for the critical mass for gravitational collapse, we can derive novel constraints on the ultra-dense matter EoS. In particular, the latter $\mCrito=\mSuprao$ case has often been used to set upper limits for the maximum mass of non-rotating NSs in the literature, $\mTOV$ \edit{\footnote{\edit{Note that in addition to these two cases, in some works (such as \cite{Shibata:2019ctb}) the authors assume the critical mass of collapse to lie in between the two limiting cases, $M_{\rm TOV,B} < M_{\rm crit,B} < M_{\rm supra,B}$. This is possible if one assumes that the merger remnant has lost angular momentum due to, e.g., GW emission.}}}. This has been done using either approximate quasi-universal relations for the ratio of the supramassive limit mass $\mSupra$ to $\mTOV$ or a small number of individual EoSs (see e.g.~\cite{Margalit:2017dij,Rezzolla:2017aly,Ruiz:2017due,Nathanail:2021tay}). However, as some quasi-universal relations have been shown to be strongly violated, e.g.,~by hybrid hadron-quark EoSs~\cite{Lau:2017qtz,Bandyopadhyay:2017dvi,Han:2018mtj,Bozzola:2019tit} and certain exotic compact star solutions \cite{Annala:2017tqz}, a comprehensive study of both limiting cases with an extensive ensemble of all viable NS-matter EoSs is clearly called for. 

In this paper, we present such an analysis using a sizable ensemble of NS matter EoSs that are randomly generated using an algorithm developed for \cite{Annala:2019puf} (see also subsection \ref{sec:EoSensemble} for details). While the generation of even millions of non-rotating stellar configurations through the TOV equations is rather straightforward, achievable on a single processor within a few days, the same task is computationally much more demanding for supramassive configurations, because it involves the iterative solution of partial differential equations. For the present work, we have created an interface for efficiently parallelizing the generation of millions of rotating configurations, such that this computational task can be completed within a few days. These calculations allow us to employ the $\mCrito=\mSuprao$ condition without having to resort to the use of universal relations, which enables us to place the most robust upper limits for the maximum TOV mass. Our novel algorithm is described in Appendix~\ref{app:BH_form_hyp}.

\begin{figure}[!t]
    \centering
        \includegraphics[height=6.3cm]{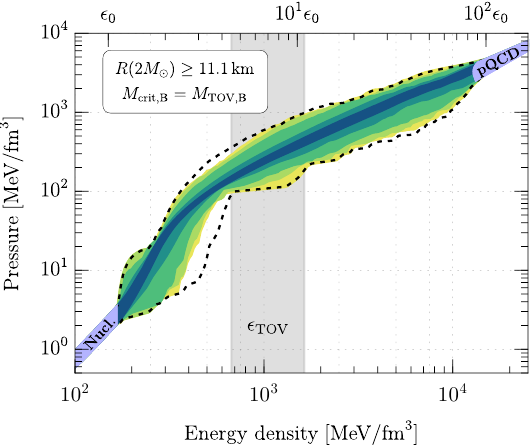}
        $\quad$
        \includegraphics[height=5.9cm]{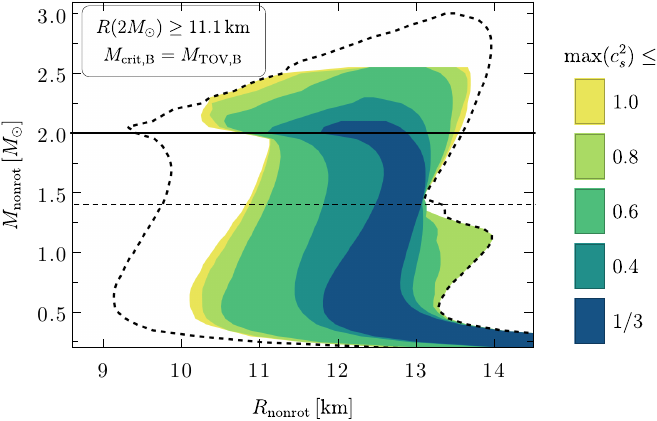} \\
    \caption{The impact on the allowed EoS (left) and MR (right) regions arising from the radius measurement of PSR J0740+6620, implemented as $\rNonrot(2M_\odot) \geq 11.1$~km, and the GW170817 BH formation hypothesis with $\mCrito=\mTOVo$ (here, ``nonrot'' indicates that we ignore the rotation of the star). The resulting ensemble supports EoSs with moderate speeds of sound, which are known to be compatible with sizeable QM cores \cite{Annala:2019puf}. The color coding used here refers to the maximal value that the speed of sound squared $c_s^2$ reaches at any density, with the lower-$c_s^2$ regions drawn on top of the higher ones. The dashed black lines correspond to the EoS and MR bands obtained by only imposing $\mTOV\geq 2.0M_\odot$ and the GW170817 tidal deformability limit $\tilde{\Lambda} < 720$ (low-spin priors). In the EoS figure, $\epsilon_0 \approx 150$~MeV/fm$^3$ represents the nuclear saturation energy density, the light blue regions illustrate the low- and high-density EoSs given by theoretical CET and pQCD calculations, \edit{and the shaded grey region shows the range of central energy densities in the cores of maximally massive non-rotating NSs (TOV stars)}.}
    \label{fig:NICERconstraintsTOV0onePanel}
\end{figure}

In addition to the above considerations concerning the merger product of the GW170817 event, recent years have witnessed some very intriguing individual NS mass and radius measurements, albeit with varying estimated uncertainties. In this context, $\mTOV$ is clearly one of the most important single parameters constraining the NS-matter EoS, and one that is bounded from below by the highest reliable individual mass measurement. While PSR~J0348$+$0432~\cite{Antoniadis:2013pzd} with gravitational mass $M=2.01\pm 0.04M_\odot$ (68\% uncertainties) remains the most accurately measured massive NS in existence, the recent observations of PSR J0740+6620 indicating a high mass of $M=2.08 \pm 0.07 M_\odot$ (68\% credible interval)~\cite{Cromartie:2019kug,Fonseca:2021wxt} point towards $\mTOV$ likely exceeding $2M_\odot$. 

At the same time, various X-ray observations have indicated that the NS radius $R$ likely falls within the interval $11\, \mathrm{km} \lesssim R \lesssim 14\, \mathrm{km}$  for a wide range of masses
\cite{Steiner:2010, Guillot:2013wu,Ozel:2015fia,Bogdanov:2016nle,Nattila:2015jra,Nattila:2017wtj,Steiner:2017vmg,Shawn:2018,Miller:2019cac,Riley:2019yda} (see also \cite{Ozel:2016oaf, Miller:2016pom} for reviews). These observations have gained additional support from the most recent radius measurement of a two-solar-mass millisecond pulsar PSR J0740+6620 by the NICER collaboration that indicated $R \gtrsim 11~\mathrm{km}$ \cite{Miller:2021qha,Riley:2021pdl,Raaijmakers:2021uju}. This kind of lower bound for the radius of a massive NS places a stringent constraint on the EoS, as has been demonstrated by several other studies~\cite{Somasundaram:2021ljr,Biswas:2021yge,Li:2021thg,Pang:2021jta} and will be discussed below. 

In Fig.~\ref{fig:NICERconstraintsTOV0onePanel} we show a sample of our results that represents the allowed EoS regions after imposing two astronomical observations in addition to the usual $\mTOV > 2M_\odot$ and $\tilde \Lambda < 720$ (low-spin priors). Firstly, we demand that the radii of the two-solar-mass NS configurations generated from our EoSs must be greater than 11.1~km, which corresponds to the 95\% lower limit of the joint NICER/XMM-Newton analysis, obtained with inflated cross-instrument-calibration uncertainties~\cite{Miller:2021qha}. Secondly, we assume that the GW170817 BH formation hypothesis holds true in conjunction with the more conservative choice $\mCrito=\mTOVo$, as current theoretical and observational knowledge suggest~\cite{GBM:2017lvd}. It is remarkable how tightly these two well-motivated and robust assumptions constrain the EoS, and moreover how nearly all EoSs that have been ruled out exhibit high maximum values of the speed of sound $c_s$ of NS matter (cf.~\fig\ref{fig:epMRNoSupra} for results without the two assumptions). More details of this analysis are presented in subsection \ref{sec:nicer}, where we discuss the impact of radius measurements of PSR J0740+6620 on our EoS ensemble, and in \ref{sec:future}, where we inspect the potential of various mass, tidal deformability, pulsar frequency, and radius measurements in constraining the EoS and the mass-radius (MR) relation.

Returning to GW measurements, in addition to GW170817 there are two other interesting and potentially relevant events reported by the LIGO/Virgo collaborations. While the binary merger GW190425~\cite{Abbott:2020uma} did not further constrain the EoS, the possible identification of the 2.6$M_\odot$ secondary binary component of the highly asymmetric event GW190814 as a NS could lead to very specific behaviour for the EoS \cite{Abbott:2020khf}. While it is unlikely that the nature of this component of GW190814 will ever be uncovered with any certainty, we will nevertheless speculate on the plausibility of the hypothesis of it having been a NS (GW190814 NS hypothesis) in subsection~\ref{sec:gw190814}. We find that although this hypothesis is in tension with the more restrictive GW170817 BH formation hypothesis assuming $\mCrito=\mSuprao$, it is compatible with the more conservative BH formation hypothesis assuming $\mCrito=\mTOVo$, although the latter may be hard to reconcile with the broader multimessenger picture of GW170817.  

Finally, within the analysis of any GW event involving NSs lies an assumed upper bound for the dimensionless spin value $\left| \chi \right| = \left| J \right|/M^2$ (in geometrized units, which we use throughout this work) of a uniformly rotating NS, where $J$ is the angular momentum of a rotating NS with mass $M$. For a BH, $|\chi| \leq 1$, but there is no such first-principles bound on NSs; one must instead postulate some prior on this quantity within, e.g.,\ a Bayesian analysis framework. The LIGO/Virgo analyses conducted thus far adopt $|\chi| < 0.89$ for their high-spin prior results for practical reasons \cite{TheLIGOScientific:2017qsa}. We investigate the maximum values of $\chi$ using our ensemble of EoSs and find that this prior is justified. More specifically, we find that $|\chi| < 0.78$ ($|\chi| < 0.81$) when the GW170817 binary tidal deformability constraint is (not) imposed. Owing to these findings, we recommend that apparent limit $|\chi| < 0.81$ (see subsection~\ref{sec:kepler} for additional details) be used in future GW analyses where NSs are suspected to be involved. Additionally, we obtain a lower limit on the tidal deformability of NSs as a function of their mass, and in Sec.~\ref{sec:nicer} provide fitting formulae that can be used to set priors for parameter estimation and to discern whether NSs or alternative compact objects, such as black holes~\cite{Yang:2017gfb,Hinderer:2018pei,Chen:2020fzm,Essick:2020ghc,Most:2020exl} or boson stars~\cite{Sennett:2017etc}, are involved in future low-mass GW events.

The organization of the present article is as follows. In Section~\ref{sec:methods}, we detail the analysis methods adopted in our work, and display the properties of both non-rotating and rotating NSs before adding additional input. In Section~\ref{sec:results}, we detail all of our results that are summarized above, as well as list new universal relations that we have uncovered between non-rotating and maximally rotating NSs with the same central energy density. Finally, in Section~\ref{sec:conclusions}, we present some brief concluding remarks.

\section{Methods\label{sec:methods}}

\subsection{Equation of state ensemble}\label{sec:EoSensemble}

Following a series of previous works focusing on the construction of model-independent EoSs for NS matter \cite{Kurkela:2014vha,Annala:2017llu,Annala:2019puf}, we build a large ensemble of randomly generated EoSs that by construction satisfy a number of robust theoretical and observational constraints. Specifically, we interpolate the EoSs between a low-density regime where the Chiral-Effective-Theory (CET) EoSs of~\cite{Tews:2012fj,Hebeler:2013nza} are applied up to 1.1 times the nuclear saturation density ($n_0 \approx 0.16$~fm$^{-3}$) and a high-density regime where we use the perturbative-Quantum-Chromodynamics (pQCD) EoSs of \cite{Kurkela:2009gj,Fraga:2013qra} for baryon chemical potentials $\mu_\text{B} \geq 2.6$~GeV, corresponding to baryonic number densities of $n_\text{B} \gtrsim 40 n_0$ \edit{\footnote{\edit{Note that the CET results build on a long series of previous works on the effective theory itself \cite{Weinberg:1990rz,Epelbaum:2008ga,Machleidt:2011zz} and the EoS in particular \cite{Hebeler:2009iv,Hebeler:2010xb}, while the pQCD EoS similarly relies on previous work performed, e.g.,~in \cite{Freedman:1976ub,Shuryak:1980tp,Vuorinen:2003fs}}}}. In the CET regime, we utilize the ``soft'' and ``hard'' EoSs of~\cite{Hebeler:2013nza}, while the pQCD EoS is continuously parameterized by the renormalization-scale variable $X \in [1,4]$ introduced in~\cite{Fraga:2013qra}. As briefly reviewed in Appendix \ref{app:interpolation}, at intermediate densities we follow the interpolation routine of Ref.~\cite{Annala:2019puf}, where we first build randomly generated piecewise-linear speed of sound squared ($c_s^2$) functions parameterized by $\mu_\text{B}$, and then integrate to obtain the pressure and energy density using thermodynamic formulae. Each resulting EoS is causal and thermodynamically stable by construction.

The intermediate-density region, where the interpolation is applied, is divided into 3--5 segments at randomly selected matching points, while the speed of sound is taken to be continuous everywhere but allowed to change in an arbitrary way, including the mimicking of a real first-order phase transition. For further details of the generation of the EoSs, we refer the reader to Appendix \ref{app:interpolation} and \cite{Annala:2019puf}. Here, we have significantly extended the ensemble so that it now contains altogether approximately~1,500,000 EoSs prior to placing any observational cuts. We have checked that the conclusions drawn in \cite{Annala:2019puf} remain unchanged with this new enlarged ensemble, and note that we use a larger number of EoSs to achieve a denser coverage of the EoS space after a larger number of observational constraints has been folded into our analysis.

\begin{figure}
    \includegraphics[height=6cm]{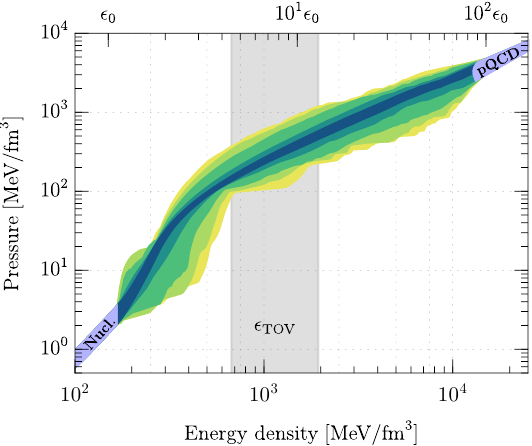}
    $\quad$
    \includegraphics[height=6cm]{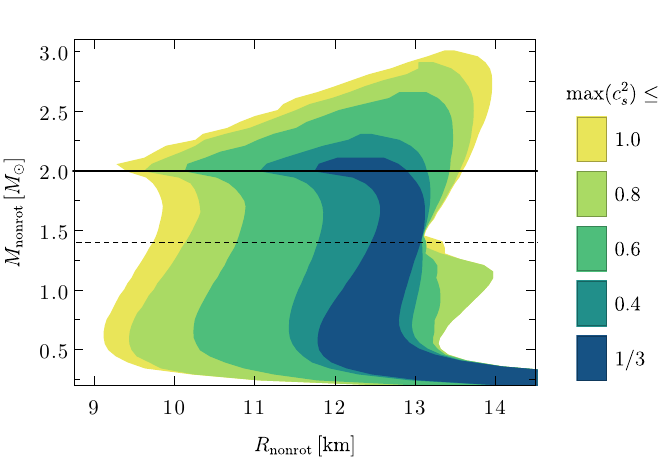} 
    \caption{The EoS (left) and non-rotating MR (right) regions corresponding to our full ensemble, where the only observational constraints enforced are the existence of two-solar-mass NSs and the GW170817 constraint $\tilde \Lambda<720$ (with low-spin priors). }
    \label{fig:epMRNoSupra}
\end{figure}

Once the EoS ensemble has been built, we proceed to place a number of observational cuts on its properties. Unless specified, all results displayed in this work assume two basic constraints similar to those applied in \cite{Annala:2019puf}:
\begin{enumerate}
    \item All EoSs must be able to support NSs of mass $2M_\odot$. The heaviest NS with an accurately known gravitational mass is the pulsar PSR~J0348$+$0432~\cite{Antoniadis:2013pzd} with 68\% error bars at $2.01\pm 0.04 M_\odot$, which prompted the use of the condition $\mTOV \geq 1.97M_\odot$  in \cite{Annala:2017llu,Annala:2019puf} and many other past works. However, the recent discovery of a NS with an even higher mass of $2.08 \pm 0.07 M_\odot$ (68\% credible interval)~\cite{Fonseca:2021wxt} motivates a conservative increase of this limit to $\mTOV \geq 2.0M_\odot$ \footnote{We will comment on the sensitivity of our results to the lower limit imposed on $\mTOV$ later.}. After imposing this constraint, approximately 500,000 EoSs survive.
    \item All EoSs must satisfy the LIGO/Virgo 90\% credible interval for the tidal deformability~$\Lambda$, inferred from the GW170817 event~\cite{TheLIGOScientific:2017qsa}. This bound is enforced through the (highest posterior density) binary-tidal-deformability constraint $\tilde{\Lambda} < 720$, corresponding to the low-spin priors ($|\chi| < 0.05$) of~\cite{Abbott:2018wiz}, instead of using the more approximate tidal deformability of a $1.4M_\odot$ NS as in~\cite{Annala:2019puf}. Note that the $\tilde\Lambda$ limit we adopt here is a little more conservative than the constraint $\tilde{\Lambda} < 700$ \edit{(symmetric credible interval)} reported in~\cite{De:2018uhw}. The binary-tidal-deformability parameter is defined as~\cite{TheLIGOScientific:2017qsa}
    \begin{equation}
        \tilde{\Lambda} \equiv \frac{16}{13} \frac{(M_1 + 12 M_2) M_1^4 \Lambda(M_1) + (M_2 + 12 M_1) M_2^4 \Lambda(M_2)}{(M_1+M_2)^5},
    \end{equation}
    where $M_1$ and $M_2$ are the gravitational masses of the two NSs,  with $M_2<M_1$
    and $\Lambda(M_1)$ and $\Lambda(M_2)$ their tidal deformabilities (imposing the same EoS), 
    determined as described in~\cite{Hinderer:2007mb}.
    Concretely, we discard a given EoS if there are no possible mass configurations, constrained by the mass ratio $q = M_2/M_1 > 0.73$ and the chirp mass $\Mch = 1.186 M_\odot$~\cite{Abbott:2018wiz}, that would lead to $\tilde{\Lambda} < 720$. After  this constraint, approx.~250,000 EoSs remain.
\end{enumerate}
The EoS family built with the above constraints is displayed in Fig.~\ref{fig:epMRNoSupra} (left), and is virtually indistinguishable from that of~\cite{Annala:2019puf}, showing that the choice of using $\Lambda(1.4 M_\odot)$ or $\tilde{\Lambda}$ does not significantly impact the allowed range of these quantities. In particular, the ``bump'' in the MR figure at large $R$ and $\mNonrot < 1.4 M_\odot$,
\edit{consisting of EoSs that approximate the maximally stiff causal EoS between the CET densities and those reached in the centers of $M\sim 1.4 M_\odot$ stars,
persists even with the $\tilde{\Lambda}$ constraint, although a recent study suggests that these NS solutions may be unstable~\cite{Jimenez:2021wil}. Finally, one should note that a limitation of our non-statistical approach is that we altogether discard solutions that fail to satisfy the above two constraints. Given the robustness of these measurements, it is very unlikely that the real NS EoS substantially deviates from the limits obtained, but our results for the absolute bounds on e.g.~$\mTOV$ or $|\chi|$ should nevertheless be interpreted with this fact in mind.}

\subsection{Neutron-star configurations}

\subsubsection{Non-rotating neutron stars}

To inspect the macroscopic properties of non-rotating NSs using our EoS ensemble, we solve the Tolman-Oppenheimer-Volkov (TOV) equations~\cite{1939PhRv...55..364T,Oppenheimer:1939ne} numerically, arriving at the MR cloud shown in Fig.~\ref{fig:epMRNoSupra} (right). Throughout the paper we use the symbol $\mNonrot$ to refer to the gravitational mass of a non-rotating star sequence, which ends at $\mTOV \equiv \max(\mNonrot)$. In this context, it should be noted that the speed-of-sound interpolation function used in our work allows for very extreme EoSs. One way to quantify this is the maximal speed of sound reached by a given EoS at any density. This value is indicated in Fig.~\ref{fig:epMRNoSupra} by color coding; in particular, equations of state that never exceed the conformal value of $c_s^2 = 1/3$ (approached at asymptotically high densities from below) are shown with the darkest shade of blue/green. 
The resulting MR cloud shown in Fig.~\ref{fig:epMRNoSupra} is again similar to its counterpart in~\cite{Annala:2019puf}. When imposing only the GW170817 tidal deformability constraint and the lower bound on the maximum mass of a NS, the radius of a 1.4$M_\odot$ star ranges from 9.6~km to 13.4~km and $\mTOV \lesssim 3.0M_\odot$. (See our final Tab.~\ref{tab:radranges} in Sec.~\ref{sec:conclusions} for a summary of the non-rotating MR clouds discussed in this work.)

\subsubsection{Rotating neutron stars}

As NSs are rigidly spun up, matter at their equators becomes less gravitationally bound. For a given gravitational mass $M$, a uniformly rotating NS in equilibrium can be spun up to a maximal frequency $\fKep(M)$, known as the Kepler frequency, before mass shedding from its equator occurs. The sequence of stars of gravitational mass $\mKep$ and rotating with maximal frequency $\fKep(\mKep)$ define the so-called Kepler sequence for the NS EoS, which ends at the maximal mass that can be supported by uniform rotation, 
denoted by $\mSupra \equiv \max(\mKep)$ for ``supramassive'' (throughout this paper, we use a subscript ``Kep'' to refer to properties along the Kepler sequence, and $\rKep$ shall always refer to the circumferential equatorial radius of stars along that sequence). 
For each EoS in our ensemble, we construct these 
Kepler sequences using the general relativistic code for hydrostationary stellar equilibrium configurations of~\cite{CST92,CST94a,CST94b}. To enable the computation of millions of Kepler sequences we developed an interface for executing the code of~\cite{CST92,CST94a,CST94b} across multiple processors. We take advantage of the fact that the Kepler sequence for any EoS is independent from that of another EoS. Therefore, the problem is embarrassingly parallelizable. Our parallelization interface executes the generation of general relativistic rotating star configurations on thousands of compute cores, allowing us to generate millions of Kepler sequences with our EoS ensemble within a few days.

NSs with gravitational masses in the range $\mTOV < M < \mSupra$ are called supramassive NSs (SMNS), while those whose mass exceeds $\mSupra$ are dubbed hypermassive~\cite{Baumgarte:1999cq}. Hypermassive NSs are able to support significantly larger gravitational mass than the TOV limit~\cite{Morrison:2004fp,Ansorg:2008pk,Espino:2019ebx,Bozzola:2019tit} because they are differentially rotating.
Both SMNSs and HMNSs eventually undergo collapse to a BH, when the star loses its excess angular-momentum support. For SMNSs this happens due to, e.g., magnetic braking on a typical timescale of the order of $1000\,\mathrm{s}$, while HMNSs can undergo collapse to a BH on a much shorter timescale, as short a few $\mathrm{ms}$. The aforementioned timescales are not precise and depend on unknown parameters (such as the star's magnetic field, outflows, spin, and internal structure)~\cite{Lehner:2014asa,Paschalidis:2016agf,Baiotti:2016qnr,Duez:2018jaf,Ciolfi:2018tal,Radice:2020ddv}.

In the case of GW170817, how close the immediate merger-product rest mass $\mRemno$ was to the TOV limit rest mass is critical: it significantly affects the timescale to collapse to a BH, which in turn can significantly influence the properties of the subsequent short gamma-ray burst and kilonova emission \cite{Ascenzi_2021}. The fact that collapse to a BH appears
to have occurred implies that the remnant rest mass $\mRemno$ was larger than a critical rest mass $\mCrito$. In particular, the immediate merger product of GW170817 must have had a rest mass at least above the TOV limit rest mass, i.e., $\mCrito=\mTOVo$. This in turn means that the immediate merger product was a SMNS at the very least. However, the fact that collapse to BH must have taken place on a timescale of order $1\,\mathrm{s}$ following merger, suggests that the immediate merger product could be an HMNS or a SMNS close to the supramassive limit. Several studies~\cite{Margalit:2017dij,Rezzolla:2017aly,Ruiz:2017due} assumed or argued that the remnant was an HMNS. Thus, we will additionally consider here $\mCrito=\mSuprao$.

\section{Results\label{sec:results}}

Continuing next to the main results of our work, we shall go through the outcomes of all the different analyses we have performed. We begin in Section~\ref{sec:kepler} by studying the properties of rapidly rotating NSs, displaying the Kepler sequences built with our EoS ensemble, and then introduce several new correlations between the properties of rotating and non-rotating stars (universal relations)  in Sec.~\ref{sec:universal}. In Section \ref{sec:gw170817}, we  systematically explore the EoS constraints under the BH formation hypothesis in connection with the GW170817 remnant, and in Sec.~\ref{sec:nicer} turn to the impact of the recent NICER observations on our EoS ensemble. Finally, in Sec.~\ref{sec:gw190814} we study the prospects for the secondary component in the GW190814 event having been a NS, and in Sec.~\ref{sec:future} detail how various future observational bounds on NS properties would impact the allowed region of the NS-matter EoS.

\subsection{Kepler sequences\label{sec:kepler}}

\begin{figure}
    \centering
    \includegraphics[height=6cm]{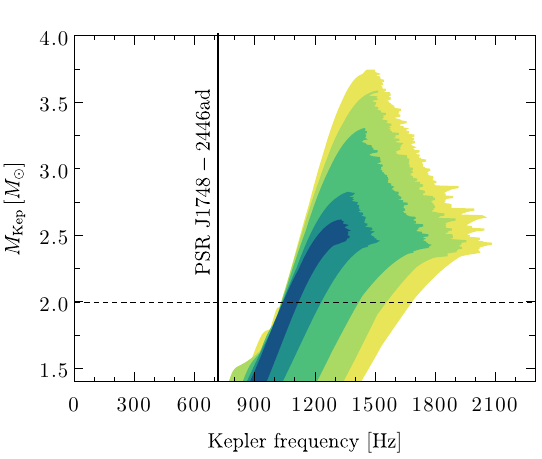}
    $\quad$
    \includegraphics[height=6cm]{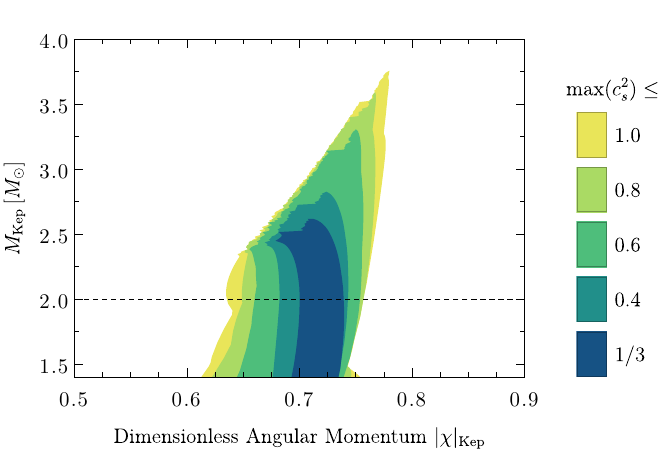}
    \caption{
    Gravitational masses of NSs on the Kepler sequence $\mKep$ as functions of the spin frequency $\fKep$ (left) and the dimensionless spin $\JMs_{\rm Kep}$ (right). In the left figure, the vertical line represents the fastest-spinning known pulsar PSR J1748$-$2446ad. The color coding follows that introduced in Fig.~\ref{fig:epMRNoSupra}, and the ensemble is the same as in that figure.
    }
    \label{fig:fKepNoSupra}
\end{figure}

We begin by discussing some results that follow from the constructed Kepler sequences. An observation of a highly spinning NS with precise (gravitational) mass measurement $M$ could be used to directly exclude any EoS for which $\fKep(M)$ is below the observed spin frequency. The Kepler sequences corresponding to the EoSs in our ensemble are shown in Fig.~\ref{fig:fKepNoSupra} (left), from which we see that all calculated Kepler frequencies for stars with masses $M > 1.4 M_\odot$ lie in excess of 716 Hz --- the frequency of the fastest-spinning known NS, PSR~J1748$-$2446ad~\cite{Hessels:2006ze}. To this end, rapidly spinning NSs do not currently place any additional constraints on the EoS for realistic NS masses, though we find that a future measurement of a rotation frequency $f(1.4M_\odot) > 775$~Hz would eliminate some allowed EoSs.

\begin{figure}
    \centering
    \includegraphics[height=8cm]{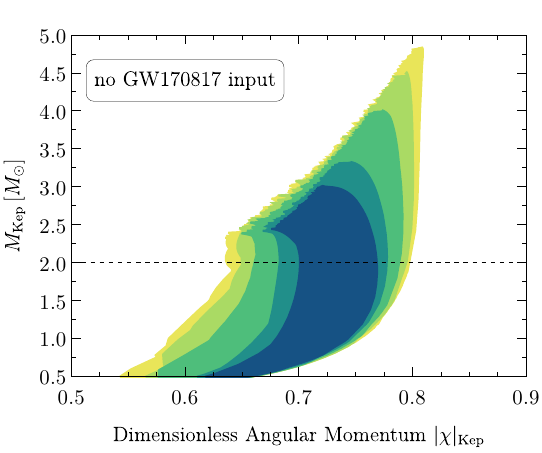}
    \caption{ \edit{The relation between gravitational masses of NSs on the Kepler sequence $\mKep$ and the dimensionless spin $\JMs_{\rm Kep}$, for the ensemble of EoSs without any input from GW170817. We provide a fit of the right boundary of this region in Eq.~\eqref{chimaxvsM}.}
    }
    \label{fig:Mchi_noGW170817}
\end{figure}

In Fig.~\ref{fig:fKepNoSupra} (right) we also display the dimensionless spin parameter along the Kepler sequences. We find that the maximal dimensionless spin that stable stars can support always satisfies $\JMs \leq 0.78$. This conclusion is, however, affected by input from GW170817, which we have implemented here by assuming the low-spin priors. Our analysis without any input from GW170817, shows that all valid EoSs satisfy $\JMs \leq 0.81$.  Furthermore, we find that the maximum possible value of $\JMs$ depends smoothly on the gravitational mass of the star ranging from $0.66$ to $0.81$ for gravitational masses in the range $0.5M_\odot$ to $4.8M_\odot$. For this reason, we also fitted the maximum allowed $\JMs_{\rm max}$ vs $M$ \edit{without any constraint from GW170817 (shown in Fig.~\ref{fig:Mchi_noGW170817})}, and found the following fitting function to be accurate to within 1 part in $10^3$
\begin{equation}\label{chimaxvsM}
    \JMs_{\rm max}( \mKep ) = 0.128502 \left(\frac{\mKep}{M_\odot}\right)-7.97109 \left(\frac{\mKep}{M_\odot}\right)^{1/3}+10.5018 \left(\frac{\mKep}{M_\odot}\right)^{1/4}-1.90573,
\end{equation}
%

where the fit is valid for $\mKep\in[0.5 M_\odot,4.8 M_\odot]$. Our upper limit $\JMs \leq 0.81$
justifies \emph{a posteriori} the choice $\JMs < 0.89$  used by the LIGO/Virgo collaboration
for high-spin priors (for technical reasons\edit{, imposed by the lack of rapid waveform models})~\cite{TheLIGOScientific:2017qsa}, but we recommend that $\JMs \leq 0.81$ or even better Eq.~\eqref{chimaxvsM} should be used in future analyses. Using Eq.~\eqref{chimaxvsM} should provide more stringent bounds on the binary NS parameters than when using a single value as an upper bound, because of degeneracies with high spin. We note here that we have verified that our bounds on $\JMs$ are insensitive to the exact choice of the minimum $\mTOV$ that we allow for our ensemble; in particular, taking $\mTOV > 1.9 M_\odot$ does not affect the reported values.

\subsection{New universal relations}
\label{sec:universal}

\begin{figure}
    \centering
    \includegraphics[height=6cm]{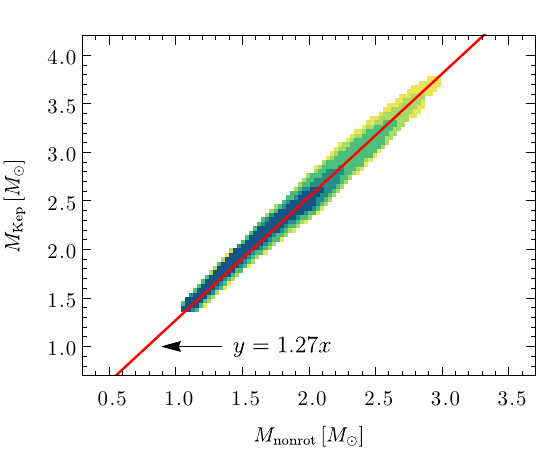}
    $\quad\;\;$
    \includegraphics[height=6cm]{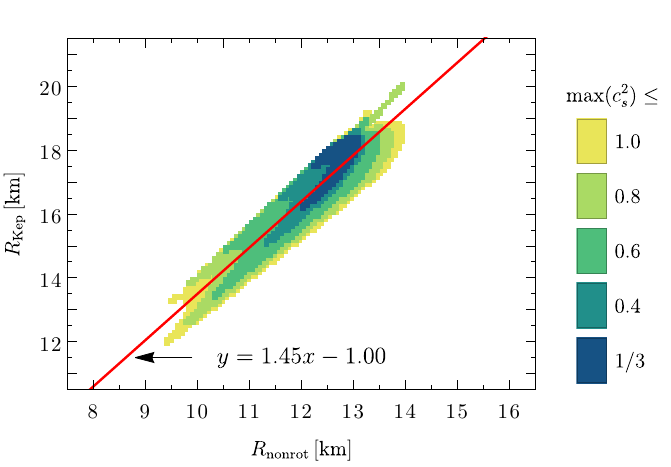} \\
    \includegraphics[height=6cm]{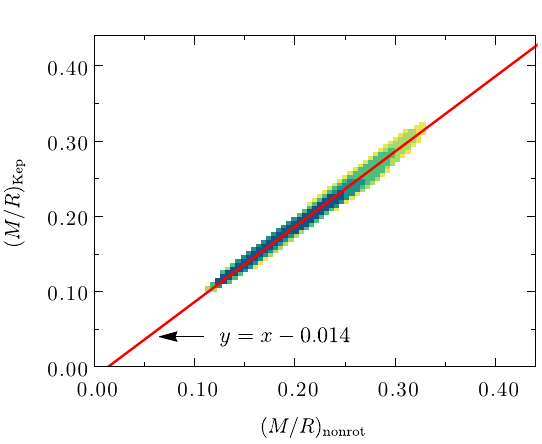}
    $\quad$
    \includegraphics[height=6cm]{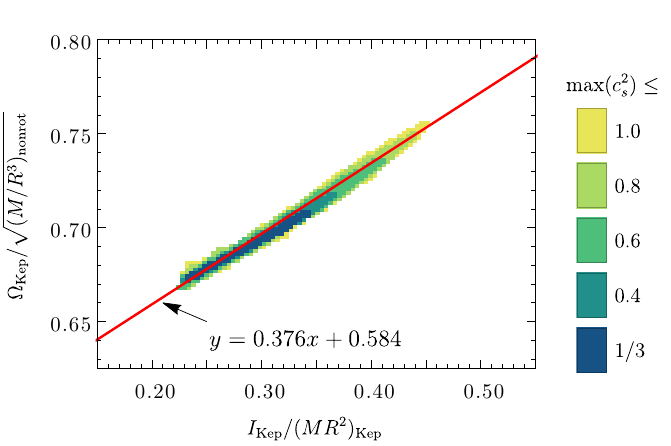} \\
    \caption{New universal relations between non-rotating and maximally rigidly rotating (Kepler) NS parameters, always comparing stars with the same central energy density. Note that only $\mTOV > 2.0 M_\odot$ and $\tilde{\Lambda}_{\rm GW170817} < 720$ have been assumed here.}
    \label{fig:new_univ_rel}
\end{figure}

By studying correlations between different physical quantities involving both rotating and non-rotating NSs constructed from our full EoS ensemble, we have discovered a number of previously unknown relations that appear to be at least approximately universal (i.e., they are followed by all of the EoSs in our ensemble). All of these relations relate quantities between rotating and non-rotating stars with the same central energy density, and the ensemble used in the fits only imposes the constraints of $\mTOV > 2.0 M_\odot$ and $\tilde{\Lambda}_{\rm GW170817} < 720$. They are displayed in Fig.~\ref{fig:new_univ_rel} and include:
    \begin{itemize}
    \item Upper row, left: A tight correlation between the non-rotating and Kepler masses, i.e.,~$\mNonrot$ and $\mKep$, holding better than to 9.3\%. We note that the correlation is different from the $\mTOV$--$\mSupra$ relation discussed in the literature, where the maximum mass configurations of the non-rotating and Kepler limits are compared~\cite{Breu:2016ufb}.
    \item Upper row, right: Another tight relation between the non-rotating and Kepler (circumferential equatorial) radii, $\rNonrot$ and $\rKep$, holding better than to 7.7\%. We point out that the extended line along the upper edge arises from the EoSs in the low-mass ``bump'' in the MR figure of \fig\ref{fig:epMRNoSupra} (those having the large $\rNonrot$ and $\mNonrot < 1.4 M_\odot$). 
    \item Lower row, left: The above correlations also imply  a correlation of the compactness of the rotating and non-rotating stars, for which the relation $\mNonrot/\rNonrot \approx \mKep/\rKep$ holds better than 17\%. The relation we list in \fig\ref{fig:new_univ_rel}, including a small offset, holds better than 8.0\%. We also find that $M/R < 0.33$ for all stellar models.
    \item Lower row, right: A strong correlation between the dimensionless angular Kepler frequency $\omegaKep \equiv 2 \pi \fKep$ and the dimensionless Kepler moment of inertia, defined as $\omegaKep/\sqrt{(M/R^3)_{\rm nonrot}}$ and $I_{\rm Kep}/(M R^2)_\text{Kep}$. This holds better than to 1.4\%.  We point out here that $\omegaKep/\sqrt{(M/R^3)_{\rm nonrot}} \approx 0.71\pm 0.04$, agreeing with previous similar expressions (see~\cite{Paschalidis:2016vmz} and references therein).
\end{itemize}

\subsection{Impact of the GW170817 BH formation hypothesis}
\label{sec:gw170817}

Based on the properties of the electromagnetic counterpart of GW170817, several works have argued that the merger remnant of the event underwent collapse to a BH, which we refer to as the BH formation hypothesis. The work of~\cite{Bauswein:2017vtn} reported constraints on the EoS by further assuming that the remnant underwent delayed (as opposed to prompt) collapse to BH. As mentioned earlier, for the immediate merger product to undergo collapse to BH, it must be at least a SMNS, i.e., $\mRemno \geq \mCrito = \mTOVo$. However, the TOV limit is the most conservative limit one can consider for $\mCrito$. 
Some works have argued that prior to the formation of the BH, the remnant was a HMNS  \cite{Margalit:2017dij,Ruiz:2017due,Rezzolla:2017aly}, and that it collapsed at an angular momentum above or near the supramassive limit. In this case, $\mRemno \geq \mCrito = \mSuprao$. In this section, we analyze the implications of these two limiting values of $\mCrito$. While reconciling the value $\mCrito = \mTOVo$ with the multimessenger picture of GW170817 is challenging, it provides the most conservative constraints on the EoS, and allows us to bracket the possible constraints on the EoS using mass information from GW170817. Moreover, as we show below, it leads to a substantially tighter constraint on the maximum TOV mass than the straightforward argument that the GW170817 total gravitational mass be greater than the TOV limit mass.

In the analysis of the merger, it is important to note that during the inspiral and merger, the gravitational mass is not a conserved quantity, since a non-negligible part of it is radiated away in GWs. By contrast, the total number of baryons is conserved, except for the ejection of a relatively small amount of matter necessary to power the observed kilonova \cite{Coulter:2017wya,Drout:2017ijr,Shappee:2017zly,Kasliwal:2017ngb, Tanaka:2017qxj,Arcavi:2017xiz,Pian:2017gtc,Smartt:2017fuw,Soares-Santos:2017lru, Nicholl:2017ahq,Cowperthwaite:2017dyu}. Therefore, to perform a proper analysis of the nature of the binary merger remnant, we must sum the total baryon masses in the initial NSs $M_{\rm 1, B}$ and $M_{\rm 2, B}$ to obtain the final baryon mass in the merger remnant $\mRemno$. To obtain the most conservative limits for the allowed NS EoSs, we may ignore the ejecta powering electromagnetic counterparts altogether, for this allows the remnant to be as massive as possible.

\begin{figure}[t]
      \centering
        \includegraphics[width=0.93\textwidth]{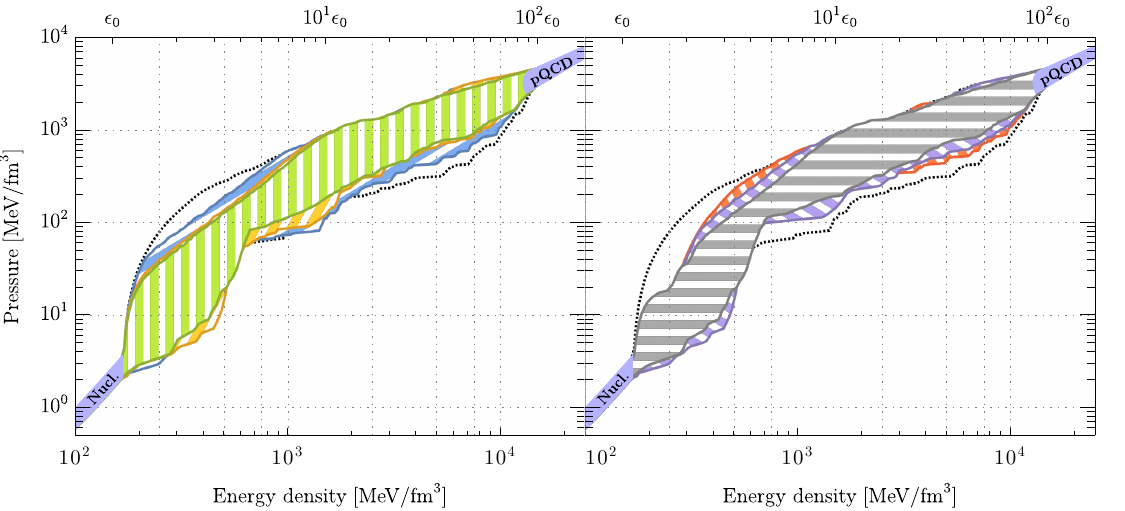}
        \includegraphics[width=0.96\textwidth]{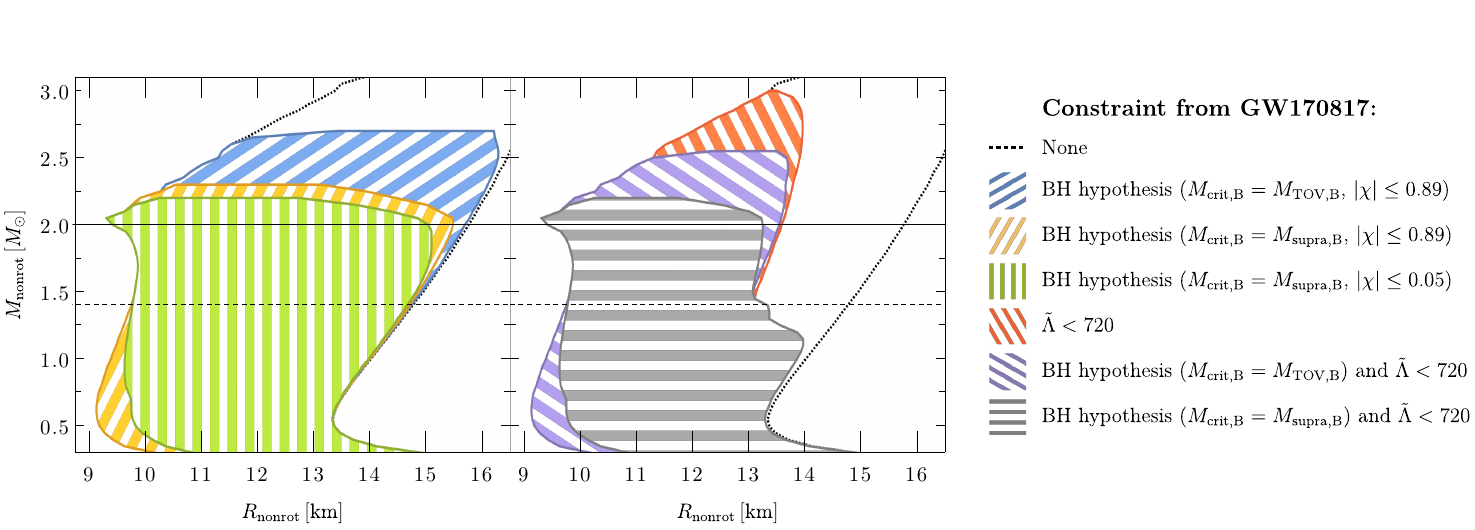}
    \caption{Upper row: the regions of allowed EoSs obtained under the different remnant assumptions discussed in the main text. The left (right) panel shows results without (with) the additional $\tilde{\Lambda}$ constraint. Note that all results corresponding to the $\tilde{\Lambda}$ constraint use the low-spin priors from \cite{Abbott:2018wiz}.  Lower row: the allowed MR regions following from the same assumptions. In all figures, the boundaries of the partially obscured regions follow the boundaries of the regions that obscure them.} 
    \label{fig:different_GW170817_implementations}
\end{figure}

We relegate the technical details of our analysis to Appendix \ref{app:BH_form_hyp}, and only point out here that the GW signal tightly constrains only the chirp mass $\Mch$, and provides a lower bound for the mass ratio $q = M_2/M_1$ (or equivalently the mass range of the one of the components). The value of $\Mch$ and range of masses for the primary are consistent with a range of binary component masses, and thus a range of baryon masses for the components $M_{\rm 1,B}$ and $M_{\rm 2,B}$ and the remnant $\mRemno = M_{\rm 1,B} + M_{\rm 2,B}$ for any given EoS. For our analysis, we construct all trial binaries consistent with the GW170817 mass parameters, and take the largest possible remnant baryon mass (arising from the most asymmetric configurations), again leading to the most conservative bounds. In this way, we can determine for a given EoS whether there exist any trial binary masses ($q$ values) consistent with GW170817 that satisfy $\mRemno > \mCrito$. If not, then this EoS is discarded.

\begin{figure}[t]
    \centering
       \includegraphics[height=6cm]{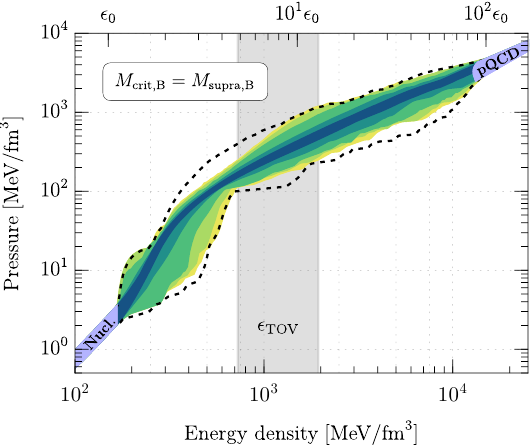}
       $\quad$
       \includegraphics[height=6cm]{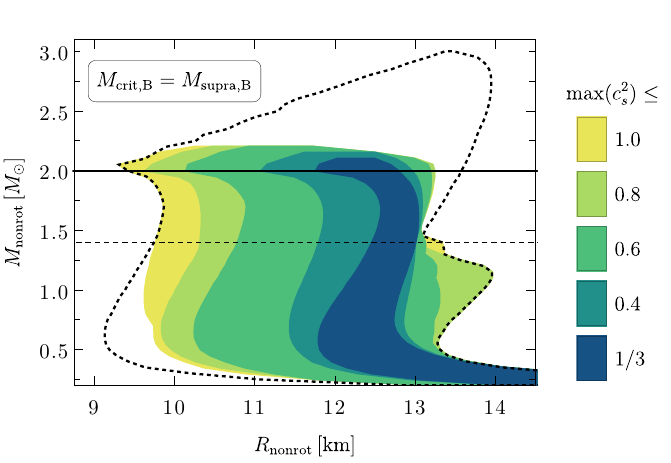}
       \medskip
       \includegraphics[height=6cm]{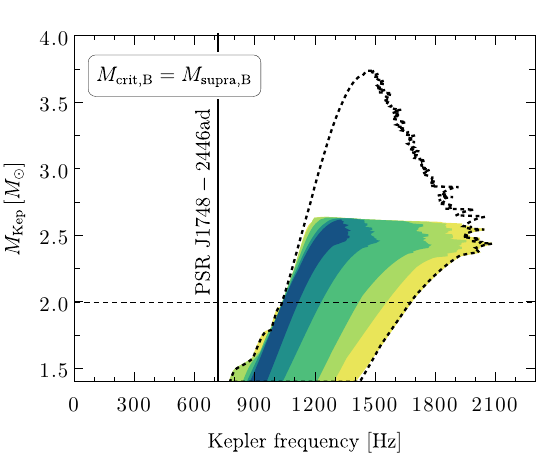}
       $\quad$
       \includegraphics[height=6cm]{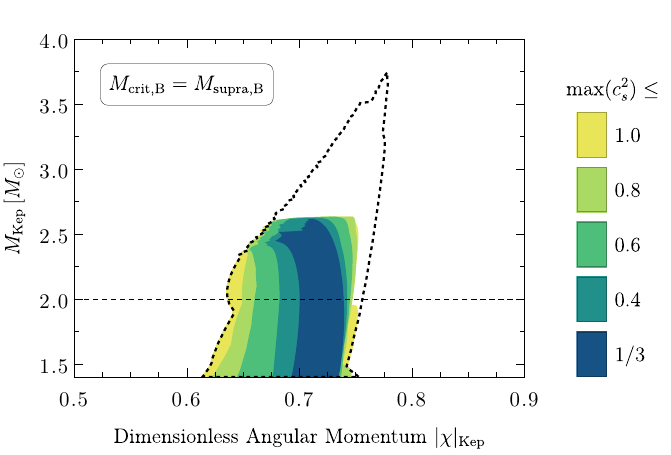}
    \caption{Upper row: the regions of allowed EoSs and (non-rotating) MR relations, obtained after implementing the BH formation hypothesis with $\mCrito=\mSuprao$ and demanding $\mTOV\geq 2.0M_\odot$ and $\tilde \Lambda_{\rm{GW170817}}<720$. Lower row: the same as above for the Kepler frequencies $\fKep(M)$ and the NS masses as a function of the dimensionless angular momentum $\JMs_{\rm Kep}$. In all panels, the dotted regions indicate the extent of the regions without the BH formation hypothesis (cf. Figs.~\ref{fig:epMRNoSupra}~and~\ref{fig:fKepNoSupra}).
    } 
    \label{fig:allowed_with_supra_ranges2}
\end{figure}

In \fig\ref{fig:different_GW170817_implementations}, we show how the value of $\mCrito$ affects the allowed EoS and MR spaces for both the high- and low-spin priors for GW170817 \cite{Abbott:2018wiz}. Where appropriate, we also display the joint results obtained by combining these hypotheses with the tidal deformability constraint $\tilde{\Lambda} < 720$, ensuring that they are both simultaneously satisfied for at least one of the trial binaries consistent with GW170817. Since the theory of NS tidal deformability only exists for slowly rotating stars, we do not impose a tidal deformability constraint paired with the high-spin priors. We note that the conservative $\mCrito=\mTOVo$ value provides the limit $\mTOV < 2.57 M_\odot$ (low-spin prior), or $\mTOV < 2.53 M_\odot$ together with $\tilde{\Lambda} < 720$, which should be considered the most robust upper limits on the TOV limit mass consistent with the GW170817 event in conjunction with the BH formation hypothesis and the low-spin prior analyses. If instead the GW170817 high-spin priors are adopted (without any tidal deformability bounds imposed), for $\mCrito=\mTOVo$ the upper bound on the TOV limit is  $2.69M_\odot$ as shown by the blue shaded region in the bottom left panel in Fig.~\ref{fig:different_GW170817_implementations}. As shown in the same figure, when $\mCrito=\mSuprao$ instead is adopted the upper limit on the maximum TOV mass is  $2.25M_\odot$ in conjunction with the GW170817 high-spin priors. These constraints are the tightest upper limits when $\mCrito=\mSuprao$ is adopted. We will return our discussion to constraints arising from the case $\mCrito=\mSuprao$ below.

Notice also that some works have used the idea that the GW170817 total gravitational mass (in conjunction with the low-spin priors) must be greater than a critical mass to constrain the TOV maximum mass. For the low-spin priors the GW170817 total gravitational mass is $M_{\rm GW170817}=2.73_{-0.01}^{+0.04}M_\odot$ (90\% credible interval)~\cite{Abbott:2018wiz}, so the straightforward argument $M_{\rm GW170817}\geq \mTOV$ implies $\mTOV\lesssim 2.7M_\odot$. This is less tight than our bound set by $\mCrito=\mTOVo$ because it does not account for the fact that it is the rest mass (baryon number) that is conserved,  not the gravitational mass. \edit{In Ref.~\cite{Rezzolla:2017aly}, however, the authors use the conservation of rest mass, the inferred total gravitational mass, and quasi-universal relations to derive the constraint $\mTOV \lesssim 2.59 M_{\odot}$. This result is compatible with our values derived above without the use of quasi-universal relations, and using the extremely well-measured chirp mass.}

Any assumption on the remnant is seen to constrain the EoS ensemble at all densities. In particular, as these hypotheses involve the properties of stars at larger central densities than those reached in two-solar-mass NSs, they can affect the EoS in the high-density region that is otherwise only constrained by the theoretical pQCD limit. The upper panels of Fig.~\ref{fig:different_GW170817_implementations} demonstrate that both hypotheses effectively exclude EoSs that remain very stiff (i.e.\ those which have larger values of $\gamma \equiv {\rm d}(\ln~p)/ {\rm d}(\ln~\epsilon)$) all the way to energy densities of order $\epsilon \lesssim 10^{3}$ MeV/fm$^3$, and impact densities higher than those affected by the $\tilde{\Lambda}$ constraint alone. This can be understood by noting that the EoSs populating the upper boundary of our EoS band support very massive rigidly rotating stars, for which the rest mass of the GW170817 remnant does not exceed $\mTOVo$ or $\mSuprao$.

\begin{figure}[!t]
    \centering
       \includegraphics[width=0.97\textwidth]{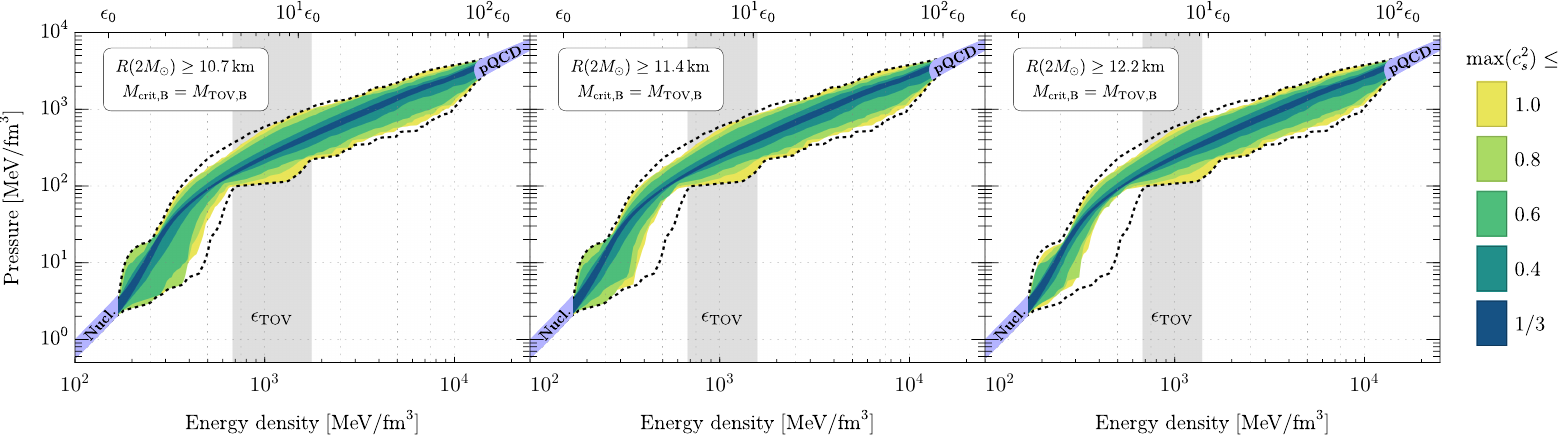} \\
     \vspace{4mm}
       \includegraphics[width=0.97\textwidth]{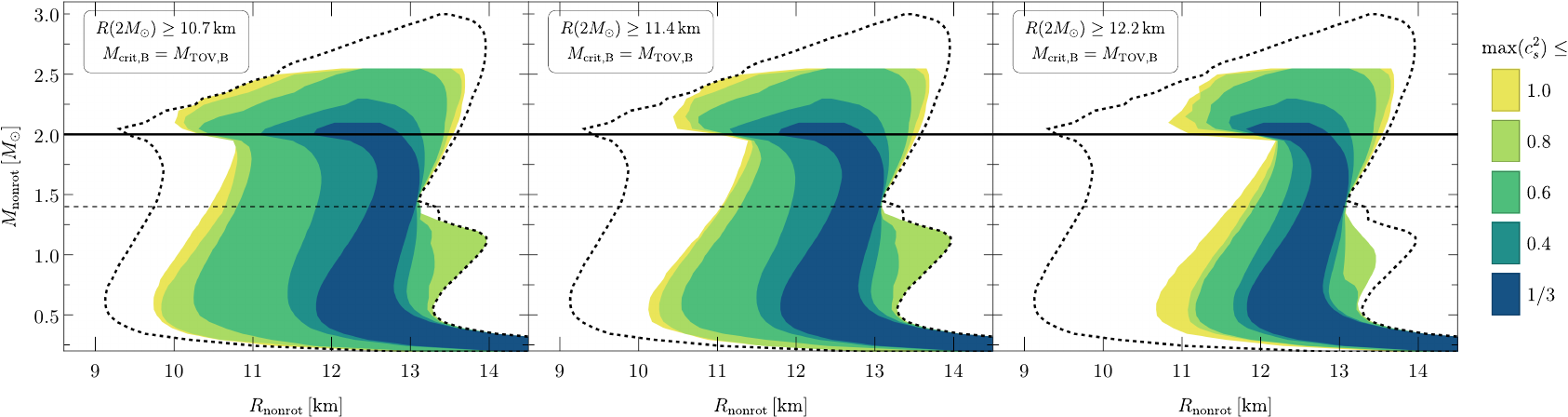} \\
    \caption{The impact on the allowed EoS band and MR region from assuming radius measurements for PSR J0740+6620, implemented as $\rNonrot(2M_\odot) \geq \edit{10.7}$~km (left), $\rNonrot(2M_\odot) \geq 11.4$~km (middle), $\rNonrot(2M_\odot) \geq 12.2$~km (right). Here, we assume the GW170817 BH formation hypothesis with $\mCrito=\mTOVo$ and the constraints $\tilde{\Lambda}_{\rm{GW170817}} < 720$ and $\mTOV\geq 2.0M_\odot$. Dotted lines correspond to the ensemble given in Fig.~\ref{fig:epMRNoSupra}.} 
    \label{fig:NICERconstraintsTOV}
\end{figure}

Next, let us consider in more detail 
the most restrictive pair of constraints, namely $\mCrito=\mSuprao$ (with low-spin priors) paired with the $\tilde{\Lambda} < 720$ condition. The effects of these constraints are shown in  \fig\ref{fig:allowed_with_supra_ranges2}, displayed differentially in $\max(c_s^2)$. The right panel in the upper row shows the effect of the hypothesis on the MR relation: In agreement with previous works finding upper bounds for $\mTOV$ in the range 2.16--2.32$M_{\odot}$ \cite{Margalit:2017dij,Rezzolla:2017aly,Ruiz:2017due}, we observe that $\mCrito=\mSuprao$ leads to the condition $\mTOV \leq 2.19 M_{\odot}$ (or $\mTOV \leq 2.22 M_{\odot}$ without the tidal deformability constraint and $\mTOV \leq 2.30 M_{\odot}$ with high-spin priors). We emphasize, however, that the absolute upper bounds of the earlier works in conjunction with $\mCrito=\mSuprao$ exceed ours, making our result the most stringent upper bound derived. In addition, all earlier works either considered a small set of EoSs or relied on universal relations observed for a limited set of model EoSs. In contrast, our result is based on studying all possible NS-matter EoSs, making them considerably more robust in comparison. We have once again verified that the maximum mass bounds reported in the above paragraphs are insensitive to decreasing the $\mTOV$ lower bound used in our analysis to $\mTOV > 1.9 M_\odot$.

In the lower row of Fig.~\ref{fig:allowed_with_supra_ranges2}, we reproduce the frequency figure of Fig.~\ref{fig:fKepNoSupra}, under the BH formation hypothesis with $\mCrito=\mSuprao$. The effect of the effective cut on maximal mass is clearly visible here, in addition to which we find a subdominant effect at lower NS masses. Finally, we briefly note that we have reproduced all the main results of \cite{Annala:2019puf} concerning the existence of quark-matter cores in massive NSs under the BH formation hypothesis and $\mCrito=\mSuprao$, but have found the main conclusions of that work to remain unaltered.

\subsection{Impact of radius measurements of PSR J0740+6620 and other NSs}\label{sec:nicer}

We turn now to considering the impact of current and future radius measurements on the NS-matter EoS within our model-independent framework. We mainly impose these constraints as lower limits for the radius of stars with various masses, recalling that the upper limits we obtain for radii from our ensemble are typically more restrictive than those associated with various individual radius measurements. 

\begin{figure}[!t]
    \centering
       \includegraphics[width=0.97\textwidth]{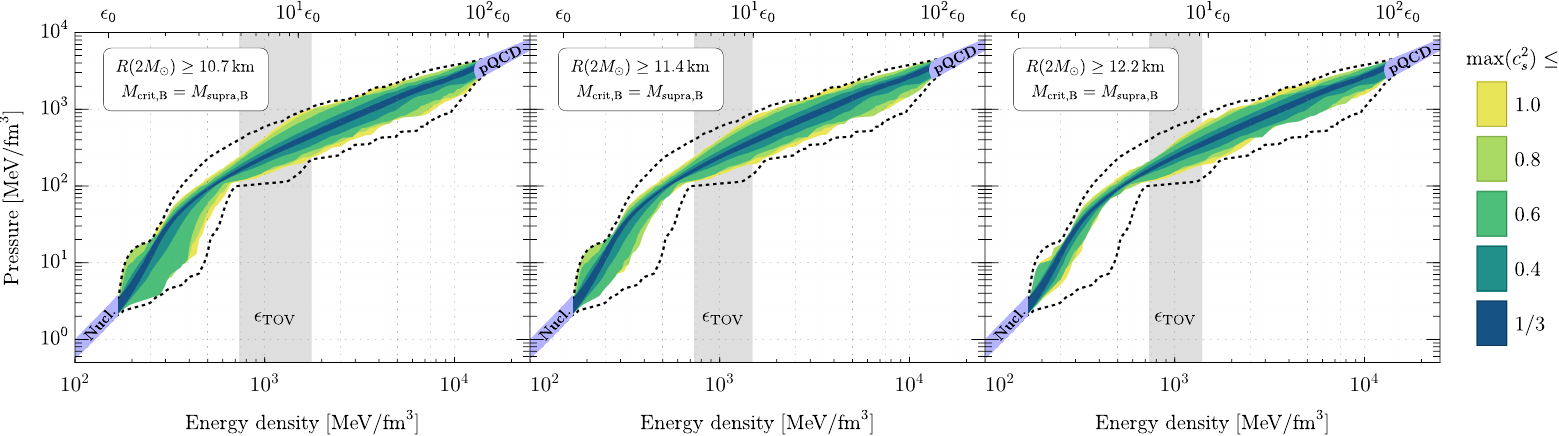} \\
     \vspace{4mm}
       \includegraphics[width=0.97\textwidth]{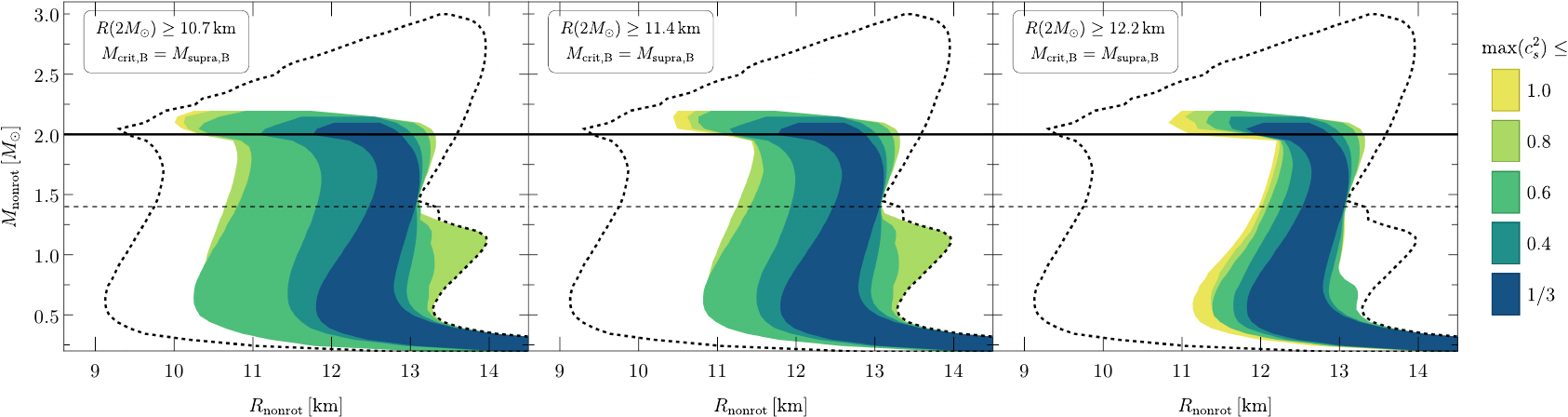} \\
    \caption{Same as \fig~\ref{fig:NICERconstraintsTOV}, but instead assuming the GW170817 BH formation hypothesis with $\mCrito=\mSuprao$.}
    \label{fig:NICERconstraintsHMNS}
\end{figure}

In this context, a particularly interesting result is the recent radius measurement of the most massive NS currently known, PSR J0740+6620, for which the NICER collaboration has presented two independent analyses \cite{Miller:2021qha,Riley:2021pdl}. 
We consider three lower limits for the radius of this star, \edit{10.7}, 11.4, and 12.2~km, which represent the minimum values of the 95\% and 68\% credible intervals of \edit{Ref.}~\cite{Riley:2021pdl} and the 68\% interval of \edit{Ref.}~\cite{Miller:2021qha}, respectively (note that a fourth lower limit of 11.1~km was already analyzed in Fig.~\ref{fig:NICERconstraintsTOV0onePanel}). 
\edit{These are by no means the only possible choices for the radius limits; 
we have chosen them merely as representative examples that have approximately equidistant spacing and correspond to lower limits (in decreasing order of confidence) resulting from usage of various analysis techniques.}
For the mass, we make the conservative  choice \footnote{Using a higher value for the mass of PSR J0740+6620 would result in a more constrained EoS ensemble, mainly because some EoSs do not support NSs of such a higher mass. For this reason, a more conservative choice is using $M = 2M_{\odot}$ as the mass parameter.}
of imposing these lower limits assuming $M = 2M_{\odot}$, approximately corresponding to  the 68\% lower limit of the measured mass of PSR J0740+6620~\cite{Cromartie:2019kug,Fonseca:2021wxt} (and the value of $\mTOV$ we impose in our analysis), and use non-rotating stellar configurations \footnote{As shown in \cite{Riley:2021pdl,Miller:2021qha}, modeling PSR J0740+6620 as non-rotating induces at most an error of 0.2~km in radius for the stiffest EoSs, with softer EoSs showing smaller errors of 0.05~km. Since the softer EoSs are those which are constrained by a lower bound on the radius, this is not a large effect with our implementation.} 
to analyze these constraints. Figs.~\ref{fig:NICERconstraintsTOV} and \ref{fig:NICERconstraintsHMNS} display the allowed EoSs and MR relations for these constraints, implemented as cuts on $\rNonrot$, with the two different BH formations hypotheses corresponding to $\mCrito = \mTOVo$ and $\mCrito = \mSuprao$, respectively. We also use the constraint $\tilde{\Lambda} < 720$ from the low-spin priors in all panels. It is worth noting that the joint mass-radius measurement of the NS in the 4U~1702$-$429 system~\cite{Nattila:2017wtj} is also fully consistent with all of our MR clouds.  Given the robustness of the assumptions that go into the results depicted in Fig.~\ref{fig:NICERconstraintsTOV0onePanel}, corresponding to $\rNonrot(2M_\odot)\geq 11.1$~km and  $\mCrito = \mTOVo$, this EoS ensemble is the one we recommend using in future applications of our results. In particular, we shall use the constraint $\rNonrot(2M_\odot) \geq 11.1$~km below when discussing possible future measurements.

As one can readily observe from Figs.~\ref{fig:NICERconstraintsTOV} and \ref{fig:NICERconstraintsHMNS}, setting hard lower limits for the radius of the massive PSR J0740+6620 has a significant effect in ruling out EoSs that are soft at low densities, i.e., those that feature smaller values of $\gamma$.  It is also worth noting that an increased lower limit for $\rNonrot$ rules out a large number of EoSs, for which the maximum value of $c_s^2$ is above 0.5. Given that the analysis of \cite{Annala:2019puf} indicated the presence of QM cores in TOV stars built from EoSs for which max$(c_s^2)\lesssim 0.7$ \footnote{In \cite{Annala:2019puf}, the onset of QM was defined to take place at the lowest density from which $\gamma$ remains below 1.75 all the way to asymptotically larger densities, where it approaches 1.}, having sizable radii for massive NSs can be seen to {be compatible with the presence of deconfined quark matter in massive NSs. As demonstrated in the rightmost column of \fig\ref{fig:NICERconstraintsHMNS}, particularly tight constraints arise in the EoS and MR clouds} under the simultaneous assumption of the GW170817 BH formation hypothesis with $\mCrito = \mSuprao$, and the highest radius limit $\rNonrot(2M_\odot) \geq 12.2$~km.  In this case, the central energy densities of the TOV limit stars are found to lie inside an interval between \edit{725 and 1390~MeV/fm$^3$}, placing them clearly above the ``kink'' in the EoS band \edit{in the upper-right panel of \fig\ref{fig:NICERconstraintsHMNS}} (located around 400--500~MeV/fm$^3$), where the EoSs qualitatively soften. Additionally, we find that the corresponding range of central energy densities for $2M_\odot$ stars is 470--970 MeV/fm$^3$. If we associate the kink structure with the deconfinement phase transition (see \cite{Annala:2019puf} for a discussion), these results indicate that massive stars such as PSR J0740+6620 and PSR J0348+0432 may well contain sizeable QM cores. \edit{Finally, we note in passing that it is curious how close our obtained lower limits for $R(1.6M_\odot)$ (see Tab.~\ref{tab:radranges} below in Sec.~\ref{sec:conclusions}) are to the results derived in \cite{Bauswein:2017vtn} using the likely absence of a prompt gravitational collapse in the GW170817 merger event. The fact that two completely different lines of reasoning lead to the same conclusion clearly highlights the robustness of this result.}

\begin{figure}[!t]
    \centering
        \includegraphics[width=0.40\textwidth]{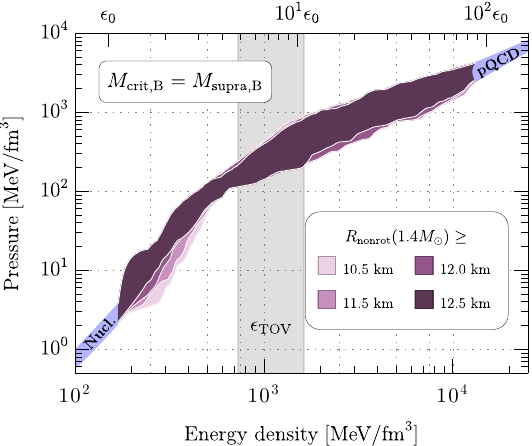}
        $\quad$
        \includegraphics[width=0.40\textwidth]{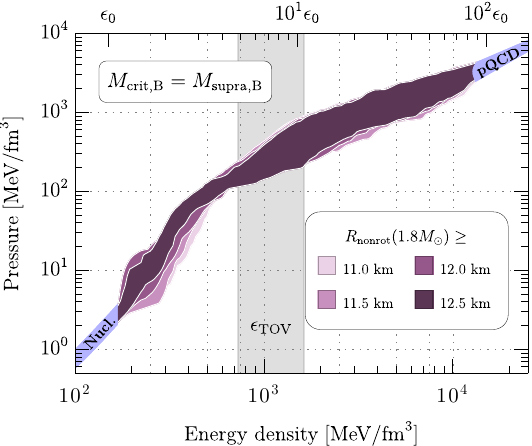}
    \caption{The effect of possible future radius measurements on the NS-matter EoS (slowly rotation stars, GW170817 BH formation hypothesis with $\mCrito = \mSuprao$ and the constraint $\rNonrot(2M_\odot) \geq 11.1$~km assumed). Here the new color scheme denotes various differential measurements.} 
    
    \label{fig:fut_meas_radii2}
\end{figure}

We additionally consider the effects of future radius measurements setting a lower bound on the radii of slowly rotating NSs of masses $1.4M_\odot$ or $1.8M_\odot$. Results from this analysis are shown in Fig.~\ref{fig:fut_meas_radii2}, where we additionally impose the BH formation hypothesis with $\mCrito = \mSuprao$ and the constraint $\rNonrot(2M_\odot) \geq 11.1$~km. Again, we see that in both cases the dominant effect occurs at low densities, where lower bounds for the radius exceeding 11~km begin to efficiently rule out EoSs that are soft at low densities. In the case of more massive stars, a lower limit for the radius is in addition seen to cut off EoSs that are very stiff at low densities. We conclude that individual radius observations with the highest potential to constrain the NS-matter EoS come from massive NSs, in agreement with~\cite{Weih:2019rzo}.

\begin{figure}
    \centering
    \includegraphics{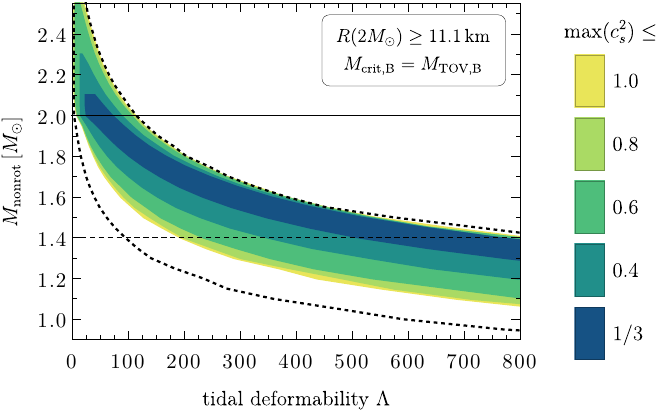}
    \caption{The range of allowed tidal deformabilities $\Lambda$ as a function of mass, assuming radius measurements for PSR J0740+6620, implemented as $\rNonrot(2M_\odot) \geq 11.1$~km. Here, we assume the GW170817 BH formation hypothesis with $\mCrito=\mTOVo$ and the constraints $\tilde{\Lambda}_{\rm{GW170817}} < 720$ and $\mTOV\geq 2.0M_\odot$. Dotted lines correspond to the ensemble given in Fig.~\ref{fig:epMRNoSupra}.}
    \label{fig:minLambda}
\end{figure}

\edit{Finally, i}n addition to the maximum spin for a given NS mass, we have also computed the minimum tidal deformability as a function of mass. Our full ranges of allowed tidal deformabilities as a function of mass are shown in \fig\ref{fig:minLambda}. If we do not fold in any information from the NICER observations (note that the $\tilde{\Lambda}$ constraint from GW170817 has no impact on the minimal tidal deformability), we find that minimum tidal deformability for a given $M = \mNonrot$ satisfies
\begin{equation}\label{eq:noLVCLambdammin}
    \log\left(\Lambda_{\rm min}(M)\right)=
    \begin{dcases}
-6.478 \log ^3 \biggl( \frac{M}{M_\odot} \biggr)-2.8228 \log ^2\biggl( \frac{M}{M_\odot} \biggr)-4.9717 \log \biggl( \frac{M}{M_\odot} \biggr)+2.7551, & 0.5 < \frac{M}{M_\odot} <1.86, \\
 -\frac{1.57765}{\log ^2\left( \frac{M}{M_\odot} \right)}+\frac{12.492}{\log \left( \frac{M}{M_\odot} \right)}-23.543, & 1.86\leq \frac{M}{M_\odot} <2.02, \\
 \log(2.84), & \frac{M}{M_\odot} \geq 2.02, 
    \end{dcases}
\end{equation}
with log denoting the base-10 logarithm. For $M=1.35 M_\odot$, this expression predicts $\Lambda_{\rm min}=110.9$.

\edit{
If we impose the GW170817 BH formation hypothesis with $\mCrito=\mTOVo$ and $\rNonrot(2M_\odot) \geq 10.7$ km, then $\Lambda_{\rm min}$ satisfies 
\begin{equation}
    \log\left(\Lambda_{\rm min}(M)\right)=
    \begin{dcases}
-7.1619 \log ^3\biggl( \frac{M}{M_\odot} \biggr)-2.6244 \log ^2\biggl( \frac{M}{M_\odot} \biggr)-4.6969 \log \left( \frac{M}{M_\odot} \right)+2.9507, & 0.5 < \frac{M}{M_\odot} <1.94, \\
 15.1833\, +\frac{1.82846}{\log ^2\left( \frac{M}{M_\odot} \right)}-\frac{10.3802}{\log \left( \frac{M}{M_\odot} \right)}, & 1.94\leq \frac{M}{M_\odot} <2.225, \\
 \log(2.84), & \frac{M}{M_\odot} \geq 2.225. 
    \end{dcases}
\end{equation}
For $M=1.35 M_\odot$, we now obtain $\Lambda_{\rm min}=189.7$.
}

Finally, if we impose the GW170817 BH formation hypothesis with $\mCrito=\mTOVo$ and $\rNonrot(2M_\odot) \geq 11.1$~km, then $\Lambda_{\rm min}$ satisfies
\begin{equation}\label{eq:LVCLambdammin}
    \log\left(\Lambda_{\rm min}(M)\right)=
    \begin{dcases}
-6.0395 \log ^3 \biggl( \frac{M}{M_\odot} \biggr) -2.4247 \log ^2\biggl( \frac{M}{M_\odot} \biggr)-4.7023 \log \biggl( \frac{M}{M_\odot} \biggr)+3.0233, & 0.5 < \frac{M}{M_\odot} <1.94, \\
 13.986\, +\frac{1.7451}{\log ^2\left( \frac{M}{M_\odot} \right)}-\frac{9.7112}{\log \left( \frac{M}{M_\odot} \right)} & 1.94,\leq \frac{M}{M_\odot} <2.29, \\
\log(2.84), & \frac{M}{M_\odot} \geq 2.29, 
    \end{dcases}
\end{equation}
which predicts $\Lambda_{\rm min}=226.9$ for $M =1.35 M_\odot$. The above expressions are accurate to within 1\%, but are always slightly below our full results. Eq.~\eqref{eq:noLVCLambdammin} approximates the lower dashed line in \fig\ref{fig:minLambda}, while  \edit{Eq.~\eqref{eq:LVCLambdammin}}  approximates the  lower edge of the colored regions in the same figure.

In all cases we find that for $M\gtrsim2.0M_\odot$ the minimum tidal deformability is independent of the mass. Our expressions can be used to set priors for future GW events where NSs are suspected to be involved. Moreover, when the high-frequency sensitivity of GW observatories becomes high enough, these expressions could be used to quickly discern whether NSs or solar-mass BHs are involved in a merger. \edit{Note that lower limits for the binary tidal deformability have also been proposed elsewhere, stemming from the EM counterparts AT2017gfo and GRB170817A of GW170817 (see e.g.~\cite{Coughlin:2018miv,Wang:2018nye,Radice:2018ozg}), which are comparable to the 11.1~km values above. Note, however, that our results in Eq.~\eqref{eq:noLVCLambdammin} are hard limits set only by the $2M_\odot$ constraint and our general interpolations, and require no input from kilonova modeling. Moreover, our limits apply for the tidal deformability of an isolated slowly rotating neutron star of any given mass, and not on the binary tidal deformability of the GW170817 event.}

\subsection{Impact of hypothesis of NS as a binary component in GW190814}
\label{sec:gw190814}

\begin{figure}[t]
    \centering
        \includegraphics[height=6.5cm]{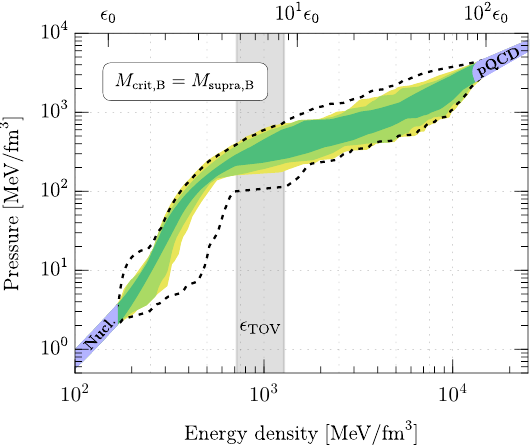}
        $\quad$
        \includegraphics[height=6.5cm]{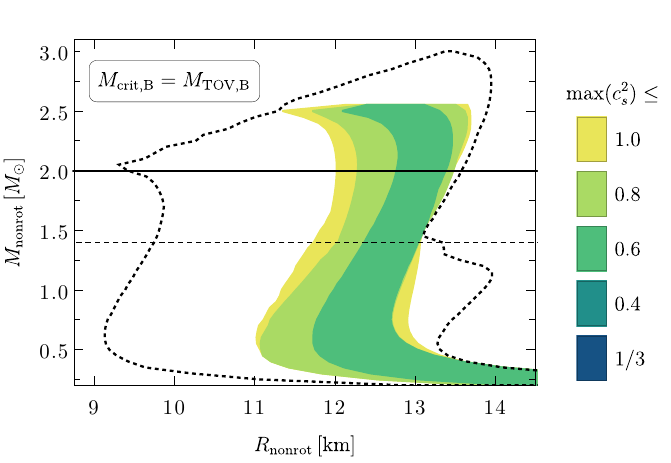}
    \caption{The impact on the allowed EoS band and MR region from assuming that the secondary in GW190814 was a slowly rotating NS, in conjunction with imposing $\tilde{\Lambda}_{\rm{GW170817}} < 720$ and the conservative BH formation hypothesis for GW170817 assuming $\mCrito = \mTOVo$. All the resulting EoSs are also consistent with the with the constraint $\rNonrot(2M_\odot) > 11.1$~km.} 
    \label{fig:gw190814constraints}
\end{figure}
Proceeding to observations of increasingly speculative nature, it is interesting to contemplate the nature of the secondary component in the highly asymmetrical event GW190814, which was measured to have a mass $M\geq 2.48M_\odot$ (90\% lower limit of the IMRPhenomPv3HM waveform model)~\cite{Abbott:2020khf}. While the most likely scenarios include the object being a BH all along, or perhaps first a rapidly spinning NS that later collapsed into a BH, it has also been suggested that this secondary may have been either a slowly or rapidly spinning NS at the time of the merger. If this was indeed the case, it is interesting to analyze the possible tension with the maximum mass limits inferred from the GW170817 event~\cite{Tsokaros:2020hli,Godzieba:2020tjn,Nathanail:2021tay,Demircik:2020jkc} (see also~\cite{Kanakis-Pegios:2020kzp} for a detailed summary of recent work on the topic).  Interestingly, there have been claims that those EoSs satisfying $M_{\rm TOV}<2.1M_\odot$ are not able to account for the 2.5$M_\odot$ even when allowing for maximal uniform rotation \cite{Tsokaros:2020hli}, in addition to which Ref.~\cite{Most:2020bba} has suggested that the secondary of GW190814 had a spin $0.49\lesssim \JMs \lesssim 0.68$ if it was a NS. 

Using our EoS ensemble, we can ask the question of how the EoS bands change if one assumes that GW190814 was a slowly rotating NS, which would imply a new lower limit for the maximum NS mass that dense nuclear EoSs can support, $\mTOV> 2.48M_\odot$. The result of this constraint in conjunction with the GW170817 constraint of $\tilde \Lambda<720$ and the fact that the GW170817 total rest-mass must have been larger than the maximum TOV rest-mass is shown in Fig.~\ref{fig:gw190814constraints}. The figure demonstrates that under the most robust constraints that observations can place, both GW170817 and the secondary of GW190814 can in principle be explained as slowly rotating NSs. However, this requires that the merger product of the event GW170817 collapsed very close to the TOV limit, which may be challenging to reconcile with the multimessenger picture of the event. Additionally, we find that this requires that the speed of sound of dense matter satisfies $c_s^2 > 0.51$ at some densities. We finally note that the GW190814 NS hypothesis is compatible with the radius limits on two-solar-mass stars discussed above.

\subsection{Impact of other possible future observations}
\label{sec:future}

Having established the constraints on our EoS ensemble derived from the GW170817 BH formation hypothesis, various radius measurements and the possibility of the GW190814 event featuring a massive NS, we now turn to the potential of different possible future measurements offering further restrictions on the allowed EoS and MR relations. To this end, we start from the ensemble obtained assuming (i) one of the two versions of the GW170817 BH formation hypothesis (as specified below) with low-spin priors, (ii) that $\tilde{\Lambda} < 720$, and (iii) that $\rNonrot(2M_\odot) \geq 11.1$~km, and further dissect it in various ways chosen to mimic realistic future observations. The latter include both firm mass measurements in excess of $2 M_\odot$; the discovery of a rapidly rotating pulsar with  $f(1.4M_\odot) > 775$~Hz; and new and more stringent constraints on the tidal deformability. We stress, however, that this part of our work is purely speculative, and is performed to inspect the types of observations that would have the largest potential impact for our knowledge of the NS-matter EoS.

\begin{figure}[t]
    \centering
        \includegraphics[width=0.32\textwidth]{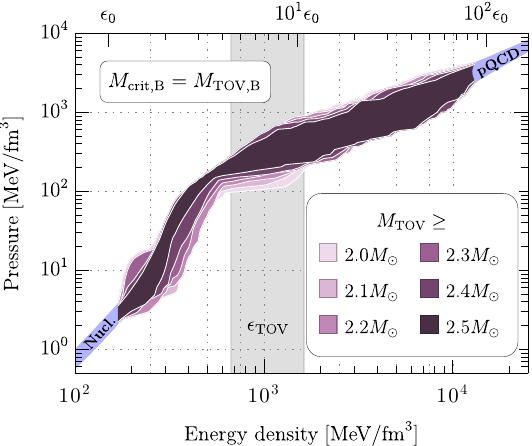}
       $\;$
       \includegraphics[width=0.32\textwidth]{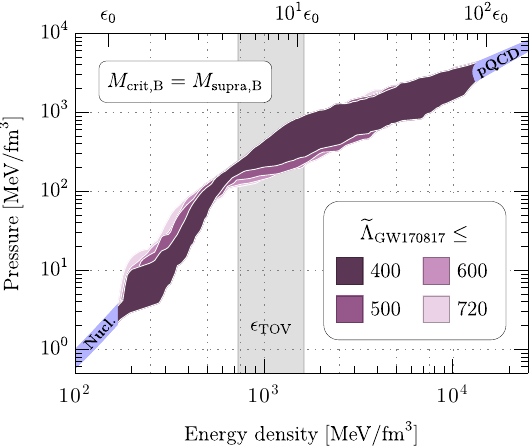}
       $\;$
        \includegraphics[width=0.32\textwidth]{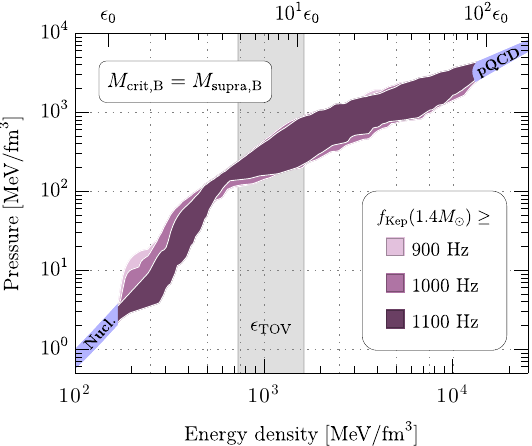}
    \caption{The effects of various hypothetical future NS measurements on the properties of viable NS-matter EoSs. All panels use the constraint $\rNonrot(2M_\odot) \geq 11.1$~km and the tidal-deformability limit $\tilde{\Lambda}_{\rm{GW170817}} < 720$. In the leftmost panel, the $\mCrito = \mTOVo$ version of the GW170817 BH formation hypothesis is assumed, while in the middle and rightmost panels, the ensemble used assumes $\mCrito = \mSuprao$.} 
    \label{fig:fut_meas_no_radii}
\end{figure}

The results of the new speculative cuts are displayed in Fig.~\ref{fig:fut_meas_no_radii} in the form of color-coded EoS bands; just as in Fig.~\ref{fig:fut_meas_radii2}, the color scheme is changed to range from purple to pink to highlight the speculative nature of these results. In the left-most figure, generated with the $\mCrito=\mTOVo$ ensemble, we first inspect the impact that increasing the TOV mass lower limit  above $2M_\odot$ would have on the EoS. 
We observe that the primary effect of a high TOV mass lower limit occurs at low densities, where EoSs with lower pressures for a given energy density become exceedingly excluded with increasing $\mTOV$, but for very high values of this parameter, the impact of the cut becomes noticeable also at very high densities, due to the constraints of causality and matching to the pQCD region. It is also worthwhile to note here that the highest mass supported by subconformal EoSs, for which $c_s^2<1/3$ at all densities, is approx.~$2.10M_\odot$.

Proceeding next to the two rightmost panels in Fig.~\ref{fig:fut_meas_no_radii}, we explore the impact of lowering the upper limit for the tidal deformability --- in anticipation of new loud GW events --- and increasing the frequency of the fastest-rotating observed NS \footnote{Note that due to the tight correlation between the NS radius and its tidal deformability, upper limits for $\tilde{\Lambda}$ effectively act as upper limits for NS radii. Enforcing e.g.~$\tilde{\Lambda}_\text{GW170817}<200$ would correspond to $1.4M_\odot$ NSs having radii well below 10~km.}. In the context of the former constraint, note that $\tilde{\Lambda}_\text{GW170817}>200$ appears to be necessary to explain the ejecta mass for the kilonova counterpart to GW170817~\cite{Kiuchi:2019lls} (see also~\cite{Radice:2017lry}), which is also seen to follow from the constraint $\rNonrot(2M_\odot) \geq 11.1$~km. Both a tighter upper bound on $\tilde{\Lambda}$ or a tighter lower bound on $\fKep$ would predominantly cut off EoSs that are stiff at low densities, \edit{with only a small, subdominant effect on the lower side of the EoS band at high densities.}

\section{Discussion and conclusions \label{sec:conclusions}}

In our study, we have discussed how different measurements and hypotheses concerning the nature of different binary merger components and merger products affect our knowledge of the properties of the cold ultra-dense matter found inside neutron stars. In order to do so, we constructed a large ensemble of model-independent interpolated EoSs, starting from a piecewise interpolation in the $\mu_\text{B}$--$c_s^2$ space, with each individual EoS required to conform to known CET results at low densities and to the pQCD EoS in the high-density limit while respecting causality. The EoS ensemble obtained was then further constrained by different NS observations, resulting in the NS properties summarized in Table~\ref{tab:radranges}. It is interesting to note the similarity of our $R_{1.4}$ ranges to the 95\% probability estimates for the same quantity obtained in the recent Bayesian analysis of \cite{Raaijmakers:2021uju}.

We studied in detail the constraints that arise from the broader multimessenger picture of GW170817, the hypothesis that the secondary binary merger component of GW190814 was a NS, and the new NICER radius measurement of PSR J0740+6620. We find that different constraints discussed are mutually compatible and indicate a great potential for reducing the current uncertainties in the NS-matter EoS over the coming years. In particular, the GW170817 BH formation hypothesis and the corresponding constraint on tidal deformability $\tilde{\Lambda} < 720$ both independently exclude EoSs that remain stiff up to very high densities; that only one of these assumptions is needed strengthens the case to exclude these EoSs. Of the hypotheses we have discussed, the only tension lies between the secondary component of the GW190814 event having been a NS and the hypothesis that the immediate binary merger product of GW170817 was a HMNS. We, however, find no direct tension between the GW190814 NS hypothesis and the more conservative assumption that a BH was formed in GW170817 (although it can be argued that the full multimessenger picture of this event is challenging to explain without the HMNS assumption). It is possible that some of these assumptions may not be compatible with others, if in addition one explicitly assumes that the remnant in GW170817 underwent delayed collapse to a black hole. We leave the exploration of this possibility for a future analysis.

\begin{table}[t]
  \setlength{\tabcolsep}{8pt}
  \centering
  \begin{tabular}{c c c c c c c  c}\hline  
    \hline
    \multicolumn{2}{c}{Assumptions} & \multicolumn{6}{c}{Resulting ensemble limits} \\
    \cmidrule(lr){1-2}\cmidrule(lr){3-8}
    BH hypothesis & $R_{2.0, \rm min}$ (km) & $R_{1.4}$~(km) & $R_{1.6}$~(km) & $R_{1.8}$~(km) & $R_{2.0}$~(km) & $\mTOV$~($M_\odot$) & \edit{$\mTOVo$~($M_\odot$)}\\
    \cmidrule(lr){1-8}
    --    &   -- &  9.6--13.4 &  9.8--13.3 &  9.7--13.5 &  9.3--13.7 & 2.98 &  \edit{3.75}\\ 
    TOV   &   -- &  9.6--13.4 &  9.8--13.2 &  9.7--13.4 &  9.3--13.6 & 2.53 &\edit{3.08}\\ 
    Supra &   -- &  9.7--13.4 &  9.8--13.2 &  9.7--13.3 &  9.3--13.3 & 2.19 & \edit{2.63}\\ 
    TOV   & \edit{10.7} & \edit{10.3}--13.2 & \edit{10.5}--13.2 & \edit{10.7}--13.4 & \edit{10.7}--13.6 & 2.53 & \edit{3.08}\\ 
    TOV   & 11.1 & 10.7--13.2 & 10.9--13.2 & 11.0--13.4 & 11.1--13.6 & 2.53 & \edit{3.08}\\ 
    TOV   & 11.4 & 10.9--13.2 & 11.1--13.2 & 11.2--13.4 & 11.4--13.6 & 2.53 & \edit{3.08}\\ 
    TOV   & 12.2 & 11.5--13.1 & 11.7--13.2 & 12.0--13.4 & 12.2--13.6 & 2.53 & \edit{3.08} \\ 
    Supra & \edit{10.7} & \edit{10.3}--13.2 & \edit{10.5}--13.2 & \edit{10.7}--13.3 & \edit{10.7}--13.3 & 2.19 & \edit{2.63} \\ 
    Supra & 11.1 & 10.8--13.2 & 11.0--13.2 & 11.1--13.3 & 11.1--13.3 & 2.19 & \edit{2.63}\\ 
    Supra & 11.4 & 11.2--13.2 & 11.3--13.2 & 11.4--13.3 & 11.4--13.3 & 2.19 & \edit{2.63}\\ 
    Supra & 12.2 & 11.9--13.1 & 12.0--13.2 & 12.1--13.3 & 12.2--13.3 & 2.19 & \edit{2.63} \\ 
    \hline\hline
\end{tabular}
  \caption{The range of radii obtained for 1.4$M_\odot$ ($R_{1.4}$), 1.6$M_\odot$ ($R_{1.6}$), 1.8$M_\odot$ ($R_{1.8}$), and 2.0$M_\odot$ ($R_{2.0}$) NSs, \edit{the upper bound on the maximum TOV gravitational mass, and the upper bound on the maximum TOV rest/baryonic mass} under various hypotheses. The two BH-formation hypotheses correspond to the $\mCrito=\mTOVo$ and $\mCrito=\mSuprao$ cases discussed in the main text, and the NICER results are implemented as lower limits for $R_{2.0}$, which we denote as $R_{2.0,\rm min}$. In addition to the assumptions explicitly listed in the columns, we assume $\mTOV\geq 2.0 M_\odot$ and $\tilde \Lambda_{\rm GW170817} < 720$ in all cases considered.  Note that the entry ``--'' indicates the absence of a constraint of the type in question. \label{tab:radranges}}
 \end{table}

One of our main results is that even the mild assumption that a BH was formed in GW170817 leads to important constraints for the NS-matter EoS. In particular, combined with the constraint on tidal deformability, we find a robust bound for the TOV mass, $\mTOV < 2.53 M_\odot$. In the case that the immediate merger product in the event was a HMNS, we further derive the more stringent constraint $\mTOV < 2.19 M_\odot$, which supports previous results in the literature. Our result can, however, be considered more robust, as it relies on fewer assumptions connecting non-rotating and rotating NSs and a large ensemble of EoSs. We have furthermore shown that all stable NSs sequences with $\mTOV \geq 2 M_\odot$ satisfy $\JMs < 0.81$ for all stable NS configurations, a value that justifies the \emph{a priori} postulated upper limit used in previous NS merger analyses by the LIGO and Virgo collaborations.

Interestingly, we find that the recent radius measurement of the massive PSR J0740+6620 by the NICER collaboration has set strong constraints on the NS matter EoS. While even larger radii are preferred by the data, the new results set a firm lower limit of approx.~11~km for the radius of a two-solar-mass NS, enough to exclude a large set of EoSs that exhibit extreme behaviors such as large speeds of sound $\max(c_s^2) > 0.6$ (see~in particular our preferred ensemble in Fig.~\ref{fig:NICERconstraintsTOV0onePanel}). We observe that such lower limits (as long as they are below 13~km) are consistent with EoSs where the speed of sound is consistently below the conformal limit $c_s^2 = 1/3$. Recalling that maximally massive stars described by EoSs featuring moderate speed-of-sound values $c_s^2\lesssim 0.5$ contain sizeable quark cores according to the results of \cite{Annala:2019puf}, we see that the radius values implied by the new NICER data are fully consistent with the presence of QM inside massive NSs.


\acknowledgments{We thank Kai Hebeler, Alex Nielsen, and Achim Schwenk for useful discussions. EA and AV were supported by the Academy of Finland grant no.~1322507, as well as by the European Research Council, grant no.~725369. VP was in part supported by NSF Grant PHY-1912619 to the University of Arizona, and TG was in part supported by the Deutsche Forschungsgemeinschaft (DFG, German Research Foundation) -- Project-Id 279384907 -- SFB 1245. Computational resources were provided by the Open Science Grid~\cite{Pordes:2007zzb,osg09}, which is supported by the National Science Foundation award 2030508, and the Extreme Science and Engineering Discovery Environment (XSEDE) under grant number TG-PHY190020. XSEDE is supported by the NSF grant No.\ ACI-1548562. Calculations were in part performed on \texttt{Stampede2}, which is funded by the NSF through award ACI-1540931. 
}

\appendix

\section{Interpolating the NS-matter EoS}
\label{app:interpolation}

\begin{figure}
    \centering
    \includegraphics[width=\textwidth]{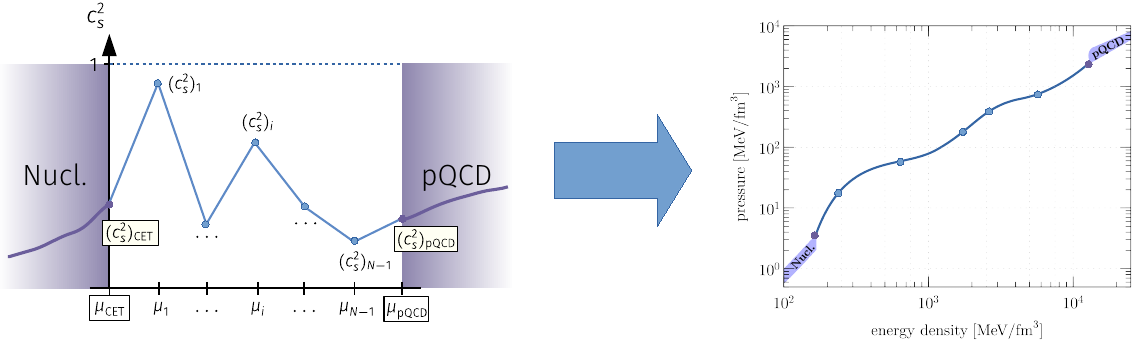}
    \caption{A cartoon of our method of EoS construction. Note that this figure is for illustration purposes only, and does not depict data for a real $c_s^2(\mu_\text{B}) \to p(\epsilon)$ construction.}
    \label{fig:eosCartoon}
\end{figure}

As briefly covered in Sec.~\ref{sec:methods}, our algorithm for generating the ensemble of interpolated NS-matter EoSs is based on interpolating the EoS between a low-density regime described by the CET EoS and a high-density regime where we use a pQCD EoS for unpaired quark matter. This follows very closely the algorithm first introduced in \cite{Annala:2019puf},  \edit{while other interpolation routines have been developed e.g.~in \cite{Read:2008iy,Ozel:2009da,Lindblom:2010bb,Landry:2018prl,Lindblom:2018rfr}}. A cartoon depicting our algorithm is shown in \fig\ref{fig:eosCartoon}. 

Between a baryon chemical potential $\mu_{\text{CET}}$ corresponding to $n_\text{B}=1.1n_0$ and $\mu_\text{B}=2.6$ GeV, we proceed as follows:
\begin{enumerate}
    \item Form a sequence of $N_{p}$ pairs 
$\{(\mu_i , c_{s,i}^2)\}_{i = 1}^{N_{p}},$ with $\mu_{1} = \mu_{\text{CET}}$, $\mu_{N_{p}} = 2.6$~GeV, and $\mu_{i-1} < \mu_{i} < \mu_{i+1}$ for other $i$. We also set $c_{s,1}^2$ and $c_{s,N_p}^2$ equal to the value $c_s^2$ takes at the end of the CET interval and at the beginning of the pQCD interval, respectively.
\item Build a piecewise-linear function connecting the above points, i.e.~for each $i = 1,\ldots,N_{p} - 1$, and for $\mu_\text{B} \in [\mu_{i} , \mu_{i + 1}]$, write
\begin{equation}
c_s^2(\mu_\text{B}) = \frac{(\mu_{i+1} - \mu_\text{B}) c_{s,i}^{2} + (\mu_\text{B} - \mu_{i}) {c_{s,i+1}^{2}}}{\mu_{i+1} - \mu_{i}}. 
\end{equation}
\item From the piecewise-linear speed of sound squared, construct the baryon density and pressure using the relations 
\begin{equation}
n_\text{B}(\mu_\text{B}) = n_\text{CET} \exp \left[ \int_{\mu_\text{CET}}^{\mu_\text{B}}\! \frac{\mathrm{d}\mu'}{\mu' c_s^2(\mu')} \right], \quad
p(\mu_\text{B}) = p_\text{CET} + n_\text{CET} \dint{\mu'}{\mu_\text{CET}}{\mu_\text{B}} n(\mu'), \label{eq:p_from_cs2}
\end{equation}
where $p_\text{CET} \equiv p(\mu_\text{CET})$.
\item Demand that there exists an $X \in [1,4]$ such that 
\begin{equation}
n_\text{B}(\mu_{N_p}) = n_{\rm pQCD}(\mu_{N_p}, X), \quad p_{\rm pQCD}(\mu_{N_p}) = p(\mu_{N_p}, X),
\end{equation}
where $n_{\rm pQCD}$ and $p_{\rm pQCD}$ are the pQCD number density and pressure, respectively, as a function of $\mu_{\rm B}$ and the pQCD $X$ parameter \cite{Fraga:2013qra}. This condition also fixes one of the $c_{s,i}^2$ points.
\end{enumerate}

The above procedure is used to generate approx.~1,500,000 EoSs with $N_p \in \{ 4, 5, 6 \}$, with the remaining matching points $\mu_i$, and speeds of sound squared $c_i^2$ randomly picked (for the CET EoS, we choose roughly the same number of ``hard'' or ``soft''  EoSs of \edit{Ref.}~\cite{Hebeler:2013nza}). Out of these, approx.~250,000 satisfy our basic constraints for the TOV mass and tidal deformability.

Finally, we note that, similar to the analysis in \cite{Annala:2019puf}, our EoSs should be understood as approximate or `smoothed' representations of real EoSs, which can have small-scale fluctuations in their $c_s(\mu_\text{B})$ (or similarly in $p(n_\text{B}))$ dependency. 
Our individual EoSs do not aim to capture these micro structures of the real EoS, which could be numerically realized by increasing $N_p \rightarrow \infty$,  but instead we want to study the general properties of the matter that can be achieved by regularizing the EoS description, similar to e.g., smoothing splines.

\section{Black hole hypothesis: our algorithm}
\label{app:BH_form_hyp}

In our analysis of the GW170817 event, we convert the gravitational masses of the binary components to rest masses in order to deduce robust limits on the NS-matter EoS. Concretely, we exclude from our analysis any EoS for which the initial binary rest mass is strictly below $\mCrito$ for all possible binary configurations consistent with the event. That multiple possible merger configurations exist follows from the constraints from GW170817: the GW signal for this event pinpointed the chirp mass $\Mch=1.186 \pm 0.001~M_{\odot}$ (90\% credible interval)~\cite{Abbott:2018wiz} but only constrained the individual component masses $M_1$ and $M_2$ to comparatively larger ranges. In particular, several different mass configurations of the components can lead to the same chirp mass.  Within our analysis, we adopt the 90\% credible range for the primary gravitational mass assuming the GW170817 low spin prior $M_1 \in [1.36M_\odot, 1.6M_\odot]$ or high-spin prior $M_1 \in [1.36M_\odot, 1.89M_\odot]$ \cite{Abbott:2018wiz}.  Here, it is worth noting that the remnants with the largest rest mass arise from the most asymmetric mass configurations. While it would seem inappropriate to use non-spinning stars with the high-spin prior analysis posteriors, the fact that for a given gravitational mass a non-spinning star has more rest mass than a spinning star makes our analysis yield the most conservative possible constraint.

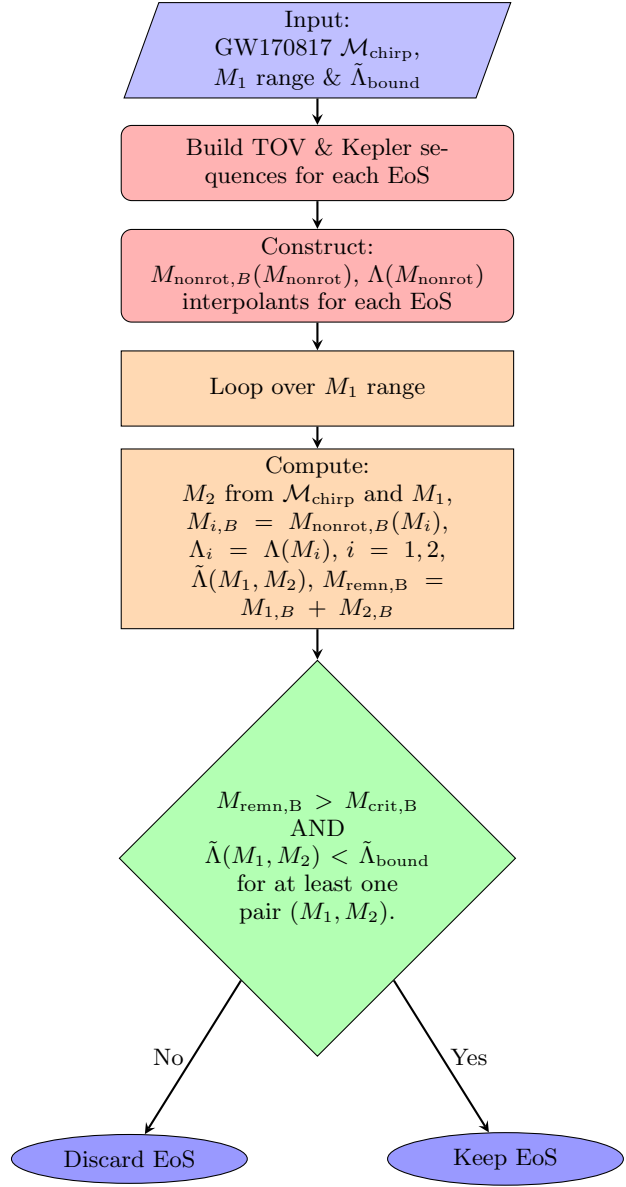
\begin{figure}[t!]
\centering
\begin{tikzpicture}[node distance=1.5cm,auto]

\node (input) [io] {Input: \\ GW170817 $\Mch$,  $M_1$ range \& $\tilde\Lambda_{\rm bound}$};

\node (start) [startstop,below of=input] {Build TOV \& Kepler sequences for each EoS};

\draw [arrow] (input) -- (start);

\node (pro1) [startstop,below of=start] {Construct:\\ $M_{{\rm nonrot},B}(M_{\rm nonrot})$, $\Lambda(M_{\rm nonrot})$ interpolants for each EoS};

\draw [arrow] (start) -- (pro1);

\node (pro2) [process,below of=pro1] {Loop over $M_1$ range};

\draw [arrow] (pro1) -- (pro2);

\node (pro3) [process,below of=pro2,yshift=-0.5cm] {Compute:\\ $M_2$ from $\Mch$ and $M_1$,\\ $M_{i,B}=M_{{\rm nonrot},B}(M_i)$, $\Lambda_i=\Lambda(M_i)$, $i=1,2$, \\ $\tilde\Lambda(M_1,M_2)$, $\mRemno=M_{1,B}+M_{2,B}$};

\draw [arrow] (pro2) -- (pro3);

\node (dec) [decision,below of=pro3,yshift=0.25cm]{ $\mRemno > \mCrito$\\ AND\\ $\tilde\Lambda(M_1,M_2)<\tilde\Lambda_{\rm bound}$ \\ for at least one pair $(M_1,M_2)$.
};

\draw [arrow] (pro3) -- (dec);

\node (pro4) [cloud,right of=dec,yshift=-4.0cm] {Keep EoS};

\draw [arrow] (dec) -- node[anchor=west] {Yes} (pro4);

\node (pro5) [cloud,left of=dec,yshift=-4.0cm] {Discard EoS};

\draw [arrow] (dec) -- node[anchor=east] {No} (pro5);

\end{tikzpicture}
\caption{Flow chart demonstrating the  process of imposing the fact that the merger product of GW170817 must have collapsed to form a BH. $M_{{\rm nonrot},B}$ is the rest mass of a non-spinning star, $M_1$ ($M_2$) the gravitational mass of the primary (secondary) of the event, and $\mCrito$ the critical baryon mass above which GW170817 would have collapsed to a black hole, the minimum value for which is the TOV-limit mass ($\mTOVo$) for a given EoS. Another value for this critical mass used in analyses found the literature is the supramassive limit mass ($\mSuprao$). \label{fig:flowchart}}
\end{figure}

To summarize, our method proceeds as follows: for any given EoS, we compute the theoretical merger remnant rest mass $\mRemno$ from the GW170817 chirp mass and the primary gravitational mass $M_1$ by using the conservation of the total baryon number before and after the merger. In particular, using the primary gravitational mass $M_1$ and the chirp mass, we can immediately determine the gravitational mass of the secondary $M_2$. Using the EoS and integrating the TOV equations we then find the baryon mass of the primary and the secondary ($M_{\rm 1,B}$ and $M_{\rm 2,B}$) from $M_1$ and $M_2$, and subsequently the total baryon mass of the remnant, $\mRemno = M_{\rm 1,B} + M_{\rm 2,B}- \mEjeo$, where $\mEjeo$ is the ejecta mass. To obtain a conservative upper mass limit, we ignore any possible ejecta within the analysis, so we set $\mEjeo=0$. If $\mRemno > \mCrito$, then the BH formation hypothesis is satisfied. A flow chart demonstrating our algorithm is shown in Fig.~\ref{fig:flowchart}.

We conduct our analysis with the choices $\mCrito = \mTOVo$ and $\mCrito=\mSuprao$, where these values must be determined separately for each EoS. We also implement two different spin priors; in the case of high-spin priors, we do not use the tidal-deformability constraint, because we do not know the spin distribution of the stars, and because at this time there exists no theory for computing tidal deformabilities for rapidly rotating stars. Thus, we implement the high-spin prior analysis by simply using the larger allowed mass range for the primary in the case of GW170817. All in all, we consider the following combination to impose joint constraints:
\begin{enumerate}
    \item $\mCrito = \mTOVo$ (primary mass range from high-spin priors),
    \item  $\mCrito = \mSuprao$ (primary mass range from high-spin priors),
    \item $\mCrito = \mSuprao$ (primary mass range from low-spin priors),
    \item $\mCrito = \mTOVo$ (primary mass range from low-spin priors) and $\tilde{\Lambda} < 720$,
    \item $\mCrito = \mSuprao$ (primary mass range from low-spin priors) and $\tilde{\Lambda} < 720$.
\end{enumerate}
The possibilities 1 and 4, which correspond to $\mCrito = \mTOVo$, are more conservative than the possibilities 2, 3, and 5, which correspond to different implementations of the $\mCrito = \mSuprao$ case. We note that when implementing multiple constraints, we always check that at least one mass configuration exists that satisfies them both simultaneously. The effects of these various joint constraints on the EoS ensemble and MR relations are shown in \fig~\ref{fig:different_GW170817_implementations} in the main text, along with the lone constraint $\tilde{\Lambda} < 720$ for reference.

\typeout{}
\bibliography{refs.bib}

\begin{thebibliography}{152}%
\makeatletter
\providecommand \@ifxundefined [1]{%
 \@ifx{#1\undefined}
}%
\providecommand \@ifnum [1]{%
 \ifnum #1\expandafter \@firstoftwo
 \else \expandafter \@secondoftwo
 \fi
}%
\providecommand \@ifx [1]{%
 \ifx #1\expandafter \@firstoftwo
 \else \expandafter \@secondoftwo
 \fi
}%
\providecommand \natexlab [1]{#1}%
\providecommand \enquote  [1]{``#1''}%
\providecommand \bibnamefont  [1]{#1}%
\providecommand \bibfnamefont [1]{#1}%
\providecommand \citenamefont [1]{#1}%
\providecommand \href@noop [0]{\@secondoftwo}%
\providecommand \href [0]{\begingroup \@sanitize@url \@href}%
\providecommand \@href[1]{\@@startlink{#1}\@@href}%
\providecommand \@@href[1]{\endgroup#1\@@endlink}%
\providecommand \@sanitize@url [0]{\catcode `\\12\catcode `\$12\catcode
  `\&12\catcode `\#12\catcode `\^12\catcode `\_12\catcode `\%12\relax}%
\providecommand \@@startlink[1]{}%
\providecommand \@@endlink[0]{}%
\providecommand \url  [0]{\begingroup\@sanitize@url \@url }%
\providecommand \@url [1]{\endgroup\@href {#1}{\urlprefix }}%
\providecommand \urlprefix  [0]{URL }%
\providecommand \Eprint [0]{\href }%
\providecommand \doibase [0]{https://doi.org/}%
\providecommand \selectlanguage [0]{\@gobble}%
\providecommand \bibinfo  [0]{\@secondoftwo}%
\providecommand \bibfield  [0]{\@secondoftwo}%
\providecommand \translation [1]{[#1]}%
\providecommand \BibitemOpen [0]{}%
\providecommand \bibitemStop [0]{}%
\providecommand \bibitemNoStop [0]{.\EOS\space}%
\providecommand \EOS [0]{\spacefactor3000\relax}%
\providecommand \BibitemShut  [1]{\csname bibitem#1\endcsname}%
\let\auto@bib@innerbib\@empty
\bibitem [{\citenamefont {Demorest}\ \emph {et~al.}(2010)\citenamefont
  {Demorest}, \citenamefont {Pennucci}, \citenamefont {Ransom}, \citenamefont
  {Roberts},\ and\ \citenamefont {Hessels}}]{Demorest:2010bx}%
  \BibitemOpen
  \bibfield  {author} {\bibinfo {author} {\bibfnamefont {P.~B.}\ \bibnamefont
  {Demorest}}, \bibinfo {author} {\bibfnamefont {T.}~\bibnamefont {Pennucci}},
  \bibinfo {author} {\bibfnamefont {S.~M.}\ \bibnamefont {Ransom}}, \bibinfo
  {author} {\bibfnamefont {M.~S.~E.}\ \bibnamefont {Roberts}},\ and\ \bibinfo
  {author} {\bibfnamefont {J.~H.~T.}\ \bibnamefont {Hessels}},\ }\bibfield
  {title} {\bibinfo {title} {{A two-solar-mass neutron star measured using
  Shapiro delay}},\ }\href {https://doi.org/10.1038/nature09466} {\bibfield
  {journal} {\bibinfo  {journal} {Nature}\ }\textbf {\bibinfo {volume} {467}},\
  \bibinfo {pages} {1081} (\bibinfo {year} {2010})},\ \Eprint
  {https://arxiv.org/abs/1010.5788} {arXiv:1010.5788 [astro-ph.HE]}
  \BibitemShut {NoStop}%
\bibitem [{\citenamefont {Antoniadis}\ \emph {et~al.}(2013)\citenamefont
  {Antoniadis} \emph {et~al.}}]{Antoniadis:2013pzd}%
  \BibitemOpen
  \bibfield  {author} {\bibinfo {author} {\bibfnamefont {J.}~\bibnamefont
  {Antoniadis}} \emph {et~al.},\ }\bibfield  {title} {\bibinfo {title} {{A
  massive pulsar in a compact relativistic binary}},\ }\href
  {https://doi.org/10.1126/science.1233232} {\bibfield  {journal} {\bibinfo
  {journal} {Science}\ }\textbf {\bibinfo {volume} {340}},\ \bibinfo {pages}
  {1233232} (\bibinfo {year} {2013})},\ \Eprint
  {https://arxiv.org/abs/1304.6875} {arXiv:1304.6875 [astro-ph.HE]}
  \BibitemShut {NoStop}%
\bibitem [{\citenamefont {Cromartie}\ \emph {et~al.}(2020)\citenamefont
  {Cromartie} \emph {et~al.}}]{Cromartie:2019kug}%
  \BibitemOpen
  \bibfield  {author} {\bibinfo {author} {\bibfnamefont {H.~T.}\ \bibnamefont
  {Cromartie}} \emph {et~al.},\ }\bibfield  {title} {\bibinfo {title}
  {{Relativistic Shapiro delay measurements of an extremely massive millisecond
  pulsar}},\ }\href {https://doi.org/10.1038/s41550-019-0880-2} {\bibfield
  {journal} {\bibinfo  {journal} {Nat. Astron.}\ }\textbf {\bibinfo {volume}
  {4}},\ \bibinfo {pages} {72} (\bibinfo {year} {2020})},\ \Eprint
  {https://arxiv.org/abs/1904.06759} {arXiv:1904.06759 [astro-ph.HE]}
  \BibitemShut {NoStop}%
\bibitem [{\citenamefont {Fonseca}\ \emph {et~al.}(2021)\citenamefont {Fonseca}
  \emph {et~al.}}]{Fonseca:2021wxt}%
  \BibitemOpen
  \bibfield  {author} {\bibinfo {author} {\bibfnamefont {E.}~\bibnamefont
  {Fonseca}} \emph {et~al.},\ }\bibfield  {title} {\bibinfo {title} {Refined
  mass and geometric measurements of the high-mass {PSR J0740+6620}},\ }\href
  {https://doi.org/10.3847/2041-8213/ac03b8} {\bibfield  {journal} {\bibinfo
  {journal} {Astrophys. J. Lett.}\ }\textbf {\bibinfo {volume} {915}},\
  \bibinfo {pages} {L12} (\bibinfo {year} {2021})},\ \Eprint
  {https://arxiv.org/abs/2104.00880} {arXiv:2104.00880 [astro-ph.HE]}
  \BibitemShut {NoStop}%
\bibitem [{\citenamefont {Fonseca}\ \emph {et~al.}(2016)\citenamefont {Fonseca}
  \emph {et~al.}}]{Fonseca:2016tux}%
  \BibitemOpen
  \bibfield  {author} {\bibinfo {author} {\bibfnamefont {E.}~\bibnamefont
  {Fonseca}} \emph {et~al.},\ }\bibfield  {title} {\bibinfo {title} {{The
  NANOGrav nine-year data set: Mass and geometric measurements of binary
  millisecond pulsars}},\ }\href {https://doi.org/10.3847/0004-637X/832/2/167}
  {\bibfield  {journal} {\bibinfo  {journal} {Astrophys. J.}\ }\textbf
  {\bibinfo {volume} {832}},\ \bibinfo {pages} {167} (\bibinfo {year}
  {2016})},\ \Eprint {https://arxiv.org/abs/1603.00545} {arXiv:1603.00545
  [astro-ph.HE]} \BibitemShut {NoStop}%
\bibitem [{\citenamefont {N\"attil\"a}\ \emph {et~al.}(2017)\citenamefont
  {N\"attil\"a}, \citenamefont {Miller}, \citenamefont {Steiner}, \citenamefont
  {Kajava}, \citenamefont {Suleimanov},\ and\ \citenamefont
  {Poutanen}}]{Nattila:2017wtj}%
  \BibitemOpen
  \bibfield  {author} {\bibinfo {author} {\bibfnamefont {J.}~\bibnamefont
  {N\"attil\"a}}, \bibinfo {author} {\bibfnamefont {M.~C.}\ \bibnamefont
  {Miller}}, \bibinfo {author} {\bibfnamefont {A.~W.}\ \bibnamefont {Steiner}},
  \bibinfo {author} {\bibfnamefont {J.~J.~E.}\ \bibnamefont {Kajava}}, \bibinfo
  {author} {\bibfnamefont {V.~F.}\ \bibnamefont {Suleimanov}},\ and\ \bibinfo
  {author} {\bibfnamefont {J.}~\bibnamefont {Poutanen}},\ }\bibfield  {title}
  {\bibinfo {title} {{Neutron star mass and radius measurements from
  atmospheric model fits to X-ray burst cooling tail spectra}},\ }\href
  {https://doi.org/10.1051/0004-6361/201731082} {\bibfield  {journal} {\bibinfo
   {journal} {Astron. Astrophys.}\ }\textbf {\bibinfo {volume} {608}},\
  \bibinfo {pages} {A31} (\bibinfo {year} {2017})},\ \Eprint
  {https://arxiv.org/abs/1709.09120} {arXiv:1709.09120 [astro-ph.HE]}
  \BibitemShut {NoStop}%
\bibitem [{\citenamefont {Riley}\ \emph {et~al.}(2019)\citenamefont {Riley}
  \emph {et~al.}}]{Riley:2019yda}%
  \BibitemOpen
  \bibfield  {author} {\bibinfo {author} {\bibfnamefont {T.~E.}\ \bibnamefont
  {Riley}} \emph {et~al.},\ }\bibfield  {title} {\bibinfo {title} {{A $NICER$
  view of PSR J0030+0451: Millisecond pulsar parameter estimation}},\ }\href
  {https://doi.org/10.3847/2041-8213/ab481c} {\bibfield  {journal} {\bibinfo
  {journal} {Astrophys. J. Lett.}\ }\textbf {\bibinfo {volume} {887}},\
  \bibinfo {pages} {L21} (\bibinfo {year} {2019})},\ \Eprint
  {https://arxiv.org/abs/1912.05702} {arXiv:1912.05702 [astro-ph.HE]}
  \BibitemShut {NoStop}%
\bibitem [{\citenamefont {Miller}\ \emph {et~al.}(2019)\citenamefont {Miller}
  \emph {et~al.}}]{Miller:2019cac}%
  \BibitemOpen
  \bibfield  {author} {\bibinfo {author} {\bibfnamefont {M.~C.}\ \bibnamefont
  {Miller}} \emph {et~al.},\ }\bibfield  {title} {\bibinfo {title} {{PSR
  J0030+0451 mass and radius from NICER data and implications for the
  properties of neutron star matter}},\ }\href
  {https://doi.org/10.3847/2041-8213/ab50c5} {\bibfield  {journal} {\bibinfo
  {journal} {Astrophys. J. Lett.}\ }\textbf {\bibinfo {volume} {887}},\
  \bibinfo {pages} {L24} (\bibinfo {year} {2019})},\ \Eprint
  {https://arxiv.org/abs/1912.05705} {arXiv:1912.05705 [astro-ph.HE]}
  \BibitemShut {NoStop}%
\bibitem [{\citenamefont {Miller}\ \emph {et~al.}(2021)\citenamefont {Miller}
  \emph {et~al.}}]{Miller:2021qha}%
  \BibitemOpen
  \bibfield  {author} {\bibinfo {author} {\bibfnamefont {M.~C.}\ \bibnamefont
  {Miller}} \emph {et~al.},\ }\bibfield  {title} {\bibinfo {title} {{The radius
  of PSR J0740+6620 from NICER and XMM-Newton data}},\ }\href
  {https://doi.org/10.3847/2041-8213/ac089b} {\bibfield  {journal} {\bibinfo
  {journal} {Astrophys. J. Lett.}\ }\textbf {\bibinfo {volume} {918}},\
  \bibinfo {pages} {L28} (\bibinfo {year} {2021})},\ \Eprint
  {https://arxiv.org/abs/2105.06979} {arXiv:2105.06979 [astro-ph.HE]}
  \BibitemShut {NoStop}%
\bibitem [{\citenamefont {Riley}\ \emph {et~al.}(2021)\citenamefont {Riley}
  \emph {et~al.}}]{Riley:2021pdl}%
  \BibitemOpen
  \bibfield  {author} {\bibinfo {author} {\bibfnamefont {T.~E.}\ \bibnamefont
  {Riley}} \emph {et~al.},\ }\bibfield  {title} {\bibinfo {title} {{A NICER
  view of the massive pulsar PSR J0740+6620 informed by radio timing and
  XMM-Newton spectroscopy}},\ }\href {https://doi.org/10.3847/2041-8213/ac0a81}
  {\bibfield  {journal} {\bibinfo  {journal} {Astrophys. J. Lett.}\ }\textbf
  {\bibinfo {volume} {918}},\ \bibinfo {pages} {L27} (\bibinfo {year}
  {2021})},\ \Eprint {https://arxiv.org/abs/2105.06980} {arXiv:2105.06980
  [astro-ph.HE]} \BibitemShut {NoStop}%
\bibitem [{\citenamefont {Abbott}\ \emph
  {et~al.}(2017{\natexlab{a}})\citenamefont {Abbott} \emph
  {et~al.}}]{TheLIGOScientific:2017qsa}%
  \BibitemOpen
  \bibfield  {author} {\bibinfo {author} {\bibfnamefont {B.~P.}\ \bibnamefont
  {Abbott}} \emph {et~al.},\ }\bibfield  {title} {\bibinfo {title} {{GW170817:
  Observation of gravitational waves from a binary neutron star inspiral}},\
  }\href {https://doi.org/10.1103/PhysRevLett.119.161101} {\bibfield  {journal}
  {\bibinfo  {journal} {Phys. Rev. Lett.}\ }\textbf {\bibinfo {volume} {119}},\
  \bibinfo {pages} {161101} (\bibinfo {year} {2017}{\natexlab{a}})},\ \Eprint
  {https://arxiv.org/abs/1710.05832} {arXiv:1710.05832 [gr-qc]} \BibitemShut
  {NoStop}%
\bibitem [{\citenamefont {Abbott}\ \emph {et~al.}(2018)\citenamefont {Abbott}
  \emph {et~al.}}]{Abbott:2018exr}%
  \BibitemOpen
  \bibfield  {author} {\bibinfo {author} {\bibfnamefont {B.~P.}\ \bibnamefont
  {Abbott}} \emph {et~al.},\ }\bibfield  {title} {\bibinfo {title} {{GW170817:
  Measurements of neutron star radii and equation of state}},\ }\href
  {https://doi.org/10.1103/PhysRevLett.121.161101} {\bibfield  {journal}
  {\bibinfo  {journal} {Phys. Rev. Lett.}\ }\textbf {\bibinfo {volume} {121}},\
  \bibinfo {pages} {161101} (\bibinfo {year} {2018})},\ \Eprint
  {https://arxiv.org/abs/1805.11581} {arXiv:1805.11581 [gr-qc]} \BibitemShut
  {NoStop}%
\bibitem [{\citenamefont {{Tolman}}(1939)}]{1939PhRv...55..364T}%
  \BibitemOpen
  \bibfield  {author} {\bibinfo {author} {\bibfnamefont {R.~C.}\ \bibnamefont
  {{Tolman}}},\ }\bibfield  {title} {\bibinfo {title} {{Static solutions of
  Einstein's field equations for spheres of fluid}},\ }\href
  {https://doi.org/10.1103/PhysRev.55.364} {\bibfield  {journal} {\bibinfo
  {journal} {Phys. Rev.}\ }\textbf {\bibinfo {volume} {55}},\ \bibinfo {pages}
  {364} (\bibinfo {year} {1939})}\BibitemShut {NoStop}%
\bibitem [{\citenamefont {Oppenheimer}\ and\ \citenamefont
  {Volkoff}(1939)}]{Oppenheimer:1939ne}%
  \BibitemOpen
  \bibfield  {author} {\bibinfo {author} {\bibfnamefont {J.~R.}\ \bibnamefont
  {Oppenheimer}}\ and\ \bibinfo {author} {\bibfnamefont {G.~M.}\ \bibnamefont
  {Volkoff}},\ }\bibfield  {title} {\bibinfo {title} {{On massive neutron
  cores}},\ }\href {https://doi.org/10.1103/PhysRev.55.374} {\bibfield
  {journal} {\bibinfo  {journal} {Phys. Rev.}\ }\textbf {\bibinfo {volume}
  {55}},\ \bibinfo {pages} {374} (\bibinfo {year} {1939})}\BibitemShut
  {NoStop}%
\bibitem [{\citenamefont {Tews}\ \emph {et~al.}(2013)\citenamefont {Tews},
  \citenamefont {Kr\"uger}, \citenamefont {Hebeler},\ and\ \citenamefont
  {Schwenk}}]{Tews:2012fj}%
  \BibitemOpen
  \bibfield  {author} {\bibinfo {author} {\bibfnamefont {I.}~\bibnamefont
  {Tews}}, \bibinfo {author} {\bibfnamefont {T.}~\bibnamefont {Kr\"uger}},
  \bibinfo {author} {\bibfnamefont {K.}~\bibnamefont {Hebeler}},\ and\ \bibinfo
  {author} {\bibfnamefont {A.}~\bibnamefont {Schwenk}},\ }\bibfield  {title}
  {\bibinfo {title} {{Neutron matter at next-to-next-to-next-to-leading order
  in chiral effective field theory}},\ }\href
  {https://doi.org/10.1103/PhysRevLett.110.032504} {\bibfield  {journal}
  {\bibinfo  {journal} {Phys. Rev. Lett.}\ }\textbf {\bibinfo {volume} {110}},\
  \bibinfo {pages} {032504} (\bibinfo {year} {2013})},\ \Eprint
  {https://arxiv.org/abs/1206.0025} {arXiv:1206.0025 [nucl-th]} \BibitemShut
  {NoStop}%
\bibitem [{\citenamefont {Drischler}\ \emph {et~al.}(2019)\citenamefont
  {Drischler}, \citenamefont {Hebeler},\ and\ \citenamefont
  {Schwenk}}]{Drischler:2017wtt}%
  \BibitemOpen
  \bibfield  {author} {\bibinfo {author} {\bibfnamefont {C.}~\bibnamefont
  {Drischler}}, \bibinfo {author} {\bibfnamefont {K.}~\bibnamefont {Hebeler}},\
  and\ \bibinfo {author} {\bibfnamefont {A.}~\bibnamefont {Schwenk}},\
  }\bibfield  {title} {\bibinfo {title} {{Chiral interactions up to
  next-to-next-to-next-to-leading order and nuclear saturation}},\ }\href
  {https://doi.org/10.1103/PhysRevLett.122.042501} {\bibfield  {journal}
  {\bibinfo  {journal} {Phys. Rev. Lett.}\ }\textbf {\bibinfo {volume} {122}},\
  \bibinfo {pages} {042501} (\bibinfo {year} {2019})},\ \Eprint
  {https://arxiv.org/abs/1710.08220} {arXiv:1710.08220 [nucl-th]} \BibitemShut
  {NoStop}%
\bibitem [{\citenamefont {Lynn}\ \emph {et~al.}(2016)\citenamefont {Lynn},
  \citenamefont {Tews}, \citenamefont {Carlson}, \citenamefont {Gandolfi},
  \citenamefont {Gezerlis}, \citenamefont {Schmidt},\ and\ \citenamefont
  {Schwenk}}]{Lynn:2015jua}%
  \BibitemOpen
  \bibfield  {author} {\bibinfo {author} {\bibfnamefont {J.~E.}\ \bibnamefont
  {Lynn}}, \bibinfo {author} {\bibfnamefont {I.}~\bibnamefont {Tews}}, \bibinfo
  {author} {\bibfnamefont {J.}~\bibnamefont {Carlson}}, \bibinfo {author}
  {\bibfnamefont {S.}~\bibnamefont {Gandolfi}}, \bibinfo {author}
  {\bibfnamefont {A.}~\bibnamefont {Gezerlis}}, \bibinfo {author}
  {\bibfnamefont {K.~E.}\ \bibnamefont {Schmidt}},\ and\ \bibinfo {author}
  {\bibfnamefont {A.}~\bibnamefont {Schwenk}},\ }\bibfield  {title} {\bibinfo
  {title} {{Chiral three-nucleon interactions in light nuclei, neutron-$\alpha$
  scattering, and neutron matter}},\ }\href
  {https://doi.org/10.1103/PhysRevLett.116.062501} {\bibfield  {journal}
  {\bibinfo  {journal} {Phys. Rev. Lett.}\ }\textbf {\bibinfo {volume} {116}},\
  \bibinfo {pages} {062501} (\bibinfo {year} {2016})},\ \Eprint
  {https://arxiv.org/abs/1509.03470} {arXiv:1509.03470 [nucl-th]} \BibitemShut
  {NoStop}%
\bibitem [{\citenamefont {Holt}\ \emph {et~al.}(2016)\citenamefont {Holt},
  \citenamefont {Rho},\ and\ \citenamefont {Weise}}]{Holt:2014hma}%
  \BibitemOpen
  \bibfield  {author} {\bibinfo {author} {\bibfnamefont {J.~W.}\ \bibnamefont
  {Holt}}, \bibinfo {author} {\bibfnamefont {M.}~\bibnamefont {Rho}},\ and\
  \bibinfo {author} {\bibfnamefont {W.}~\bibnamefont {Weise}},\ }\bibfield
  {title} {\bibinfo {title} {{Chiral symmetry and effective field theories for
  hadronic, nuclear and stellar matter}},\ }\href
  {https://doi.org/10.1016/j.physrep.2015.10.011} {\bibfield  {journal}
  {\bibinfo  {journal} {Phys. Rept.}\ }\textbf {\bibinfo {volume} {621}},\
  \bibinfo {pages} {2} (\bibinfo {year} {2016})},\ \Eprint
  {https://arxiv.org/abs/1411.6681} {arXiv:1411.6681 [nucl-th]} \BibitemShut
  {NoStop}%
\bibitem [{\citenamefont {Kurkela}\ \emph {et~al.}(2010)\citenamefont
  {Kurkela}, \citenamefont {Romatschke},\ and\ \citenamefont
  {Vuorinen}}]{Kurkela:2009gj}%
  \BibitemOpen
  \bibfield  {author} {\bibinfo {author} {\bibfnamefont {A.}~\bibnamefont
  {Kurkela}}, \bibinfo {author} {\bibfnamefont {P.}~\bibnamefont
  {Romatschke}},\ and\ \bibinfo {author} {\bibfnamefont {A.}~\bibnamefont
  {Vuorinen}},\ }\bibfield  {title} {\bibinfo {title} {{Cold quark matter}},\
  }\href {https://doi.org/10.1103/PhysRevD.81.105021} {\bibfield  {journal}
  {\bibinfo  {journal} {Phys. Rev. D}\ }\textbf {\bibinfo {volume} {81}},\
  \bibinfo {pages} {105021} (\bibinfo {year} {2010})},\ \Eprint
  {https://arxiv.org/abs/0912.1856} {arXiv:0912.1856 [hep-ph]} \BibitemShut
  {NoStop}%
\bibitem [{\citenamefont {Gorda}\ \emph {et~al.}(2018)\citenamefont {Gorda},
  \citenamefont {Kurkela}, \citenamefont {Romatschke}, \citenamefont
  {S\"appi},\ and\ \citenamefont {Vuorinen}}]{Gorda:2018gpy}%
  \BibitemOpen
  \bibfield  {author} {\bibinfo {author} {\bibfnamefont {T.}~\bibnamefont
  {Gorda}}, \bibinfo {author} {\bibfnamefont {A.}~\bibnamefont {Kurkela}},
  \bibinfo {author} {\bibfnamefont {P.}~\bibnamefont {Romatschke}}, \bibinfo
  {author} {\bibfnamefont {S.}~\bibnamefont {S\"appi}},\ and\ \bibinfo {author}
  {\bibfnamefont {A.}~\bibnamefont {Vuorinen}},\ }\bibfield  {title} {\bibinfo
  {title} {{Next-to-next-to-next-to-leading order pressure of cold quark
  matter: Leading logarithm}},\ }\href
  {https://doi.org/10.1103/PhysRevLett.121.202701} {\bibfield  {journal}
  {\bibinfo  {journal} {Phys. Rev. Lett.}\ }\textbf {\bibinfo {volume} {121}},\
  \bibinfo {pages} {202701} (\bibinfo {year} {2018})},\ \Eprint
  {https://arxiv.org/abs/1807.04120} {arXiv:1807.04120 [hep-ph]} \BibitemShut
  {NoStop}%
\bibitem [{\citenamefont {Gorda}\ \emph
  {et~al.}(2021{\natexlab{a}})\citenamefont {Gorda}, \citenamefont {Kurkela},
  \citenamefont {Paatelainen}, \citenamefont {S\"appi},\ and\ \citenamefont
  {Vuorinen}}]{Gorda:2021znl}%
  \BibitemOpen
  \bibfield  {author} {\bibinfo {author} {\bibfnamefont {T.}~\bibnamefont
  {Gorda}}, \bibinfo {author} {\bibfnamefont {A.}~\bibnamefont {Kurkela}},
  \bibinfo {author} {\bibfnamefont {R.}~\bibnamefont {Paatelainen}}, \bibinfo
  {author} {\bibfnamefont {S.}~\bibnamefont {S\"appi}},\ and\ \bibinfo {author}
  {\bibfnamefont {A.}~\bibnamefont {Vuorinen}},\ }\bibfield  {title} {\bibinfo
  {title} {{Soft interactions in cold quark matter}},\ }\href
  {https://doi.org/10.1103/PhysRevLett.127.162003} {\bibfield  {journal}
  {\bibinfo  {journal} {Phys. Rev. Lett.}\ }\textbf {\bibinfo {volume} {127}},\
  \bibinfo {pages} {162003} (\bibinfo {year} {2021}{\natexlab{a}})},\ \Eprint
  {https://arxiv.org/abs/2103.05658} {arXiv:2103.05658 [hep-ph]} \BibitemShut
  {NoStop}%
\bibitem [{\citenamefont {Gorda}\ \emph
  {et~al.}(2021{\natexlab{b}})\citenamefont {Gorda}, \citenamefont {Kurkela},
  \citenamefont {Paatelainen}, \citenamefont {S\"appi},\ and\ \citenamefont
  {Vuorinen}}]{Gorda:2021kme}%
  \BibitemOpen
  \bibfield  {author} {\bibinfo {author} {\bibfnamefont {T.}~\bibnamefont
  {Gorda}}, \bibinfo {author} {\bibfnamefont {A.}~\bibnamefont {Kurkela}},
  \bibinfo {author} {\bibfnamefont {R.}~\bibnamefont {Paatelainen}}, \bibinfo
  {author} {\bibfnamefont {S.}~\bibnamefont {S\"appi}},\ and\ \bibinfo {author}
  {\bibfnamefont {A.}~\bibnamefont {Vuorinen}},\ }\bibfield  {title} {\bibinfo
  {title} {Cold quark matter at $\mathrm{N}^{3}\mathrm{LO}$: Soft
  contributions},\ }\href {https://doi.org/10.1103/PhysRevD.104.074015}
  {\bibfield  {journal} {\bibinfo  {journal} {Phys. Rev. D}\ }\textbf {\bibinfo
  {volume} {104}},\ \bibinfo {pages} {074015} (\bibinfo {year}
  {2021}{\natexlab{b}})},\ \Eprint {https://arxiv.org/abs/2103.07427}
  {arXiv:2103.07427 [hep-ph]} \BibitemShut {NoStop}%
\bibitem [{\citenamefont {Annala}\ \emph
  {et~al.}(2018{\natexlab{a}})\citenamefont {Annala}, \citenamefont {Gorda},
  \citenamefont {Kurkela},\ and\ \citenamefont {Vuorinen}}]{Annala:2017llu}%
  \BibitemOpen
  \bibfield  {author} {\bibinfo {author} {\bibfnamefont {E.}~\bibnamefont
  {Annala}}, \bibinfo {author} {\bibfnamefont {T.}~\bibnamefont {Gorda}},
  \bibinfo {author} {\bibfnamefont {A.}~\bibnamefont {Kurkela}},\ and\ \bibinfo
  {author} {\bibfnamefont {A.}~\bibnamefont {Vuorinen}},\ }\bibfield  {title}
  {\bibinfo {title} {{Gravitational-wave constraints on the neutron-star-matter
  equation of state}},\ }\href {https://doi.org/10.1103/PhysRevLett.120.172703}
  {\bibfield  {journal} {\bibinfo  {journal} {Phys. Rev. Lett.}\ }\textbf
  {\bibinfo {volume} {120}},\ \bibinfo {pages} {172703} (\bibinfo {year}
  {2018}{\natexlab{a}})},\ \Eprint {https://arxiv.org/abs/1711.02644}
  {arXiv:1711.02644 [astro-ph.HE]} \BibitemShut {NoStop}%
\bibitem [{\citenamefont {Margalit}\ and\ \citenamefont
  {Metzger}(2017)}]{Margalit:2017dij}%
  \BibitemOpen
  \bibfield  {author} {\bibinfo {author} {\bibfnamefont {B.}~\bibnamefont
  {Margalit}}\ and\ \bibinfo {author} {\bibfnamefont {B.~D.}\ \bibnamefont
  {Metzger}},\ }\bibfield  {title} {\bibinfo {title} {{Constraining the maximum
  mass of neutron stars from multi-messenger observations of GW170817}},\
  }\href {https://doi.org/10.3847/2041-8213/aa991c} {\bibfield  {journal}
  {\bibinfo  {journal} {Astrophys. J. Lett.}\ }\textbf {\bibinfo {volume}
  {850}},\ \bibinfo {pages} {L19} (\bibinfo {year} {2017})},\ \Eprint
  {https://arxiv.org/abs/1710.05938} {arXiv:1710.05938 [astro-ph.HE]}
  \BibitemShut {NoStop}%
\bibitem [{\citenamefont {Rezzolla}\ \emph {et~al.}(2018)\citenamefont
  {Rezzolla}, \citenamefont {Most},\ and\ \citenamefont
  {Weih}}]{Rezzolla:2017aly}%
  \BibitemOpen
  \bibfield  {author} {\bibinfo {author} {\bibfnamefont {L.}~\bibnamefont
  {Rezzolla}}, \bibinfo {author} {\bibfnamefont {E.~R.}\ \bibnamefont {Most}},\
  and\ \bibinfo {author} {\bibfnamefont {L.~R.}\ \bibnamefont {Weih}},\
  }\bibfield  {title} {\bibinfo {title} {{Using gravitational-wave observations
  and quasi-universal relations to constrain the maximum mass of neutron
  stars}},\ }\href {https://doi.org/10.3847/2041-8213/aaa401} {\bibfield
  {journal} {\bibinfo  {journal} {Astrophys. J. Lett.}\ }\textbf {\bibinfo
  {volume} {852}},\ \bibinfo {pages} {L25} (\bibinfo {year} {2018})},\ \Eprint
  {https://arxiv.org/abs/1711.00314} {arXiv:1711.00314 [astro-ph.HE]}
  \BibitemShut {NoStop}%
\bibitem [{\citenamefont {Ruiz}\ \emph {et~al.}(2018)\citenamefont {Ruiz},
  \citenamefont {Shapiro},\ and\ \citenamefont {Tsokaros}}]{Ruiz:2017due}%
  \BibitemOpen
  \bibfield  {author} {\bibinfo {author} {\bibfnamefont {M.}~\bibnamefont
  {Ruiz}}, \bibinfo {author} {\bibfnamefont {S.~L.}\ \bibnamefont {Shapiro}},\
  and\ \bibinfo {author} {\bibfnamefont {A.}~\bibnamefont {Tsokaros}},\
  }\bibfield  {title} {\bibinfo {title} {{GW170817, general relativistic
  magnetohydrodynamic simulations, and the neutron star maximum mass}},\ }\href
  {https://doi.org/10.1103/PhysRevD.97.021501} {\bibfield  {journal} {\bibinfo
  {journal} {Phys. Rev. D}\ }\textbf {\bibinfo {volume} {97}},\ \bibinfo
  {pages} {021501(R)} (\bibinfo {year} {2018})},\ \Eprint
  {https://arxiv.org/abs/1711.00473} {arXiv:1711.00473 [astro-ph.HE]}
  \BibitemShut {NoStop}%
\bibitem [{\citenamefont {Bauswein}\ \emph {et~al.}(2017)\citenamefont
  {Bauswein}, \citenamefont {Just}, \citenamefont {Janka},\ and\ \citenamefont
  {Stergioulas}}]{Bauswein:2017vtn}%
  \BibitemOpen
  \bibfield  {author} {\bibinfo {author} {\bibfnamefont {A.}~\bibnamefont
  {Bauswein}}, \bibinfo {author} {\bibfnamefont {O.}~\bibnamefont {Just}},
  \bibinfo {author} {\bibfnamefont {H.-T.}\ \bibnamefont {Janka}},\ and\
  \bibinfo {author} {\bibfnamefont {N.}~\bibnamefont {Stergioulas}},\
  }\bibfield  {title} {\bibinfo {title} {{Neutron-star radius constraints from
  GW170817 and future detections}},\ }\href
  {https://doi.org/10.3847/2041-8213/aa9994} {\bibfield  {journal} {\bibinfo
  {journal} {Astrophys. J. Lett.}\ }\textbf {\bibinfo {volume} {850}},\
  \bibinfo {pages} {L34} (\bibinfo {year} {2017})},\ \Eprint
  {https://arxiv.org/abs/1710.06843} {arXiv:1710.06843 [astro-ph.HE]}
  \BibitemShut {NoStop}%
\bibitem [{\citenamefont {Radice}\ \emph {et~al.}(2018)\citenamefont {Radice},
  \citenamefont {Perego}, \citenamefont {Zappa},\ and\ \citenamefont
  {Bernuzzi}}]{Radice:2017lry}%
  \BibitemOpen
  \bibfield  {author} {\bibinfo {author} {\bibfnamefont {D.}~\bibnamefont
  {Radice}}, \bibinfo {author} {\bibfnamefont {A.}~\bibnamefont {Perego}},
  \bibinfo {author} {\bibfnamefont {F.}~\bibnamefont {Zappa}},\ and\ \bibinfo
  {author} {\bibfnamefont {S.}~\bibnamefont {Bernuzzi}},\ }\bibfield  {title}
  {\bibinfo {title} {{GW170817: Joint constraint on the neutron star equation
  of state from multimessenger observations}},\ }\href
  {https://doi.org/10.3847/2041-8213/aaa402} {\bibfield  {journal} {\bibinfo
  {journal} {Astrophys. J. Lett.}\ }\textbf {\bibinfo {volume} {852}},\
  \bibinfo {pages} {L29} (\bibinfo {year} {2018})},\ \Eprint
  {https://arxiv.org/abs/1711.03647} {arXiv:1711.03647 [astro-ph.HE]}
  \BibitemShut {NoStop}%
\bibitem [{\citenamefont {Most}\ \emph {et~al.}(2018)\citenamefont {Most},
  \citenamefont {Weih}, \citenamefont {Rezzolla},\ and\ \citenamefont
  {Schaffner-Bielich}}]{Most:2018hfd}%
  \BibitemOpen
  \bibfield  {author} {\bibinfo {author} {\bibfnamefont {E.~R.}\ \bibnamefont
  {Most}}, \bibinfo {author} {\bibfnamefont {L.~R.}\ \bibnamefont {Weih}},
  \bibinfo {author} {\bibfnamefont {L.}~\bibnamefont {Rezzolla}},\ and\
  \bibinfo {author} {\bibfnamefont {J.}~\bibnamefont {Schaffner-Bielich}},\
  }\bibfield  {title} {\bibinfo {title} {{New constraints on radii and tidal
  deformabilities of neutron stars from GW170817}},\ }\href
  {https://doi.org/10.1103/PhysRevLett.120.261103} {\bibfield  {journal}
  {\bibinfo  {journal} {Phys. Rev. Lett.}\ }\textbf {\bibinfo {volume} {120}},\
  \bibinfo {pages} {261103} (\bibinfo {year} {2018})},\ \Eprint
  {https://arxiv.org/abs/1803.00549} {arXiv:1803.00549 [gr-qc]} \BibitemShut
  {NoStop}%
\bibitem [{\citenamefont {Dietrich}\ \emph {et~al.}(2020)\citenamefont
  {Dietrich}, \citenamefont {Coughlin}, \citenamefont {Pang}, \citenamefont
  {Bulla}, \citenamefont {Heinzel}, \citenamefont {Issa}, \citenamefont
  {Tews},\ and\ \citenamefont {Antier}}]{Dietrich:2020efo}%
  \BibitemOpen
  \bibfield  {author} {\bibinfo {author} {\bibfnamefont {T.}~\bibnamefont
  {Dietrich}}, \bibinfo {author} {\bibfnamefont {M.~W.}\ \bibnamefont
  {Coughlin}}, \bibinfo {author} {\bibfnamefont {P.~T.~H.}\ \bibnamefont
  {Pang}}, \bibinfo {author} {\bibfnamefont {M.}~\bibnamefont {Bulla}},
  \bibinfo {author} {\bibfnamefont {J.}~\bibnamefont {Heinzel}}, \bibinfo
  {author} {\bibfnamefont {L.}~\bibnamefont {Issa}}, \bibinfo {author}
  {\bibfnamefont {I.}~\bibnamefont {Tews}},\ and\ \bibinfo {author}
  {\bibfnamefont {S.}~\bibnamefont {Antier}},\ }\bibfield  {title} {\bibinfo
  {title} {{Multimessenger constraints on the neutron-star equation of state
  and the Hubble constant}},\ }\href {https://doi.org/10.1126/science.abb4317}
  {\bibfield  {journal} {\bibinfo  {journal} {Science}\ }\textbf {\bibinfo
  {volume} {370}},\ \bibinfo {pages} {1450} (\bibinfo {year} {2020})},\ \Eprint
  {https://arxiv.org/abs/2002.11355} {arXiv:2002.11355 [astro-ph.HE]}
  \BibitemShut {NoStop}%
\bibitem [{\citenamefont {Capano}\ \emph {et~al.}(2020)\citenamefont {Capano},
  \citenamefont {Tews}, \citenamefont {Brown}, \citenamefont {Margalit},
  \citenamefont {De}, \citenamefont {Kumar}, \citenamefont {Brown},
  \citenamefont {Krishnan},\ and\ \citenamefont {Reddy}}]{Capano:2019eae}%
  \BibitemOpen
  \bibfield  {author} {\bibinfo {author} {\bibfnamefont {C.~D.}\ \bibnamefont
  {Capano}}, \bibinfo {author} {\bibfnamefont {I.}~\bibnamefont {Tews}},
  \bibinfo {author} {\bibfnamefont {S.~M.}\ \bibnamefont {Brown}}, \bibinfo
  {author} {\bibfnamefont {B.}~\bibnamefont {Margalit}}, \bibinfo {author}
  {\bibfnamefont {S.}~\bibnamefont {De}}, \bibinfo {author} {\bibfnamefont
  {S.}~\bibnamefont {Kumar}}, \bibinfo {author} {\bibfnamefont {D.~A.}\
  \bibnamefont {Brown}}, \bibinfo {author} {\bibfnamefont {B.}~\bibnamefont
  {Krishnan}},\ and\ \bibinfo {author} {\bibfnamefont {S.}~\bibnamefont
  {Reddy}},\ }\bibfield  {title} {\bibinfo {title} {{Stringent constraints on
  neutron-star radii from multimessenger observations and nuclear theory}},\
  }\href {https://doi.org/10.1038/s41550-020-1014-6} {\bibfield  {journal}
  {\bibinfo  {journal} {Nat. Astron.}\ }\textbf {\bibinfo {volume} {4}},\
  \bibinfo {pages} {625} (\bibinfo {year} {2020})},\ \Eprint
  {https://arxiv.org/abs/1908.10352} {arXiv:1908.10352 [astro-ph.HE]}
  \BibitemShut {NoStop}%
\bibitem [{\citenamefont {Landry}\ and\ \citenamefont
  {Essick}(2019)}]{Landry:2018prl}%
  \BibitemOpen
  \bibfield  {author} {\bibinfo {author} {\bibfnamefont {P.}~\bibnamefont
  {Landry}}\ and\ \bibinfo {author} {\bibfnamefont {R.}~\bibnamefont
  {Essick}},\ }\bibfield  {title} {\bibinfo {title} {{Nonparametric inference
  of the neutron star equation of state from gravitational wave
  observations}},\ }\href {https://doi.org/10.1103/PhysRevD.99.084049}
  {\bibfield  {journal} {\bibinfo  {journal} {Phys. Rev. D}\ }\textbf {\bibinfo
  {volume} {99}},\ \bibinfo {pages} {084049} (\bibinfo {year} {2019})},\
  \Eprint {https://arxiv.org/abs/1811.12529} {arXiv:1811.12529 [gr-qc]}
  \BibitemShut {NoStop}%
\bibitem [{\citenamefont {Raithel}\ \emph {et~al.}(2018)\citenamefont
  {Raithel}, \citenamefont {\"Ozel},\ and\ \citenamefont
  {Psaltis}}]{Raithel:2018ncd}%
  \BibitemOpen
  \bibfield  {author} {\bibinfo {author} {\bibfnamefont {C.~A.}\ \bibnamefont
  {Raithel}}, \bibinfo {author} {\bibfnamefont {F.}~\bibnamefont {\"Ozel}},\
  and\ \bibinfo {author} {\bibfnamefont {D.}~\bibnamefont {Psaltis}},\
  }\bibfield  {title} {\bibinfo {title} {{Tidal deformability from GW170817 as
  a direct probe of the neutron star radius}},\ }\href
  {https://doi.org/10.3847/2041-8213/aabcbf} {\bibfield  {journal} {\bibinfo
  {journal} {Astrophys. J. Lett.}\ }\textbf {\bibinfo {volume} {857}},\
  \bibinfo {pages} {L23} (\bibinfo {year} {2018})},\ \Eprint
  {https://arxiv.org/abs/1803.07687} {arXiv:1803.07687 [astro-ph.HE]}
  \BibitemShut {NoStop}%
\bibitem [{\citenamefont {Raithel}\ and\ \citenamefont
  {\"Ozel}(2019)}]{Raithel:2019ejc}%
  \BibitemOpen
  \bibfield  {author} {\bibinfo {author} {\bibfnamefont {C.~A.}\ \bibnamefont
  {Raithel}}\ and\ \bibinfo {author} {\bibfnamefont {F.}~\bibnamefont
  {\"Ozel}},\ }\bibfield  {title} {\bibinfo {title} {{Measurement of the
  nuclear symmetry energy parameters from gravitational-wave events}},\ }\href
  {https://doi.org/10.3847/1538-4357/ab48e6} {\bibfield  {journal} {\bibinfo
  {journal} {Astrophys. J.}\ }\textbf {\bibinfo {volume} {885}},\ \bibinfo
  {pages} {121} (\bibinfo {year} {2019})},\ \Eprint
  {https://arxiv.org/abs/1908.00018} {arXiv:1908.00018 [astro-ph.HE]}
  \BibitemShut {NoStop}%
\bibitem [{\citenamefont {Raaijmakers}\ \emph {et~al.}(2020)\citenamefont
  {Raaijmakers} \emph {et~al.}}]{Raaijmakers:2019dks}%
  \BibitemOpen
  \bibfield  {author} {\bibinfo {author} {\bibfnamefont {G.}~\bibnamefont
  {Raaijmakers}} \emph {et~al.},\ }\bibfield  {title} {\bibinfo {title}
  {{Constraining the dense matter equation of state with joint analysis of
  NICER and LIGO/Virgo measurements}},\ }\href
  {https://doi.org/10.3847/2041-8213/ab822f} {\bibfield  {journal} {\bibinfo
  {journal} {Astrophys. J. Lett.}\ }\textbf {\bibinfo {volume} {893}},\
  \bibinfo {pages} {L21} (\bibinfo {year} {2020})},\ \Eprint
  {https://arxiv.org/abs/1912.11031} {arXiv:1912.11031 [astro-ph.HE]}
  \BibitemShut {NoStop}%
\bibitem [{\citenamefont {Essick}\ \emph {et~al.}(2020)\citenamefont {Essick},
  \citenamefont {Landry},\ and\ \citenamefont {Holz}}]{Essick:2019ldf}%
  \BibitemOpen
  \bibfield  {author} {\bibinfo {author} {\bibfnamefont {R.}~\bibnamefont
  {Essick}}, \bibinfo {author} {\bibfnamefont {P.}~\bibnamefont {Landry}},\
  and\ \bibinfo {author} {\bibfnamefont {D.~E.}\ \bibnamefont {Holz}},\
  }\bibfield  {title} {\bibinfo {title} {{Nonparametric inference of neutron
  star composition, equation of state, and maximum mass with GW170817}},\
  }\href {https://doi.org/10.1103/PhysRevD.101.063007} {\bibfield  {journal}
  {\bibinfo  {journal} {Phys. Rev. D}\ }\textbf {\bibinfo {volume} {101}},\
  \bibinfo {pages} {063007} (\bibinfo {year} {2020})},\ \Eprint
  {https://arxiv.org/abs/1910.09740} {arXiv:1910.09740 [astro-ph.HE]}
  \BibitemShut {NoStop}%
\bibitem [{\citenamefont {Jokela}\ \emph {et~al.}(2021)\citenamefont {Jokela},
  \citenamefont {J\"arvinen}, \citenamefont {Nijs},\ and\ \citenamefont
  {Remes}}]{Jokela:2020piw}%
  \BibitemOpen
  \bibfield  {author} {\bibinfo {author} {\bibfnamefont {N.}~\bibnamefont
  {Jokela}}, \bibinfo {author} {\bibfnamefont {M.}~\bibnamefont {J\"arvinen}},
  \bibinfo {author} {\bibfnamefont {G.}~\bibnamefont {Nijs}},\ and\ \bibinfo
  {author} {\bibfnamefont {J.}~\bibnamefont {Remes}},\ }\bibfield  {title}
  {\bibinfo {title} {{Unified weak and strong coupling framework for nuclear
  matter and neutron stars}},\ }\href
  {https://doi.org/10.1103/PhysRevD.103.086004} {\bibfield  {journal} {\bibinfo
   {journal} {Phys. Rev. D}\ }\textbf {\bibinfo {volume} {103}},\ \bibinfo
  {pages} {086004} (\bibinfo {year} {2021})},\ \Eprint
  {https://arxiv.org/abs/2006.01141} {arXiv:2006.01141 [hep-ph]} \BibitemShut
  {NoStop}%
\bibitem [{\citenamefont {Al-Mamun}\ \emph {et~al.}(2021)\citenamefont
  {Al-Mamun}, \citenamefont {Steiner}, \citenamefont {N\"attil\"a},
  \citenamefont {Lange}, \citenamefont {O'Shaughnessy}, \citenamefont {Tews},
  \citenamefont {Gandolfi}, \citenamefont {Heinke},\ and\ \citenamefont
  {Han}}]{Al-Mamun:2020vzu}%
  \BibitemOpen
  \bibfield  {author} {\bibinfo {author} {\bibfnamefont {M.}~\bibnamefont
  {Al-Mamun}}, \bibinfo {author} {\bibfnamefont {A.~W.}\ \bibnamefont
  {Steiner}}, \bibinfo {author} {\bibfnamefont {J.}~\bibnamefont
  {N\"attil\"a}}, \bibinfo {author} {\bibfnamefont {J.}~\bibnamefont {Lange}},
  \bibinfo {author} {\bibfnamefont {R.}~\bibnamefont {O'Shaughnessy}}, \bibinfo
  {author} {\bibfnamefont {I.}~\bibnamefont {Tews}}, \bibinfo {author}
  {\bibfnamefont {S.}~\bibnamefont {Gandolfi}}, \bibinfo {author}
  {\bibfnamefont {C.}~\bibnamefont {Heinke}},\ and\ \bibinfo {author}
  {\bibfnamefont {S.}~\bibnamefont {Han}},\ }\bibfield  {title} {\bibinfo
  {title} {{Combining electromagnetic and gravitational-wave constraints on
  neutron-star masses and radii}},\ }\href
  {https://doi.org/10.1103/PhysRevLett.126.061101} {\bibfield  {journal}
  {\bibinfo  {journal} {Phys. Rev. Lett.}\ }\textbf {\bibinfo {volume} {126}},\
  \bibinfo {pages} {061101} (\bibinfo {year} {2021})},\ \Eprint
  {https://arxiv.org/abs/2008.12817} {arXiv:2008.12817 [astro-ph.HE]}
  \BibitemShut {NoStop}%
\bibitem [{\citenamefont {Essick}\ \emph {et~al.}(2021)\citenamefont {Essick},
  \citenamefont {Tews}, \citenamefont {Landry},\ and\ \citenamefont
  {Schwenk}}]{Essick:2021kjb}%
  \BibitemOpen
  \bibfield  {author} {\bibinfo {author} {\bibfnamefont {R.}~\bibnamefont
  {Essick}}, \bibinfo {author} {\bibfnamefont {I.}~\bibnamefont {Tews}},
  \bibinfo {author} {\bibfnamefont {P.}~\bibnamefont {Landry}},\ and\ \bibinfo
  {author} {\bibfnamefont {A.}~\bibnamefont {Schwenk}},\ }\bibfield  {title}
  {\bibinfo {title} {{Astrophysical constraints on the symmetry energy and the
  neutron skin of {$^{208}$Pb} with minimal modeling assumptions}},\ }\href
  {https://doi.org/10.1103/PhysRevLett.127.192701} {\bibfield  {journal}
  {\bibinfo  {journal} {Phys. Rev. Lett.}\ }\textbf {\bibinfo {volume} {127}},\
  \bibinfo {pages} {192701} (\bibinfo {year} {2021})},\ \Eprint
  {https://arxiv.org/abs/2102.10074} {arXiv:2102.10074 [nucl-th]} \BibitemShut
  {NoStop}%
\bibitem [{\citenamefont {Baym}\ \emph {et~al.}(2018)\citenamefont {Baym},
  \citenamefont {Hatsuda}, \citenamefont {Kojo}, \citenamefont {Powell},
  \citenamefont {Song},\ and\ \citenamefont {Takatsuka}}]{Baym:2017whm}%
  \BibitemOpen
  \bibfield  {author} {\bibinfo {author} {\bibfnamefont {G.}~\bibnamefont
  {Baym}}, \bibinfo {author} {\bibfnamefont {T.}~\bibnamefont {Hatsuda}},
  \bibinfo {author} {\bibfnamefont {T.}~\bibnamefont {Kojo}}, \bibinfo {author}
  {\bibfnamefont {P.~D.}\ \bibnamefont {Powell}}, \bibinfo {author}
  {\bibfnamefont {Y.}~\bibnamefont {Song}},\ and\ \bibinfo {author}
  {\bibfnamefont {T.}~\bibnamefont {Takatsuka}},\ }\bibfield  {title} {\bibinfo
  {title} {{From hadrons to quarks in neutron stars: a review}},\ }\href
  {https://doi.org/10.1088/1361-6633/aaae14} {\bibfield  {journal} {\bibinfo
  {journal} {Rept. Prog. Phys.}\ }\textbf {\bibinfo {volume} {81}},\ \bibinfo
  {pages} {056902} (\bibinfo {year} {2018})},\ \Eprint
  {https://arxiv.org/abs/1707.04966} {arXiv:1707.04966 [astro-ph.HE]}
  \BibitemShut {NoStop}%
\bibitem [{\citenamefont {Gandolfi}\ \emph {et~al.}(2019)\citenamefont
  {Gandolfi}, \citenamefont {Lippuner}, \citenamefont {Steiner}, \citenamefont
  {Tews}, \citenamefont {Du},\ and\ \citenamefont
  {Al-Mamun}}]{Gandolfi:2019zpj}%
  \BibitemOpen
  \bibfield  {author} {\bibinfo {author} {\bibfnamefont {S.}~\bibnamefont
  {Gandolfi}}, \bibinfo {author} {\bibfnamefont {J.}~\bibnamefont {Lippuner}},
  \bibinfo {author} {\bibfnamefont {A.~W.}\ \bibnamefont {Steiner}}, \bibinfo
  {author} {\bibfnamefont {I.}~\bibnamefont {Tews}}, \bibinfo {author}
  {\bibfnamefont {X.}~\bibnamefont {Du}},\ and\ \bibinfo {author}
  {\bibfnamefont {M.}~\bibnamefont {Al-Mamun}},\ }\bibfield  {title} {\bibinfo
  {title} {{From the microscopic to the macroscopic world: from nucleons to
  neutron stars}},\ }\href {https://doi.org/10.1088/1361-6471/ab29b3}
  {\bibfield  {journal} {\bibinfo  {journal} {J. Phys. G}\ }\textbf {\bibinfo
  {volume} {46}},\ \bibinfo {pages} {103001} (\bibinfo {year} {2019})},\
  \Eprint {https://arxiv.org/abs/1903.06730} {arXiv:1903.06730 [nucl-th]}
  \BibitemShut {NoStop}%
\bibitem [{\citenamefont {Raithel}(2019)}]{Raithel:2019uzi}%
  \BibitemOpen
  \bibfield  {author} {\bibinfo {author} {\bibfnamefont {C.~A.}\ \bibnamefont
  {Raithel}},\ }\bibfield  {title} {\bibinfo {title} {{Constraints on the
  neutron star equation of state from GW170817}},\ }\href
  {https://doi.org/10.1140/epja/i2019-12759-5} {\bibfield  {journal} {\bibinfo
  {journal} {Eur. Phys. J. A}\ }\textbf {\bibinfo {volume} {55}},\ \bibinfo
  {pages} {80} (\bibinfo {year} {2019})},\ \Eprint
  {https://arxiv.org/abs/1904.10002} {arXiv:1904.10002 [astro-ph.HE]}
  \BibitemShut {NoStop}%
\bibitem [{\citenamefont {Horowitz}(2019)}]{Horowitz:2019piw}%
  \BibitemOpen
  \bibfield  {author} {\bibinfo {author} {\bibfnamefont {C.~J.}\ \bibnamefont
  {Horowitz}},\ }\bibfield  {title} {\bibinfo {title} {{Neutron rich matter in
  the laboratory and in the heavens after GW170817}},\ }\href
  {https://doi.org/10.1016/j.aop.2019.167992} {\bibfield  {journal} {\bibinfo
  {journal} {Annals Phys.}\ }\textbf {\bibinfo {volume} {411}},\ \bibinfo
  {pages} {167992} (\bibinfo {year} {2019})},\ \Eprint
  {https://arxiv.org/abs/1911.00411} {arXiv:1911.00411 [astro-ph.HE]}
  \BibitemShut {NoStop}%
\bibitem [{\citenamefont {Baiotti}(2019)}]{Baiotti:2019sew}%
  \BibitemOpen
  \bibfield  {author} {\bibinfo {author} {\bibfnamefont {L.}~\bibnamefont
  {Baiotti}},\ }\bibfield  {title} {\bibinfo {title} {{Gravitational waves from
  neutron star mergers and their relation to the nuclear equation of state}},\
  }\href {https://doi.org/10.1016/j.ppnp.2019.103714} {\bibfield  {journal}
  {\bibinfo  {journal} {Prog. Part. Nucl. Phys.}\ }\textbf {\bibinfo {volume}
  {109}},\ \bibinfo {pages} {103714} (\bibinfo {year} {2019})},\ \Eprint
  {https://arxiv.org/abs/1907.08534} {arXiv:1907.08534 [astro-ph.HE]}
  \BibitemShut {NoStop}%
\bibitem [{\citenamefont {Chatziioannou}(2020)}]{Chatziioannou:2020pqz}%
  \BibitemOpen
  \bibfield  {author} {\bibinfo {author} {\bibfnamefont {K.}~\bibnamefont
  {Chatziioannou}},\ }\bibfield  {title} {\bibinfo {title} {{Neutron star tidal
  deformability and equation-of-state constraints}},\ }\href
  {https://doi.org/10.1007/s10714-020-02754-3} {\bibfield  {journal} {\bibinfo
  {journal} {Gen. Rel. Grav.}\ }\textbf {\bibinfo {volume} {52}},\ \bibinfo
  {pages} {109} (\bibinfo {year} {2020})},\ \Eprint
  {https://arxiv.org/abs/2006.03168} {arXiv:2006.03168 [gr-qc]} \BibitemShut
  {NoStop}%
\bibitem [{\citenamefont {Radice}\ \emph {et~al.}(2020)\citenamefont {Radice},
  \citenamefont {Bernuzzi},\ and\ \citenamefont {Perego}}]{Radice:2020ddv}%
  \BibitemOpen
  \bibfield  {author} {\bibinfo {author} {\bibfnamefont {D.}~\bibnamefont
  {Radice}}, \bibinfo {author} {\bibfnamefont {S.}~\bibnamefont {Bernuzzi}},\
  and\ \bibinfo {author} {\bibfnamefont {A.}~\bibnamefont {Perego}},\
  }\bibfield  {title} {\bibinfo {title} {{The dynamics of binary neutron star
  mergers and GW170817}},\ }\href
  {https://doi.org/10.1146/annurev-nucl-013120-114541} {\bibfield  {journal}
  {\bibinfo  {journal} {Ann. Rev. Nucl. Part. Sci.}\ }\textbf {\bibinfo
  {volume} {70}},\ \bibinfo {pages} {95} (\bibinfo {year} {2020})},\ \Eprint
  {https://arxiv.org/abs/2002.03863} {arXiv:2002.03863 [astro-ph.HE]}
  \BibitemShut {NoStop}%
\bibitem [{\citenamefont {Paschalidis}\ \emph {et~al.}(2018)\citenamefont
  {Paschalidis}, \citenamefont {Yagi}, \citenamefont {Alvarez-Castillo},
  \citenamefont {Blaschke},\ and\ \citenamefont
  {Sedrakian}}]{Paschalidis:2017qmb}%
  \BibitemOpen
  \bibfield  {author} {\bibinfo {author} {\bibfnamefont {V.}~\bibnamefont
  {Paschalidis}}, \bibinfo {author} {\bibfnamefont {K.}~\bibnamefont {Yagi}},
  \bibinfo {author} {\bibfnamefont {D.}~\bibnamefont {Alvarez-Castillo}},
  \bibinfo {author} {\bibfnamefont {D.~B.}\ \bibnamefont {Blaschke}},\ and\
  \bibinfo {author} {\bibfnamefont {A.}~\bibnamefont {Sedrakian}},\ }\bibfield
  {title} {\bibinfo {title} {{Implications from GW170817 and I-Love-Q relations
  for relativistic hybrid stars}},\ }\href
  {https://doi.org/10.1103/PhysRevD.97.084038} {\bibfield  {journal} {\bibinfo
  {journal} {Phys. Rev. D}\ }\textbf {\bibinfo {volume} {97}},\ \bibinfo
  {pages} {084038} (\bibinfo {year} {2018})},\ \Eprint
  {https://arxiv.org/abs/1712.00451} {arXiv:1712.00451 [astro-ph.HE]}
  \BibitemShut {NoStop}%
\bibitem [{\citenamefont {Annala}\ \emph {et~al.}(2020)\citenamefont {Annala},
  \citenamefont {Gorda}, \citenamefont {Kurkela}, \citenamefont {N\"attil\"a},\
  and\ \citenamefont {Vuorinen}}]{Annala:2019puf}%
  \BibitemOpen
  \bibfield  {author} {\bibinfo {author} {\bibfnamefont {E.}~\bibnamefont
  {Annala}}, \bibinfo {author} {\bibfnamefont {T.}~\bibnamefont {Gorda}},
  \bibinfo {author} {\bibfnamefont {A.}~\bibnamefont {Kurkela}}, \bibinfo
  {author} {\bibfnamefont {J.}~\bibnamefont {N\"attil\"a}},\ and\ \bibinfo
  {author} {\bibfnamefont {A.}~\bibnamefont {Vuorinen}},\ }\bibfield  {title}
  {\bibinfo {title} {{Evidence for quark-matter cores in massive neutron
  stars}},\ }\href {https://doi.org/10.1038/s41567-020-0914-9} {\bibfield
  {journal} {\bibinfo  {journal} {Nat. Phys.}\ }\textbf {\bibinfo {volume}
  {16}},\ \bibinfo {pages} {907} (\bibinfo {year} {2020})},\ \Eprint
  {https://arxiv.org/abs/1903.09121} {arXiv:1903.09121 [astro-ph.HE]}
  \BibitemShut {NoStop}%
\bibitem [{\citenamefont {Ferreira}\ \emph {et~al.}(2020)\citenamefont
  {Ferreira}, \citenamefont {C\^amara~Pereira},\ and\ \citenamefont
  {Provid\^encia}}]{Ferreira:2020kvu}%
  \BibitemOpen
  \bibfield  {author} {\bibinfo {author} {\bibfnamefont {M.}~\bibnamefont
  {Ferreira}}, \bibinfo {author} {\bibfnamefont {R.}~\bibnamefont
  {C\^amara~Pereira}},\ and\ \bibinfo {author} {\bibfnamefont {C.}~\bibnamefont
  {Provid\^encia}},\ }\bibfield  {title} {\bibinfo {title} {{Quark matter in
  light neutron stars}},\ }\href {https://doi.org/10.1103/PhysRevD.102.083030}
  {\bibfield  {journal} {\bibinfo  {journal} {Phys. Rev. D}\ }\textbf {\bibinfo
  {volume} {102}},\ \bibinfo {pages} {083030} (\bibinfo {year} {2020})},\
  \Eprint {https://arxiv.org/abs/2008.12563} {arXiv:2008.12563 [nucl-th]}
  \BibitemShut {NoStop}%
\bibitem [{\citenamefont {Minamikawa}\ \emph {et~al.}(2021)\citenamefont
  {Minamikawa}, \citenamefont {Kojo},\ and\ \citenamefont
  {Harada}}]{Minamikawa:2020jfj}%
  \BibitemOpen
  \bibfield  {author} {\bibinfo {author} {\bibfnamefont {T.}~\bibnamefont
  {Minamikawa}}, \bibinfo {author} {\bibfnamefont {T.}~\bibnamefont {Kojo}},\
  and\ \bibinfo {author} {\bibfnamefont {M.}~\bibnamefont {Harada}},\
  }\bibfield  {title} {\bibinfo {title} {{Quark-hadron crossover equations of
  state for neutron stars: Constraining the chiral invariant mass in a parity
  doublet model}},\ }\href {https://doi.org/10.1103/PhysRevC.103.045205}
  {\bibfield  {journal} {\bibinfo  {journal} {Phys. Rev. C}\ }\textbf {\bibinfo
  {volume} {103}},\ \bibinfo {pages} {045205} (\bibinfo {year} {2021})},\
  \Eprint {https://arxiv.org/abs/2011.13684} {arXiv:2011.13684 [nucl-th]}
  \BibitemShut {NoStop}%
\bibitem [{\citenamefont {Blacker}\ \emph {et~al.}(2020)\citenamefont
  {Blacker}, \citenamefont {Bastian}, \citenamefont {Bauswein}, \citenamefont
  {Blaschke}, \citenamefont {Fischer}, \citenamefont {Oertel}, \citenamefont
  {Soultanis},\ and\ \citenamefont {Typel}}]{Blacker:2020nlq}%
  \BibitemOpen
  \bibfield  {author} {\bibinfo {author} {\bibfnamefont {S.}~\bibnamefont
  {Blacker}}, \bibinfo {author} {\bibfnamefont {N.-U.~F.}\ \bibnamefont
  {Bastian}}, \bibinfo {author} {\bibfnamefont {A.}~\bibnamefont {Bauswein}},
  \bibinfo {author} {\bibfnamefont {D.~B.}\ \bibnamefont {Blaschke}}, \bibinfo
  {author} {\bibfnamefont {T.}~\bibnamefont {Fischer}}, \bibinfo {author}
  {\bibfnamefont {M.}~\bibnamefont {Oertel}}, \bibinfo {author} {\bibfnamefont
  {T.}~\bibnamefont {Soultanis}},\ and\ \bibinfo {author} {\bibfnamefont
  {S.}~\bibnamefont {Typel}},\ }\bibfield  {title} {\bibinfo {title}
  {{Constraining the onset density of the hadron-quark phase transition with
  gravitational-wave observations}},\ }\href
  {https://doi.org/10.1103/PhysRevD.102.123023} {\bibfield  {journal} {\bibinfo
   {journal} {Phys. Rev. D}\ }\textbf {\bibinfo {volume} {102}},\ \bibinfo
  {pages} {123023} (\bibinfo {year} {2020})},\ \Eprint
  {https://arxiv.org/abs/2006.03789} {arXiv:2006.03789 [astro-ph.HE]}
  \BibitemShut {NoStop}%
\bibitem [{\citenamefont {Shibata}\ \emph {et~al.}(2017)\citenamefont
  {Shibata}, \citenamefont {Fujibayashi}, \citenamefont {Hotokezaka},
  \citenamefont {Kiuchi}, \citenamefont {Kyutoku}, \citenamefont {Sekiguchi},\
  and\ \citenamefont {Tanaka}}]{Shibata:2017xdx}%
  \BibitemOpen
  \bibfield  {author} {\bibinfo {author} {\bibfnamefont {M.}~\bibnamefont
  {Shibata}}, \bibinfo {author} {\bibfnamefont {S.}~\bibnamefont
  {Fujibayashi}}, \bibinfo {author} {\bibfnamefont {K.}~\bibnamefont
  {Hotokezaka}}, \bibinfo {author} {\bibfnamefont {K.}~\bibnamefont {Kiuchi}},
  \bibinfo {author} {\bibfnamefont {K.}~\bibnamefont {Kyutoku}}, \bibinfo
  {author} {\bibfnamefont {Y.}~\bibnamefont {Sekiguchi}},\ and\ \bibinfo
  {author} {\bibfnamefont {M.}~\bibnamefont {Tanaka}},\ }\bibfield  {title}
  {\bibinfo {title} {{Modeling GW170817 based on numerical relativity and its
  implications}},\ }\href {https://doi.org/10.1103/PhysRevD.96.123012}
  {\bibfield  {journal} {\bibinfo  {journal} {Phys. Rev. D}\ }\textbf {\bibinfo
  {volume} {96}},\ \bibinfo {pages} {123012} (\bibinfo {year} {2017})},\
  \Eprint {https://arxiv.org/abs/1710.07579} {arXiv:1710.07579 [astro-ph.HE]}
  \BibitemShut {NoStop}%
\bibitem [{\citenamefont {Shibata}\ \emph {et~al.}(2019)\citenamefont
  {Shibata}, \citenamefont {Zhou}, \citenamefont {Kiuchi},\ and\ \citenamefont
  {Fujibayashi}}]{Shibata:2019ctb}%
  \BibitemOpen
  \bibfield  {author} {\bibinfo {author} {\bibfnamefont {M.}~\bibnamefont
  {Shibata}}, \bibinfo {author} {\bibfnamefont {E.}~\bibnamefont {Zhou}},
  \bibinfo {author} {\bibfnamefont {K.}~\bibnamefont {Kiuchi}},\ and\ \bibinfo
  {author} {\bibfnamefont {S.}~\bibnamefont {Fujibayashi}},\ }\bibfield
  {title} {\bibinfo {title} {{Constraint on the maximum mass of neutron stars
  using GW170817 event}},\ }\href {https://doi.org/10.1103/PhysRevD.100.023015}
  {\bibfield  {journal} {\bibinfo  {journal} {Phys. Rev. D}\ }\textbf {\bibinfo
  {volume} {100}},\ \bibinfo {pages} {023015} (\bibinfo {year} {2019})},\
  \Eprint {https://arxiv.org/abs/1905.03656} {arXiv:1905.03656 [astro-ph.HE]}
  \BibitemShut {NoStop}%
\bibitem [{Note1()}]{Note1}%
  \BibitemOpen
  \bibinfo {note} {A differentially rotating stellar configuration has an
  angular velocity that is a function of the distance from the axis of
  rotation.}\BibitemShut {Stop}%
\bibitem [{\citenamefont {Abbott}\ \emph
  {et~al.}(2017{\natexlab{b}})\citenamefont {Abbott} \emph
  {et~al.}}]{GBM:2017lvd}%
  \BibitemOpen
  \bibfield  {author} {\bibinfo {author} {\bibfnamefont {B.~P.}\ \bibnamefont
  {Abbott}} \emph {et~al.},\ }\bibfield  {title} {\bibinfo {title}
  {{Multi-messenger observations of a binary neutron star merger}},\ }\href
  {https://doi.org/10.3847/2041-8213/aa91c9} {\bibfield  {journal} {\bibinfo
  {journal} {Astrophys. J. Lett.}\ }\textbf {\bibinfo {volume} {848}},\
  \bibinfo {pages} {L12} (\bibinfo {year} {2017}{\natexlab{b}})},\ \Eprint
  {https://arxiv.org/abs/1710.05833} {arXiv:1710.05833 [astro-ph.HE]}
  \BibitemShut {NoStop}%
\bibitem [{Note2()}]{Note2}%
  \BibitemOpen
  \bibinfo {note} {The definition of the rest or baryon mass of a star is
  $M_{\protect \rm B}=\DOTSI \intop \ilimits@ {\protect \rm d}^3x\protect \sqrt
  {-g}\rho _{\protect \rm B}u^0$, where $g$ is the determinant of the spacetime
  metric, $\rho _{\protect \rm B}$ the rest-mass density, and $u^0$ the time
  component of the perfect fluid four-velocity. The rest-mass density is the
  product of the baryon number density $n_{\protect \rm B}$ times a mean baryon
  mass. Strictly speaking, the mean baryon mass depends on the composition of
  matter, but for the calculations performed in this work we take it to be the
  neutron rest mass. This choice leads to rest mass values that are accurate to
  within roughly 1 part in $10^3$, but since we use this definition
  consistently throughout our work, our results are insensitive to the mean
  baryon mass value.}\BibitemShut {Stop}%
\bibitem [{\citenamefont {Bozzola}\ \emph {et~al.}(2018)\citenamefont
  {Bozzola}, \citenamefont {Stergioulas},\ and\ \citenamefont
  {Bauswein}}]{Bozzola:2017qbu}%
  \BibitemOpen
  \bibfield  {author} {\bibinfo {author} {\bibfnamefont {G.}~\bibnamefont
  {Bozzola}}, \bibinfo {author} {\bibfnamefont {N.}~\bibnamefont
  {Stergioulas}},\ and\ \bibinfo {author} {\bibfnamefont {A.}~\bibnamefont
  {Bauswein}},\ }\bibfield  {title} {\bibinfo {title} {{Universal relations for
  differentially rotating relativistic stars at the threshold to collapse}},\
  }\href {https://doi.org/10.1093/mnras/stx3002} {\bibfield  {journal}
  {\bibinfo  {journal} {Mon. Not. R. Astron. Soc.}\ }\textbf {\bibinfo {volume}
  {474}},\ \bibinfo {pages} {3557} (\bibinfo {year} {2018})},\ \Eprint
  {https://arxiv.org/abs/1709.02787} {arXiv:1709.02787 [gr-qc]} \BibitemShut
  {NoStop}%
\bibitem [{\citenamefont {Weih}\ \emph {et~al.}(2018)\citenamefont {Weih},
  \citenamefont {Most},\ and\ \citenamefont {Rezzolla}}]{Weih:2017mcw}%
  \BibitemOpen
  \bibfield  {author} {\bibinfo {author} {\bibfnamefont {L.~R.}\ \bibnamefont
  {Weih}}, \bibinfo {author} {\bibfnamefont {E.~R.}\ \bibnamefont {Most}},\
  and\ \bibinfo {author} {\bibfnamefont {L.}~\bibnamefont {Rezzolla}},\
  }\bibfield  {title} {\bibinfo {title} {{On the stability and maximum mass of
  differentially rotating relativistic stars}},\ }\href
  {https://doi.org/10.1093/mnrasl/slx178} {\bibfield  {journal} {\bibinfo
  {journal} {Mon. Not. R. Astron. Soc. Lett.}\ }\textbf {\bibinfo {volume}
  {473}},\ \bibinfo {pages} {L126} (\bibinfo {year} {2018})},\ \Eprint
  {https://arxiv.org/abs/1709.06058} {arXiv:1709.06058 [gr-qc]} \BibitemShut
  {NoStop}%
\bibitem [{\citenamefont {{Cook}}\ \emph {et~al.}(1992)\citenamefont {{Cook}},
  \citenamefont {{Shapiro}},\ and\ \citenamefont
  {{Teukolsky}}}]{1992ApJ...398..203C}%
  \BibitemOpen
  \bibfield  {author} {\bibinfo {author} {\bibfnamefont {G.~B.}\ \bibnamefont
  {{Cook}}}, \bibinfo {author} {\bibfnamefont {S.~L.}\ \bibnamefont
  {{Shapiro}}},\ and\ \bibinfo {author} {\bibfnamefont {S.~A.}\ \bibnamefont
  {{Teukolsky}}},\ }\bibfield  {title} {\bibinfo {title} {Spin-up of a rapidly
  rotating star by angular momentum loss: Effects of general relativity},\
  }\href {https://doi.org/10.1086/171849} {\bibfield  {journal} {\bibinfo
  {journal} {\apj}\ }\textbf {\bibinfo {volume} {398}},\ \bibinfo {pages} {203}
  (\bibinfo {year} {1992})}\BibitemShut {NoStop}%
\bibitem [{Note3()}]{Note3}%
  \BibitemOpen
  \bibinfo {note} {\protect \leavevmode {\protect \color {black}Note that in
  addition to these two cases, in some works (such as \cite {Shibata:2019ctb})
  the authors assume the critical mass of collapse to lie in between the two
  limiting cases, $M_{\protect \rm TOV,B} < M_{\protect \rm crit,B} <
  M_{\protect \rm supra,B}$. This is possible if one assumes that the merger
  remnant has lost angular momentum due to, e.g., GW emission.}}\BibitemShut
  {Stop}%
\bibitem [{\citenamefont {Nathanail}\ \emph {et~al.}(2021)\citenamefont
  {Nathanail}, \citenamefont {Most},\ and\ \citenamefont
  {Rezzolla}}]{Nathanail:2021tay}%
  \BibitemOpen
  \bibfield  {author} {\bibinfo {author} {\bibfnamefont {A.}~\bibnamefont
  {Nathanail}}, \bibinfo {author} {\bibfnamefont {E.~R.}\ \bibnamefont
  {Most}},\ and\ \bibinfo {author} {\bibfnamefont {L.}~\bibnamefont
  {Rezzolla}},\ }\bibfield  {title} {\bibinfo {title} {{GW170817 and GW190814:
  tension on the maximum mass}},\ }\href
  {https://doi.org/10.3847/2041-8213/abdfc6} {\bibfield  {journal} {\bibinfo
  {journal} {Astrophys. J. Lett.}\ }\textbf {\bibinfo {volume} {908}},\
  \bibinfo {pages} {L28} (\bibinfo {year} {2021})},\ \Eprint
  {https://arxiv.org/abs/2101.01735} {arXiv:2101.01735 [astro-ph.HE]}
  \BibitemShut {NoStop}%
\bibitem [{\citenamefont {Lau}\ \emph {et~al.}(2017)\citenamefont {Lau},
  \citenamefont {Leung},\ and\ \citenamefont {Lin}}]{Lau:2017qtz}%
  \BibitemOpen
  \bibfield  {author} {\bibinfo {author} {\bibfnamefont {S.~Y.}\ \bibnamefont
  {Lau}}, \bibinfo {author} {\bibfnamefont {P.~T.}\ \bibnamefont {Leung}},\
  and\ \bibinfo {author} {\bibfnamefont {L.-M.}\ \bibnamefont {Lin}},\
  }\bibfield  {title} {\bibinfo {title} {{Tidal deformations of compact stars
  with crystalline quark matter}},\ }\href
  {https://doi.org/10.1103/PhysRevD.95.101302} {\bibfield  {journal} {\bibinfo
  {journal} {Phys. Rev. D}\ }\textbf {\bibinfo {volume} {95}},\ \bibinfo
  {pages} {101302(R)} (\bibinfo {year} {2017})},\ \Eprint
  {https://arxiv.org/abs/1705.01710} {arXiv:1705.01710 [astro-ph.HE]}
  \BibitemShut {NoStop}%
\bibitem [{\citenamefont {Bandyopadhyay}\ \emph {et~al.}(2018)\citenamefont
  {Bandyopadhyay}, \citenamefont {Bhat}, \citenamefont {Char},\ and\
  \citenamefont {Chatterjee}}]{Bandyopadhyay:2017dvi}%
  \BibitemOpen
  \bibfield  {author} {\bibinfo {author} {\bibfnamefont {D.}~\bibnamefont
  {Bandyopadhyay}}, \bibinfo {author} {\bibfnamefont {S.~A.}\ \bibnamefont
  {Bhat}}, \bibinfo {author} {\bibfnamefont {P.}~\bibnamefont {Char}},\ and\
  \bibinfo {author} {\bibfnamefont {D.}~\bibnamefont {Chatterjee}},\ }\bibfield
   {title} {\bibinfo {title} {{Moment of inertia, quadrupole moment, Love
  number of neutron star and their relations with strange-matter equations of
  state}},\ }\href {https://doi.org/10.1140/epja/i2018-12456-y} {\bibfield
  {journal} {\bibinfo  {journal} {Eur. Phys. J. A}\ }\textbf {\bibinfo {volume}
  {54}},\ \bibinfo {pages} {26} (\bibinfo {year} {2018})},\ \Eprint
  {https://arxiv.org/abs/1712.01715} {arXiv:1712.01715 [astro-ph.HE]}
  \BibitemShut {NoStop}%
\bibitem [{\citenamefont {Han}\ and\ \citenamefont
  {Steiner}(2019)}]{Han:2018mtj}%
  \BibitemOpen
  \bibfield  {author} {\bibinfo {author} {\bibfnamefont {S.}~\bibnamefont
  {Han}}\ and\ \bibinfo {author} {\bibfnamefont {A.~W.}\ \bibnamefont
  {Steiner}},\ }\bibfield  {title} {\bibinfo {title} {{Tidal deformability with
  sharp phase transitions in binary neutron stars}},\ }\href
  {https://doi.org/10.1103/PhysRevD.99.083014} {\bibfield  {journal} {\bibinfo
  {journal} {Phys. Rev. D}\ }\textbf {\bibinfo {volume} {99}},\ \bibinfo
  {pages} {083014} (\bibinfo {year} {2019})},\ \Eprint
  {https://arxiv.org/abs/1810.10967} {arXiv:1810.10967 [nucl-th]} \BibitemShut
  {NoStop}%
\bibitem [{\citenamefont {Bozzola}\ \emph {et~al.}(2019)\citenamefont
  {Bozzola}, \citenamefont {Espino}, \citenamefont {Lewin},\ and\ \citenamefont
  {Paschalidis}}]{Bozzola:2019tit}%
  \BibitemOpen
  \bibfield  {author} {\bibinfo {author} {\bibfnamefont {G.}~\bibnamefont
  {Bozzola}}, \bibinfo {author} {\bibfnamefont {P.~L.}\ \bibnamefont {Espino}},
  \bibinfo {author} {\bibfnamefont {C.~D.}\ \bibnamefont {Lewin}},\ and\
  \bibinfo {author} {\bibfnamefont {V.}~\bibnamefont {Paschalidis}},\
  }\bibfield  {title} {\bibinfo {title} {{Maximum mass and universal relations
  of rotating relativistic hybrid hadron-quark stars}},\ }\href
  {https://doi.org/10.1140/epja/i2019-12831-2} {\bibfield  {journal} {\bibinfo
  {journal} {Eur. Phys. J. A}\ }\textbf {\bibinfo {volume} {55}},\ \bibinfo
  {pages} {149} (\bibinfo {year} {2019})},\ \Eprint
  {https://arxiv.org/abs/1905.00028} {arXiv:1905.00028 [astro-ph.HE]}
  \BibitemShut {NoStop}%
\bibitem [{\citenamefont {Annala}\ \emph
  {et~al.}(2018{\natexlab{b}})\citenamefont {Annala}, \citenamefont {Ecker},
  \citenamefont {Hoyos}, \citenamefont {Jokela}, \citenamefont
  {Rodr\'iguez~Fern\'andez},\ and\ \citenamefont {Vuorinen}}]{Annala:2017tqz}%
  \BibitemOpen
  \bibfield  {author} {\bibinfo {author} {\bibfnamefont {E.}~\bibnamefont
  {Annala}}, \bibinfo {author} {\bibfnamefont {C.}~\bibnamefont {Ecker}},
  \bibinfo {author} {\bibfnamefont {C.}~\bibnamefont {Hoyos}}, \bibinfo
  {author} {\bibfnamefont {N.}~\bibnamefont {Jokela}}, \bibinfo {author}
  {\bibfnamefont {D.}~\bibnamefont {Rodr\'iguez~Fern\'andez}},\ and\ \bibinfo
  {author} {\bibfnamefont {A.}~\bibnamefont {Vuorinen}},\ }\bibfield  {title}
  {\bibinfo {title} {Holographic compact stars meet gravitational wave
  constraints},\ }\href {https://doi.org/10.1007/JHEP12(2018)078} {\bibfield
  {journal} {\bibinfo  {journal} {{J. High Energy Phys.}}\ }\textbf {\bibinfo
  {volume} {2018}},\ \bibinfo {pages} {78} (\bibinfo {year}
  {2018}{\natexlab{b}})},\ \Eprint {https://arxiv.org/abs/1711.06244}
  {arXiv:1711.06244 [astro-ph.HE]} \BibitemShut {NoStop}%
\bibitem [{\citenamefont {{Steiner}}\ \emph {et~al.}(2010)\citenamefont
  {{Steiner}}, \citenamefont {{Lattimer}},\ and\ \citenamefont
  {{Brown}}}]{Steiner:2010}%
  \BibitemOpen
  \bibfield  {author} {\bibinfo {author} {\bibfnamefont {A.~W.}\ \bibnamefont
  {{Steiner}}}, \bibinfo {author} {\bibfnamefont {J.~M.}\ \bibnamefont
  {{Lattimer}}},\ and\ \bibinfo {author} {\bibfnamefont {E.~F.}\ \bibnamefont
  {{Brown}}},\ }\bibfield  {title} {\bibinfo {title} {{The equation of state
  from observed masses and radii of neutron stars}},\ }\href
  {https://doi.org/10.1088/0004-637X/722/1/33} {\bibfield  {journal} {\bibinfo
  {journal} {Astrophys. J.}\ }\textbf {\bibinfo {volume} {722}},\ \bibinfo
  {pages} {33} (\bibinfo {year} {2010})},\ \Eprint
  {https://arxiv.org/abs/1005.0811} {arXiv:1005.0811 [astro-ph.HE]}
  \BibitemShut {NoStop}%
\bibitem [{\citenamefont {Guillot}\ \emph {et~al.}(2013)\citenamefont
  {Guillot}, \citenamefont {Servillat}, \citenamefont {Webb},\ and\
  \citenamefont {Rutledge}}]{Guillot:2013wu}%
  \BibitemOpen
  \bibfield  {author} {\bibinfo {author} {\bibfnamefont {S.}~\bibnamefont
  {Guillot}}, \bibinfo {author} {\bibfnamefont {M.}~\bibnamefont {Servillat}},
  \bibinfo {author} {\bibfnamefont {N.~A.}\ \bibnamefont {Webb}},\ and\
  \bibinfo {author} {\bibfnamefont {R.~E.}\ \bibnamefont {Rutledge}},\
  }\bibfield  {title} {\bibinfo {title} {{Measurement of the radius of neutron
  stars with high signal-to-noise quiescent low-mass X-ray binaries in globular
  clusters}},\ }\href {https://doi.org/10.1088/0004-637X/772/1/7} {\bibfield
  {journal} {\bibinfo  {journal} {Astrophys. J.}\ }\textbf {\bibinfo {volume}
  {772}},\ \bibinfo {pages} {7} (\bibinfo {year} {2013})},\ \Eprint
  {https://arxiv.org/abs/1302.0023} {arXiv:1302.0023 [astro-ph.HE]}
  \BibitemShut {NoStop}%
\bibitem [{\citenamefont {{\"O}zel}\ \emph {et~al.}(2016)\citenamefont
  {{\"O}zel}, \citenamefont {Psaltis}, \citenamefont {G{\"u}ver}, \citenamefont
  {Baym}, \citenamefont {Heinke},\ and\ \citenamefont
  {Guillot}}]{Ozel:2015fia}%
  \BibitemOpen
  \bibfield  {author} {\bibinfo {author} {\bibfnamefont {F.}~\bibnamefont
  {{\"O}zel}}, \bibinfo {author} {\bibfnamefont {D.}~\bibnamefont {Psaltis}},
  \bibinfo {author} {\bibfnamefont {T.}~\bibnamefont {G{\"u}ver}}, \bibinfo
  {author} {\bibfnamefont {G.}~\bibnamefont {Baym}}, \bibinfo {author}
  {\bibfnamefont {C.}~\bibnamefont {Heinke}},\ and\ \bibinfo {author}
  {\bibfnamefont {S.}~\bibnamefont {Guillot}},\ }\bibfield  {title} {\bibinfo
  {title} {{The dense matter equation of state from neutron star radius and
  mass measurements}},\ }\href {https://doi.org/10.3847/0004-637X/820/1/28}
  {\bibfield  {journal} {\bibinfo  {journal} {Astrophys. J.}\ }\textbf
  {\bibinfo {volume} {820}},\ \bibinfo {pages} {28} (\bibinfo {year} {2016})},\
  \Eprint {https://arxiv.org/abs/1505.05155} {arXiv:1505.05155 [astro-ph.HE]}
  \BibitemShut {NoStop}%
\bibitem [{\citenamefont {Bogdanov}\ \emph {et~al.}(2016)\citenamefont
  {Bogdanov}, \citenamefont {Heinke}, \citenamefont {{\"O}zel},\ and\
  \citenamefont {Güver}}]{Bogdanov:2016nle}%
  \BibitemOpen
  \bibfield  {author} {\bibinfo {author} {\bibfnamefont {S.}~\bibnamefont
  {Bogdanov}}, \bibinfo {author} {\bibfnamefont {C.~O.}\ \bibnamefont
  {Heinke}}, \bibinfo {author} {\bibfnamefont {F.}~\bibnamefont {{\"O}zel}},\
  and\ \bibinfo {author} {\bibfnamefont {T.}~\bibnamefont {Güver}},\
  }\bibfield  {title} {\bibinfo {title} {{Neutron star mass-radius constraints
  of the quiescent low-mass X-ray binaries X7 and X5 in the globular cluster 47
  Tuc}},\ }\href {https://doi.org/10.3847/0004-637X/831/2/184} {\bibfield
  {journal} {\bibinfo  {journal} {Astrophys. J.}\ }\textbf {\bibinfo {volume}
  {831}},\ \bibinfo {pages} {184} (\bibinfo {year} {2016})},\ \Eprint
  {https://arxiv.org/abs/1603.01630} {arXiv:1603.01630 [astro-ph.HE]}
  \BibitemShut {NoStop}%
\bibitem [{\citenamefont {N{\"a}ttil{\"a}}\ \emph {et~al.}(2016)\citenamefont
  {N{\"a}ttil{\"a}}, \citenamefont {Steiner}, \citenamefont {Kajava},
  \citenamefont {Suleimanov},\ and\ \citenamefont
  {Poutanen}}]{Nattila:2015jra}%
  \BibitemOpen
  \bibfield  {author} {\bibinfo {author} {\bibfnamefont {J.}~\bibnamefont
  {N{\"a}ttil{\"a}}}, \bibinfo {author} {\bibfnamefont {A.~W.}\ \bibnamefont
  {Steiner}}, \bibinfo {author} {\bibfnamefont {J.~J.~E.}\ \bibnamefont
  {Kajava}}, \bibinfo {author} {\bibfnamefont {V.~F.}\ \bibnamefont
  {Suleimanov}},\ and\ \bibinfo {author} {\bibfnamefont {J.}~\bibnamefont
  {Poutanen}},\ }\bibfield  {title} {\bibinfo {title} {{Equation of state
  constraints for the cold dense matter inside neutron stars using the cooling
  tail method}},\ }\href {https://doi.org/10.1051/0004-6361/201527416}
  {\bibfield  {journal} {\bibinfo  {journal} {Astron. Astrophys.}\ }\textbf
  {\bibinfo {volume} {591}},\ \bibinfo {pages} {A25} (\bibinfo {year}
  {2016})},\ \Eprint {https://arxiv.org/abs/1509.06561} {arXiv:1509.06561
  [astro-ph.HE]} \BibitemShut {NoStop}%
\bibitem [{\citenamefont {Steiner}\ \emph {et~al.}(2018)\citenamefont
  {Steiner}, \citenamefont {Heinke}, \citenamefont {Bogdanov}, \citenamefont
  {Li}, \citenamefont {Ho}, \citenamefont {Bahramian},\ and\ \citenamefont
  {Han}}]{Steiner:2017vmg}%
  \BibitemOpen
  \bibfield  {author} {\bibinfo {author} {\bibfnamefont {A.~W.}\ \bibnamefont
  {Steiner}}, \bibinfo {author} {\bibfnamefont {C.~O.}\ \bibnamefont {Heinke}},
  \bibinfo {author} {\bibfnamefont {S.}~\bibnamefont {Bogdanov}}, \bibinfo
  {author} {\bibfnamefont {C.~K.}\ \bibnamefont {Li}}, \bibinfo {author}
  {\bibfnamefont {W.~C.~G.}\ \bibnamefont {Ho}}, \bibinfo {author}
  {\bibfnamefont {A.}~\bibnamefont {Bahramian}},\ and\ \bibinfo {author}
  {\bibfnamefont {S.}~\bibnamefont {Han}},\ }\bibfield  {title} {\bibinfo
  {title} {{Constraining the mass and radius of neutron stars in globular
  clusters}},\ }\href {https://doi.org/10.1093/mnras/sty215} {\bibfield
  {journal} {\bibinfo  {journal} {Mon. Not. R. Astron. Soc.}\ }\textbf
  {\bibinfo {volume} {476}},\ \bibinfo {pages} {421} (\bibinfo {year}
  {2018})},\ \Eprint {https://arxiv.org/abs/1709.05013} {arXiv:1709.05013
  [astro-ph.HE]} \BibitemShut {NoStop}%
\bibitem [{\citenamefont {{Shaw}}\ \emph {et~al.}(2018)\citenamefont {{Shaw}},
  \citenamefont {{Heinke}}, \citenamefont {{Steiner}}, \citenamefont
  {{Campana}}, \citenamefont {{Cohn}}, \citenamefont {{Ho}}, \citenamefont
  {{Lugger}},\ and\ \citenamefont {{Servillat}}}]{Shawn:2018}%
  \BibitemOpen
  \bibfield  {author} {\bibinfo {author} {\bibfnamefont {A.~W.}\ \bibnamefont
  {{Shaw}}}, \bibinfo {author} {\bibfnamefont {C.~O.}\ \bibnamefont
  {{Heinke}}}, \bibinfo {author} {\bibfnamefont {A.~W.}\ \bibnamefont
  {{Steiner}}}, \bibinfo {author} {\bibfnamefont {S.}~\bibnamefont
  {{Campana}}}, \bibinfo {author} {\bibfnamefont {H.~N.}\ \bibnamefont
  {{Cohn}}}, \bibinfo {author} {\bibfnamefont {W.~C.~G.}\ \bibnamefont {{Ho}}},
  \bibinfo {author} {\bibfnamefont {P.~M.}\ \bibnamefont {{Lugger}}},\ and\
  \bibinfo {author} {\bibfnamefont {M.}~\bibnamefont {{Servillat}}},\
  }\bibfield  {title} {\bibinfo {title} {{The radius of the quiescent neutron
  star in the globular cluster M13}},\ }\href
  {https://doi.org/10.1093/mnras/sty582} {\bibfield  {journal} {\bibinfo
  {journal} {Mon. Not. R. Astron. Soc.}\ }\textbf {\bibinfo {volume} {476}},\
  \bibinfo {pages} {4713} (\bibinfo {year} {2018})},\ \Eprint
  {https://arxiv.org/abs/1803.00029} {arXiv:1803.00029 [astro-ph.HE]}
  \BibitemShut {NoStop}%
\bibitem [{\citenamefont {\"Ozel}\ and\ \citenamefont
  {Freire}(2016)}]{Ozel:2016oaf}%
  \BibitemOpen
  \bibfield  {author} {\bibinfo {author} {\bibfnamefont {F.}~\bibnamefont
  {\"Ozel}}\ and\ \bibinfo {author} {\bibfnamefont {P.}~\bibnamefont
  {Freire}},\ }\bibfield  {title} {\bibinfo {title} {{Masses, radii, and the
  equation of state of neutron stars}},\ }\href
  {https://doi.org/10.1146/annurev-astro-081915-023322} {\bibfield  {journal}
  {\bibinfo  {journal} {Ann. Rev. Astron. Astrophys.}\ }\textbf {\bibinfo
  {volume} {54}},\ \bibinfo {pages} {401} (\bibinfo {year} {2016})},\ \Eprint
  {https://arxiv.org/abs/1603.02698} {arXiv:1603.02698 [astro-ph.HE]}
  \BibitemShut {NoStop}%
\bibitem [{\citenamefont {Miller}\ and\ \citenamefont
  {Lamb}(2016)}]{Miller:2016pom}%
  \BibitemOpen
  \bibfield  {author} {\bibinfo {author} {\bibfnamefont {M.~C.}\ \bibnamefont
  {Miller}}\ and\ \bibinfo {author} {\bibfnamefont {F.~K.}\ \bibnamefont
  {Lamb}},\ }\bibfield  {title} {\bibinfo {title} {{Observational constraints
  on neutron star masses and radii}},\ }\href
  {https://doi.org/10.1140/epja/i2016-16063-8} {\bibfield  {journal} {\bibinfo
  {journal} {Eur. Phys. J. A}\ }\textbf {\bibinfo {volume} {52}},\ \bibinfo
  {pages} {63} (\bibinfo {year} {2016})},\ \Eprint
  {https://arxiv.org/abs/1604.03894} {arXiv:1604.03894 [astro-ph.HE]}
  \BibitemShut {NoStop}%
\bibitem [{\citenamefont {Raaijmakers}\ \emph {et~al.}(2021)\citenamefont
  {Raaijmakers}, \citenamefont {Greif}, \citenamefont {Hebeler}, \citenamefont
  {Hinderer}, \citenamefont {Nissanke}, \citenamefont {Schwenk}, \citenamefont
  {Riley}, \citenamefont {Watts}, \citenamefont {Lattimer},\ and\ \citenamefont
  {Ho}}]{Raaijmakers:2021uju}%
  \BibitemOpen
  \bibfield  {author} {\bibinfo {author} {\bibfnamefont {G.}~\bibnamefont
  {Raaijmakers}}, \bibinfo {author} {\bibfnamefont {S.~K.}\ \bibnamefont
  {Greif}}, \bibinfo {author} {\bibfnamefont {K.}~\bibnamefont {Hebeler}},
  \bibinfo {author} {\bibfnamefont {T.}~\bibnamefont {Hinderer}}, \bibinfo
  {author} {\bibfnamefont {S.}~\bibnamefont {Nissanke}}, \bibinfo {author}
  {\bibfnamefont {A.}~\bibnamefont {Schwenk}}, \bibinfo {author} {\bibfnamefont
  {T.~E.}\ \bibnamefont {Riley}}, \bibinfo {author} {\bibfnamefont {A.~L.}\
  \bibnamefont {Watts}}, \bibinfo {author} {\bibfnamefont {J.~M.}\ \bibnamefont
  {Lattimer}},\ and\ \bibinfo {author} {\bibfnamefont {W.~C.~G.}\ \bibnamefont
  {Ho}},\ }\bibfield  {title} {\bibinfo {title} {{Constraints on the dense
  matter equation of state and neutron star properties from NICER's mass-radius
  estimate of PSR J0740+6620 and multimessenger observations}},\ }\href
  {https://doi.org/10.3847/2041-8213/ac089a} {\bibfield  {journal} {\bibinfo
  {journal} {Astrophys. J. Lett.}\ }\textbf {\bibinfo {volume} {918}},\
  \bibinfo {pages} {L29} (\bibinfo {year} {2021})},\ \Eprint
  {https://arxiv.org/abs/2105.06981} {arXiv:2105.06981 [astro-ph.HE]}
  \BibitemShut {NoStop}%
\bibitem [{\citenamefont {Somasundaram}\ and\ \citenamefont
  {Margueron}(2021)}]{Somasundaram:2021ljr}%
  \BibitemOpen
  \bibfield  {author} {\bibinfo {author} {\bibfnamefont {R.}~\bibnamefont
  {Somasundaram}}\ and\ \bibinfo {author} {\bibfnamefont {J.}~\bibnamefont
  {Margueron}},\ }\bibfield  {title} {\bibinfo {title} {{Impact of massive
  neutron star radii on the nature of phase transitions in dense matter}},\
  }\Eprint {https://arxiv.org/abs/2104.13612} {arXiv:2104.13612 [astro-ph.HE]}
  (\bibinfo {year} {2021})\BibitemShut {NoStop}%
\bibitem [{\citenamefont {Biswas}(2021)}]{Biswas:2021yge}%
  \BibitemOpen
  \bibfield  {author} {\bibinfo {author} {\bibfnamefont {B.}~\bibnamefont
  {Biswas}},\ }\bibfield  {title} {\bibinfo {title} {{Impact of PREX-II and
  combined radio/NICER/XMM-Newton\textquoteright{}s mass\textendash{}radius
  measurement of PSR J0740+6620 on the dense-matter equation of state}},\
  }\href {https://doi.org/10.3847/1538-4357/ac1c72} {\bibfield  {journal}
  {\bibinfo  {journal} {Astrophys. J.}\ }\textbf {\bibinfo {volume} {921}},\
  \bibinfo {pages} {63} (\bibinfo {year} {2021})},\ \Eprint
  {https://arxiv.org/abs/2105.02886} {arXiv:2105.02886 [astro-ph.HE]}
  \BibitemShut {NoStop}%
\bibitem [{\citenamefont {Li}\ \emph {et~al.}(2021)\citenamefont {Li},
  \citenamefont {Cai}, \citenamefont {Xie},\ and\ \citenamefont
  {Zhang}}]{Li:2021thg}%
  \BibitemOpen
  \bibfield  {author} {\bibinfo {author} {\bibfnamefont {B.-A.}\ \bibnamefont
  {Li}}, \bibinfo {author} {\bibfnamefont {B.-J.}\ \bibnamefont {Cai}},
  \bibinfo {author} {\bibfnamefont {W.-J.}\ \bibnamefont {Xie}},\ and\ \bibinfo
  {author} {\bibfnamefont {N.-B.}\ \bibnamefont {Zhang}},\ }\bibfield  {title}
  {\bibinfo {title} {Progress in constraining nuclear symmetry energy using
  neutron star observables since {GW170817}},\ }\href
  {https://doi.org/10.3390/universe7060182} {\bibfield  {journal} {\bibinfo
  {journal} {Universe}\ }\textbf {\bibinfo {volume} {7}},\ \bibinfo {pages}
  {182} (\bibinfo {year} {2021})},\ \Eprint {https://arxiv.org/abs/2105.04629}
  {arXiv:2105.04629 [nucl-th]} \BibitemShut {NoStop}%
\bibitem [{\citenamefont {Pang}\ \emph {et~al.}(2021)\citenamefont {Pang},
  \citenamefont {Tews}, \citenamefont {Coughlin}, \citenamefont {Bulla},
  \citenamefont {Van Den~Broeck},\ and\ \citenamefont
  {Dietrich}}]{Pang:2021jta}%
  \BibitemOpen
  \bibfield  {author} {\bibinfo {author} {\bibfnamefont {P.~T.~H.}\
  \bibnamefont {Pang}}, \bibinfo {author} {\bibfnamefont {I.}~\bibnamefont
  {Tews}}, \bibinfo {author} {\bibfnamefont {M.~W.}\ \bibnamefont {Coughlin}},
  \bibinfo {author} {\bibfnamefont {M.}~\bibnamefont {Bulla}}, \bibinfo
  {author} {\bibfnamefont {C.}~\bibnamefont {Van Den~Broeck}},\ and\ \bibinfo
  {author} {\bibfnamefont {T.}~\bibnamefont {Dietrich}},\ }\bibfield  {title}
  {\bibinfo {title} {{Nuclear Physics Multimessenger Astrophysics Constraints
  on the Neutron Star Equation of State: Adding NICER\textquoteright{}s PSR
  J0740+6620 Measurement}},\ }\href {https://doi.org/10.3847/1538-4357/ac19ab}
  {\bibfield  {journal} {\bibinfo  {journal} {Astrophys. J.}\ }\textbf
  {\bibinfo {volume} {922}},\ \bibinfo {pages} {14} (\bibinfo {year} {2021})},\
  \Eprint {https://arxiv.org/abs/2105.08688} {arXiv:2105.08688 [astro-ph.HE]}
  \BibitemShut {NoStop}%
\bibitem [{\citenamefont {Abbott}\ \emph
  {et~al.}(2020{\natexlab{a}})\citenamefont {Abbott} \emph
  {et~al.}}]{Abbott:2020uma}%
  \BibitemOpen
  \bibfield  {author} {\bibinfo {author} {\bibfnamefont {B.~P.}\ \bibnamefont
  {Abbott}} \emph {et~al.},\ }\bibfield  {title} {\bibinfo {title} {{GW190425:
  Observation of a compact binary coalescence with total mass $\sim 3.4
  M_{\odot}$}},\ }\href {https://doi.org/10.3847/2041-8213/ab75f5} {\bibfield
  {journal} {\bibinfo  {journal} {Astrophys. J. Lett.}\ }\textbf {\bibinfo
  {volume} {892}},\ \bibinfo {pages} {L3} (\bibinfo {year}
  {2020}{\natexlab{a}})},\ \Eprint {https://arxiv.org/abs/2001.01761}
  {arXiv:2001.01761 [astro-ph.HE]} \BibitemShut {NoStop}%
\bibitem [{\citenamefont {Abbott}\ \emph
  {et~al.}(2020{\natexlab{b}})\citenamefont {Abbott} \emph
  {et~al.}}]{Abbott:2020khf}%
  \BibitemOpen
  \bibfield  {author} {\bibinfo {author} {\bibfnamefont {R.}~\bibnamefont
  {Abbott}} \emph {et~al.},\ }\bibfield  {title} {\bibinfo {title} {{GW190814:
  Gravitational waves from the coalescence of a 23 solar mass black hole with a
  2.6 solar mass compact object}},\ }\href
  {https://doi.org/10.3847/2041-8213/ab960f} {\bibfield  {journal} {\bibinfo
  {journal} {Astrophys. J. Lett.}\ }\textbf {\bibinfo {volume} {896}},\
  \bibinfo {pages} {L44} (\bibinfo {year} {2020}{\natexlab{b}})},\ \Eprint
  {https://arxiv.org/abs/2006.12611} {arXiv:2006.12611 [astro-ph.HE]}
  \BibitemShut {NoStop}%
\bibitem [{\citenamefont {Yang}\ \emph {et~al.}(2018)\citenamefont {Yang},
  \citenamefont {East},\ and\ \citenamefont {Lehner}}]{Yang:2017gfb}%
  \BibitemOpen
  \bibfield  {author} {\bibinfo {author} {\bibfnamefont {H.}~\bibnamefont
  {Yang}}, \bibinfo {author} {\bibfnamefont {W.~E.}\ \bibnamefont {East}},\
  and\ \bibinfo {author} {\bibfnamefont {L.}~\bibnamefont {Lehner}},\
  }\bibfield  {title} {\bibinfo {title} {{Can we distinguish low-mass black
  holes in neutron star binaries?}},\ }\href
  {https://doi.org/10.3847/1538-4357/aab2b0} {\bibfield  {journal} {\bibinfo
  {journal} {Astrophys. J.}\ }\textbf {\bibinfo {volume} {856}},\ \bibinfo
  {pages} {110} (\bibinfo {year} {2018})},\ \bibinfo {note} {[Erratum:
  Astrophys. J. \textbf{870}, 139 (2019)]},\ \Eprint
  {https://arxiv.org/abs/1710.05891} {arXiv:1710.05891 [gr-qc]} \BibitemShut
  {NoStop}%
\bibitem [{\citenamefont {Hinderer}\ \emph {et~al.}(2019)\citenamefont
  {Hinderer} \emph {et~al.}}]{Hinderer:2018pei}%
  \BibitemOpen
  \bibfield  {author} {\bibinfo {author} {\bibfnamefont {T.}~\bibnamefont
  {Hinderer}} \emph {et~al.},\ }\bibfield  {title} {\bibinfo {title}
  {{Distinguishing the nature of comparable-mass neutron star binary systems
  with multimessenger observations: GW170817 case study}},\ }\href
  {https://doi.org/10.1103/PhysRevD.100.063021} {\bibfield  {journal} {\bibinfo
   {journal} {Phys. Rev. D}\ }\textbf {\bibinfo {volume} {100}},\ \bibinfo
  {pages} {06321} (\bibinfo {year} {2019})},\ \Eprint
  {https://arxiv.org/abs/1808.03836} {arXiv:1808.03836 [astro-ph.HE]}
  \BibitemShut {NoStop}%
\bibitem [{\citenamefont {Chen}\ \emph {et~al.}(2020)\citenamefont {Chen},
  \citenamefont {Johnson-McDaniel}, \citenamefont {Dietrich},\ and\
  \citenamefont {Dudi}}]{Chen:2020fzm}%
  \BibitemOpen
  \bibfield  {author} {\bibinfo {author} {\bibfnamefont {A.}~\bibnamefont
  {Chen}}, \bibinfo {author} {\bibfnamefont {N.~K.}\ \bibnamefont
  {Johnson-McDaniel}}, \bibinfo {author} {\bibfnamefont {T.}~\bibnamefont
  {Dietrich}},\ and\ \bibinfo {author} {\bibfnamefont {R.}~\bibnamefont
  {Dudi}},\ }\bibfield  {title} {\bibinfo {title} {{Distinguishing high-mass
  binary neutron stars from binary black holes with second- and
  third-generation gravitational wave observatories}},\ }\href
  {https://doi.org/10.1103/PhysRevD.101.103008} {\bibfield  {journal} {\bibinfo
   {journal} {Phys. Rev. D}\ }\textbf {\bibinfo {volume} {101}},\ \bibinfo
  {pages} {103008} (\bibinfo {year} {2020})},\ \Eprint
  {https://arxiv.org/abs/2001.11470} {arXiv:2001.11470 [astro-ph.HE]}
  \BibitemShut {NoStop}%
\bibitem [{\citenamefont {Essick}\ and\ \citenamefont
  {Landry}(2020)}]{Essick:2020ghc}%
  \BibitemOpen
  \bibfield  {author} {\bibinfo {author} {\bibfnamefont {R.}~\bibnamefont
  {Essick}}\ and\ \bibinfo {author} {\bibfnamefont {P.}~\bibnamefont
  {Landry}},\ }\bibfield  {title} {\bibinfo {title} {{Discriminating between
  neutron stars and black holes with imperfect knowledge of the maximum neutron
  star mass}},\ }\href {https://doi.org/10.3847/1538-4357/abbd3b} {\bibfield
  {journal} {\bibinfo  {journal} {Astrophys. J.}\ }\textbf {\bibinfo {volume}
  {904}},\ \bibinfo {pages} {80} (\bibinfo {year} {2020})},\ \Eprint
  {https://arxiv.org/abs/2007.01372} {arXiv:2007.01372 [astro-ph.HE]}
  \BibitemShut {NoStop}%
\bibitem [{\citenamefont {Most}\ \emph {et~al.}(2021)\citenamefont {Most},
  \citenamefont {Papenfort}, \citenamefont {Tootle},\ and\ \citenamefont
  {Rezzolla}}]{Most:2020exl}%
  \BibitemOpen
  \bibfield  {author} {\bibinfo {author} {\bibfnamefont {E.~R.}\ \bibnamefont
  {Most}}, \bibinfo {author} {\bibfnamefont {L.~J.}\ \bibnamefont {Papenfort}},
  \bibinfo {author} {\bibfnamefont {S.}~\bibnamefont {Tootle}},\ and\ \bibinfo
  {author} {\bibfnamefont {L.}~\bibnamefont {Rezzolla}},\ }\bibfield  {title}
  {\bibinfo {title} {{Fast ejecta as a potential way to distinguish black holes
  from neutron stars in high-mass gravitational-wave events}},\ }\href
  {https://doi.org/10.3847/1538-4357/abf0a5} {\bibfield  {journal} {\bibinfo
  {journal} {Astrophys. J.}\ }\textbf {\bibinfo {volume} {912}},\ \bibinfo
  {pages} {80} (\bibinfo {year} {2021})},\ \Eprint
  {https://arxiv.org/abs/2012.03896} {arXiv:2012.03896 [astro-ph.HE]}
  \BibitemShut {NoStop}%
\bibitem [{\citenamefont {Sennett}\ \emph {et~al.}(2017)\citenamefont
  {Sennett}, \citenamefont {Hinderer}, \citenamefont {Steinhoff}, \citenamefont
  {Buonanno},\ and\ \citenamefont {Ossokine}}]{Sennett:2017etc}%
  \BibitemOpen
  \bibfield  {author} {\bibinfo {author} {\bibfnamefont {N.}~\bibnamefont
  {Sennett}}, \bibinfo {author} {\bibfnamefont {T.}~\bibnamefont {Hinderer}},
  \bibinfo {author} {\bibfnamefont {J.}~\bibnamefont {Steinhoff}}, \bibinfo
  {author} {\bibfnamefont {A.}~\bibnamefont {Buonanno}},\ and\ \bibinfo
  {author} {\bibfnamefont {S.}~\bibnamefont {Ossokine}},\ }\bibfield  {title}
  {\bibinfo {title} {{Distinguishing Boson Stars from Black Holes and Neutron
  Stars from Tidal Interactions in Inspiraling Binary Systems}},\ }\href
  {https://doi.org/10.1103/PhysRevD.96.024002} {\bibfield  {journal} {\bibinfo
  {journal} {Phys. Rev. D}\ }\textbf {\bibinfo {volume} {96}},\ \bibinfo
  {pages} {024002} (\bibinfo {year} {2017})},\ \Eprint
  {https://arxiv.org/abs/1704.08651} {arXiv:1704.08651 [gr-qc]} \BibitemShut
  {NoStop}%
\bibitem [{\citenamefont {Kurkela}\ \emph {et~al.}(2014)\citenamefont
  {Kurkela}, \citenamefont {Fraga}, \citenamefont {Schaffner-Bielich},\ and\
  \citenamefont {Vuorinen}}]{Kurkela:2014vha}%
  \BibitemOpen
  \bibfield  {author} {\bibinfo {author} {\bibfnamefont {A.}~\bibnamefont
  {Kurkela}}, \bibinfo {author} {\bibfnamefont {E.~S.}\ \bibnamefont {Fraga}},
  \bibinfo {author} {\bibfnamefont {J.}~\bibnamefont {Schaffner-Bielich}},\
  and\ \bibinfo {author} {\bibfnamefont {A.}~\bibnamefont {Vuorinen}},\
  }\bibfield  {title} {\bibinfo {title} {{Constraining neutron star matter with
  Quantum Chromodynamics}},\ }\href
  {https://doi.org/10.1088/0004-637X/789/2/127} {\bibfield  {journal} {\bibinfo
   {journal} {Astrophys. J.}\ }\textbf {\bibinfo {volume} {789}},\ \bibinfo
  {pages} {127} (\bibinfo {year} {2014})},\ \Eprint
  {https://arxiv.org/abs/1402.6618} {arXiv:1402.6618 [astro-ph.HE]}
  \BibitemShut {NoStop}%
\bibitem [{\citenamefont {Hebeler}\ \emph {et~al.}(2013)\citenamefont
  {Hebeler}, \citenamefont {Lattimer}, \citenamefont {Pethick},\ and\
  \citenamefont {Schwenk}}]{Hebeler:2013nza}%
  \BibitemOpen
  \bibfield  {author} {\bibinfo {author} {\bibfnamefont {K.}~\bibnamefont
  {Hebeler}}, \bibinfo {author} {\bibfnamefont {J.~M.}\ \bibnamefont
  {Lattimer}}, \bibinfo {author} {\bibfnamefont {C.~J.}\ \bibnamefont
  {Pethick}},\ and\ \bibinfo {author} {\bibfnamefont {A.}~\bibnamefont
  {Schwenk}},\ }\bibfield  {title} {\bibinfo {title} {{Equation of state and
  neutron star properties constrained by nuclear physics and observation}},\
  }\href {https://doi.org/10.1088/0004-637X/773/1/11} {\bibfield  {journal}
  {\bibinfo  {journal} {Astrophys. J.}\ }\textbf {\bibinfo {volume} {773}},\
  \bibinfo {pages} {11} (\bibinfo {year} {2013})},\ \Eprint
  {https://arxiv.org/abs/1303.4662} {arXiv:1303.4662 [astro-ph.SR]}
  \BibitemShut {NoStop}%
\bibitem [{\citenamefont {Fraga}\ \emph {et~al.}(2014)\citenamefont {Fraga},
  \citenamefont {Kurkela},\ and\ \citenamefont {Vuorinen}}]{Fraga:2013qra}%
  \BibitemOpen
  \bibfield  {author} {\bibinfo {author} {\bibfnamefont {E.~S.}\ \bibnamefont
  {Fraga}}, \bibinfo {author} {\bibfnamefont {A.}~\bibnamefont {Kurkela}},\
  and\ \bibinfo {author} {\bibfnamefont {A.}~\bibnamefont {Vuorinen}},\
  }\bibfield  {title} {\bibinfo {title} {{Interacting quark matter equation of
  state for compact stars}},\ }\href
  {https://doi.org/10.1088/2041-8205/781/2/L25} {\bibfield  {journal} {\bibinfo
   {journal} {Astrophys. J. Lett.}\ }\textbf {\bibinfo {volume} {781}},\
  \bibinfo {pages} {L25} (\bibinfo {year} {2014})},\ \Eprint
  {https://arxiv.org/abs/1311.5154} {arXiv:1311.5154 [nucl-th]} \BibitemShut
  {NoStop}%
\bibitem [{Note4()}]{Note4}%
  \BibitemOpen
  \bibinfo {note} {\protect \leavevmode {\protect \color {black}Note that the
  CET results build on a long series of previous works on the effective theory
  itself \cite {Weinberg:1990rz,Epelbaum:2008ga,Machleidt:2011zz} and the EoS
  in particular \cite {Hebeler:2009iv,Hebeler:2010xb}, while the pQCD EoS
  similarly relies on previous work performed, e.g.,~in \cite
  {Freedman:1976ub,Shuryak:1980tp,Vuorinen:2003fs}}}\BibitemShut {NoStop}%
\bibitem [{Note5()}]{Note5}%
  \BibitemOpen
  \bibinfo {note} {We will comment on the sensitivity of our results to the
  lower limit imposed on $M_{\protect \rm TOV}$ later.}\BibitemShut {Stop}%
\bibitem [{\citenamefont {Abbott}\ \emph {et~al.}(2019)\citenamefont {Abbott}
  \emph {et~al.}}]{Abbott:2018wiz}%
  \BibitemOpen
  \bibfield  {author} {\bibinfo {author} {\bibfnamefont {B.~P.}\ \bibnamefont
  {Abbott}} \emph {et~al.},\ }\bibfield  {title} {\bibinfo {title} {{Properties
  of the binary neutron star merger GW170817}},\ }\href
  {https://doi.org/10.1103/PhysRevX.9.011001} {\bibfield  {journal} {\bibinfo
  {journal} {Phys. Rev. X}\ }\textbf {\bibinfo {volume} {9}},\ \bibinfo {pages}
  {011001} (\bibinfo {year} {2019})},\ \Eprint
  {https://arxiv.org/abs/1805.11579} {arXiv:1805.11579 [gr-qc]} \BibitemShut
  {NoStop}%
\bibitem [{\citenamefont {De}\ \emph {et~al.}(2018)\citenamefont {De},
  \citenamefont {Finstad}, \citenamefont {Lattimer}, \citenamefont {Brown},
  \citenamefont {Berger},\ and\ \citenamefont {Biwer}}]{De:2018uhw}%
  \BibitemOpen
  \bibfield  {author} {\bibinfo {author} {\bibfnamefont {S.}~\bibnamefont
  {De}}, \bibinfo {author} {\bibfnamefont {D.}~\bibnamefont {Finstad}},
  \bibinfo {author} {\bibfnamefont {J.~M.}\ \bibnamefont {Lattimer}}, \bibinfo
  {author} {\bibfnamefont {D.~A.}\ \bibnamefont {Brown}}, \bibinfo {author}
  {\bibfnamefont {E.}~\bibnamefont {Berger}},\ and\ \bibinfo {author}
  {\bibfnamefont {C.~M.}\ \bibnamefont {Biwer}},\ }\bibfield  {title} {\bibinfo
  {title} {{Tidal deformabilities and radii of neutron stars from the
  observation of GW170817}},\ }\href
  {https://doi.org/10.1103/PhysRevLett.121.091102} {\bibfield  {journal}
  {\bibinfo  {journal} {Phys. Rev. Lett.}\ }\textbf {\bibinfo {volume} {121}},\
  \bibinfo {pages} {091102} (\bibinfo {year} {2018})},\ \bibinfo {note}
  {[Erratum: Phys. Rev. Lett. \textbf{121}, 259902 (2018)]},\ \Eprint
  {https://arxiv.org/abs/1804.08583} {arXiv:1804.08583 [astro-ph.HE]}
  \BibitemShut {NoStop}%
\bibitem [{\citenamefont {Hinderer}(2008)}]{Hinderer:2007mb}%
  \BibitemOpen
  \bibfield  {author} {\bibinfo {author} {\bibfnamefont {T.}~\bibnamefont
  {Hinderer}},\ }\bibfield  {title} {\bibinfo {title} {{Tidal Love numbers of
  neutron stars}},\ }\href {https://doi.org/10.1086/533487} {\bibfield
  {journal} {\bibinfo  {journal} {Astrophys. J.}\ }\textbf {\bibinfo {volume}
  {677}},\ \bibinfo {pages} {1216} (\bibinfo {year} {2008})},\ \bibinfo {note}
  {[Erratum: Astrophys. J. \textbf{697}, 964 (2009)]},\ \Eprint
  {https://arxiv.org/abs/0711.2420} {arXiv:0711.2420 [astro-ph]} \BibitemShut
  {NoStop}%
\bibitem [{\citenamefont {Jim\'enez}\ and\ \citenamefont
  {Fraga}(2021)}]{Jimenez:2021wil}%
  \BibitemOpen
  \bibfield  {author} {\bibinfo {author} {\bibfnamefont {J.~C.}\ \bibnamefont
  {Jim\'enez}}\ and\ \bibinfo {author} {\bibfnamefont {E.~S.}\ \bibnamefont
  {Fraga}},\ }\bibfield  {title} {\bibinfo {title} {Radial oscillations in
  neutron stars from {QCD}},\ }\href
  {https://doi.org/10.1103/PhysRevD.104.014002} {\bibfield  {journal} {\bibinfo
   {journal} {Phys. Rev. D}\ }\textbf {\bibinfo {volume} {104}},\ \bibinfo
  {pages} {014002} (\bibinfo {year} {2021})},\ \Eprint
  {https://arxiv.org/abs/2104.13480} {arXiv:2104.13480 [hep-ph]} \BibitemShut
  {NoStop}%
\bibitem [{\citenamefont {Cook}\ \emph {et~al.}(1992)\citenamefont {Cook},
  \citenamefont {Shapiro},\ and\ \citenamefont {Teukolsky}}]{CST92}%
  \BibitemOpen
  \bibfield  {author} {\bibinfo {author} {\bibfnamefont {G.~B.}\ \bibnamefont
  {Cook}}, \bibinfo {author} {\bibfnamefont {S.~L.}\ \bibnamefont {Shapiro}},\
  and\ \bibinfo {author} {\bibfnamefont {S.~A.}\ \bibnamefont {Teukolsky}},\
  }\bibfield  {title} {\bibinfo {title} {{Spin-up of a rapidly rotating star by
  angular momentum loss: Effects of general relativity}},\ }\href
  {https://doi.org/10.1086/171849} {\bibfield  {journal} {\bibinfo  {journal}
  {Astrophys. J.}\ }\textbf {\bibinfo {volume} {398}},\ \bibinfo {pages} {203}
  (\bibinfo {year} {1992})}\BibitemShut {NoStop}%
\bibitem [{\citenamefont {Cook}\ \emph
  {et~al.}(1994{\natexlab{a}})\citenamefont {Cook}, \citenamefont {Shapiro},\
  and\ \citenamefont {Teukolsky}}]{CST94a}%
  \BibitemOpen
  \bibfield  {author} {\bibinfo {author} {\bibfnamefont {G.~B.}\ \bibnamefont
  {Cook}}, \bibinfo {author} {\bibfnamefont {S.~L.}\ \bibnamefont {Shapiro}},\
  and\ \bibinfo {author} {\bibfnamefont {S.~A.}\ \bibnamefont {Teukolsky}},\
  }\bibfield  {title} {\bibinfo {title} {{Rapidly rotating polytropes in
  general relativity}},\ }\href {https://doi.org/10.1086/173721} {\bibfield
  {journal} {\bibinfo  {journal} {Astrophys. J.}\ }\textbf {\bibinfo {volume}
  {422}},\ \bibinfo {pages} {227} (\bibinfo {year}
  {1994}{\natexlab{a}})}\BibitemShut {NoStop}%
\bibitem [{\citenamefont {Cook}\ \emph
  {et~al.}(1994{\natexlab{b}})\citenamefont {Cook}, \citenamefont {Shapiro},\
  and\ \citenamefont {Teukolsky}}]{CST94b}%
  \BibitemOpen
  \bibfield  {author} {\bibinfo {author} {\bibfnamefont {G.~B.}\ \bibnamefont
  {Cook}}, \bibinfo {author} {\bibfnamefont {S.~L.}\ \bibnamefont {Shapiro}},\
  and\ \bibinfo {author} {\bibfnamefont {S.~A.}\ \bibnamefont {Teukolsky}},\
  }\bibfield  {title} {\bibinfo {title} {{Rapidly rotating neutron stars in
  general relativity: Realistic equations of state}},\ }\href
  {https://doi.org/10.1086/173934} {\bibfield  {journal} {\bibinfo  {journal}
  {Astrophys. J.}\ }\textbf {\bibinfo {volume} {424}},\ \bibinfo {pages} {823}
  (\bibinfo {year} {1994}{\natexlab{b}})}\BibitemShut {NoStop}%
\bibitem [{\citenamefont {Baumgarte}\ \emph {et~al.}(2000)\citenamefont
  {Baumgarte}, \citenamefont {Shapiro},\ and\ \citenamefont
  {Shibata}}]{Baumgarte:1999cq}%
  \BibitemOpen
  \bibfield  {author} {\bibinfo {author} {\bibfnamefont {T.~W.}\ \bibnamefont
  {Baumgarte}}, \bibinfo {author} {\bibfnamefont {S.~L.}\ \bibnamefont
  {Shapiro}},\ and\ \bibinfo {author} {\bibfnamefont {M.}~\bibnamefont
  {Shibata}},\ }\bibfield  {title} {\bibinfo {title} {{On the maximum mass of
  differentially rotating neutron stars}},\ }\href
  {https://doi.org/10.1086/312425} {\bibfield  {journal} {\bibinfo  {journal}
  {Astrophys. J. Lett.}\ }\textbf {\bibinfo {volume} {528}},\ \bibinfo {pages}
  {L29} (\bibinfo {year} {2000})},\ \Eprint
  {https://arxiv.org/abs/astro-ph/9910565} {arXiv:astro-ph/9910565}
  \BibitemShut {NoStop}%
\bibitem [{\citenamefont {Morrison}\ \emph {et~al.}(2004)\citenamefont
  {Morrison}, \citenamefont {Baumgarte},\ and\ \citenamefont
  {Shapiro}}]{Morrison:2004fp}%
  \BibitemOpen
  \bibfield  {author} {\bibinfo {author} {\bibfnamefont {I.~A.}\ \bibnamefont
  {Morrison}}, \bibinfo {author} {\bibfnamefont {T.~W.}\ \bibnamefont
  {Baumgarte}},\ and\ \bibinfo {author} {\bibfnamefont {S.~L.}\ \bibnamefont
  {Shapiro}},\ }\bibfield  {title} {\bibinfo {title} {{Effect of differential
  rotation on the maximum mass of neutron stars: Realistic nuclear equations of
  state}},\ }\href {https://doi.org/10.1086/421897} {\bibfield  {journal}
  {\bibinfo  {journal} {Astrophys. J.}\ }\textbf {\bibinfo {volume} {610}},\
  \bibinfo {pages} {941} (\bibinfo {year} {2004})},\ \Eprint
  {https://arxiv.org/abs/astro-ph/0401581} {arXiv:astro-ph/0401581}
  \BibitemShut {NoStop}%
\bibitem [{\citenamefont {Ansorg}\ \emph {et~al.}(2009)\citenamefont {Ansorg},
  \citenamefont {Gondek-Rosi\'{n}ska},\ and\ \citenamefont
  {Villain}}]{Ansorg:2008pk}%
  \BibitemOpen
  \bibfield  {author} {\bibinfo {author} {\bibfnamefont {M.}~\bibnamefont
  {Ansorg}}, \bibinfo {author} {\bibfnamefont {D.}~\bibnamefont
  {Gondek-Rosi\'{n}ska}},\ and\ \bibinfo {author} {\bibfnamefont
  {L.}~\bibnamefont {Villain}},\ }\bibfield  {title} {\bibinfo {title} {{On the
  solution space of differentially rotating neutron stars in general
  relativity}},\ }\href {https://doi.org/10.1111/j.1365-2966.2009.14904.x}
  {\bibfield  {journal} {\bibinfo  {journal} {Mon. Not. R. Astron. Soc.}\
  }\textbf {\bibinfo {volume} {396}},\ \bibinfo {pages} {2359} (\bibinfo {year}
  {2009})},\ \Eprint {https://arxiv.org/abs/0812.3347} {arXiv:0812.3347
  [gr-qc]} \BibitemShut {NoStop}%
\bibitem [{\citenamefont {Espino}\ and\ \citenamefont
  {Paschalidis}(2019)}]{Espino:2019ebx}%
  \BibitemOpen
  \bibfield  {author} {\bibinfo {author} {\bibfnamefont {P.~L.}\ \bibnamefont
  {Espino}}\ and\ \bibinfo {author} {\bibfnamefont {V.}~\bibnamefont
  {Paschalidis}},\ }\bibfield  {title} {\bibinfo {title} {{Revisiting the
  maximum mass of differentially rotating neutron stars in general relativity
  with realistic equations of state}},\ }\href
  {https://doi.org/10.1103/PhysRevD.99.083017} {\bibfield  {journal} {\bibinfo
  {journal} {Phys. Rev. D}\ }\textbf {\bibinfo {volume} {99}},\ \bibinfo
  {pages} {083017} (\bibinfo {year} {2019})},\ \Eprint
  {https://arxiv.org/abs/1901.05479} {arXiv:1901.05479 [astro-ph.HE]}
  \BibitemShut {NoStop}%
\bibitem [{\citenamefont {Lehner}\ and\ \citenamefont
  {Pretorius}(2014)}]{Lehner:2014asa}%
  \BibitemOpen
  \bibfield  {author} {\bibinfo {author} {\bibfnamefont {L.}~\bibnamefont
  {Lehner}}\ and\ \bibinfo {author} {\bibfnamefont {F.}~\bibnamefont
  {Pretorius}},\ }\bibfield  {title} {\bibinfo {title} {{Numerical relativity
  and astrophysics}},\ }\href
  {https://doi.org/10.1146/annurev-astro-081913-040031} {\bibfield  {journal}
  {\bibinfo  {journal} {Ann. Rev. Astron. Astrophys.}\ }\textbf {\bibinfo
  {volume} {52}},\ \bibinfo {pages} {661} (\bibinfo {year} {2014})},\ \Eprint
  {https://arxiv.org/abs/1405.4840} {arXiv:1405.4840 [astro-ph.HE]}
  \BibitemShut {NoStop}%
\bibitem [{\citenamefont {Paschalidis}(2017)}]{Paschalidis:2016agf}%
  \BibitemOpen
  \bibfield  {author} {\bibinfo {author} {\bibfnamefont {V.}~\bibnamefont
  {Paschalidis}},\ }\bibfield  {title} {\bibinfo {title} {{General relativistic
  simulations of compact binary mergers as engines for short gamma-ray
  bursts}},\ }\href {https://doi.org/10.1088/1361-6382/aa61ce} {\bibfield
  {journal} {\bibinfo  {journal} {Class. Quant. Grav.}\ }\textbf {\bibinfo
  {volume} {34}},\ \bibinfo {pages} {084002} (\bibinfo {year} {2017})},\
  \Eprint {https://arxiv.org/abs/1611.01519} {arXiv:1611.01519 [astro-ph.HE]}
  \BibitemShut {NoStop}%
\bibitem [{\citenamefont {Baiotti}\ and\ \citenamefont
  {Rezzolla}(2017)}]{Baiotti:2016qnr}%
  \BibitemOpen
  \bibfield  {author} {\bibinfo {author} {\bibfnamefont {L.}~\bibnamefont
  {Baiotti}}\ and\ \bibinfo {author} {\bibfnamefont {L.}~\bibnamefont
  {Rezzolla}},\ }\bibfield  {title} {\bibinfo {title} {Binary neutron star
  mergers: a review of {Einstein's} richest laboratory},\ }\href
  {https://doi.org/10.1088/1361-6633/aa67bb} {\bibfield  {journal} {\bibinfo
  {journal} {Rept. Prog. Phys.}\ }\textbf {\bibinfo {volume} {80}},\ \bibinfo
  {pages} {096901} (\bibinfo {year} {2017})},\ \Eprint
  {https://arxiv.org/abs/1607.03540} {arXiv:1607.03540 [gr-qc]} \BibitemShut
  {NoStop}%
\bibitem [{\citenamefont {Duez}\ and\ \citenamefont
  {Zlochower}(2019)}]{Duez:2018jaf}%
  \BibitemOpen
  \bibfield  {author} {\bibinfo {author} {\bibfnamefont {M.~D.}\ \bibnamefont
  {Duez}}\ and\ \bibinfo {author} {\bibfnamefont {Y.}~\bibnamefont
  {Zlochower}},\ }\bibfield  {title} {\bibinfo {title} {{Numerical relativity
  of compact binaries in the 21st century}},\ }\href
  {https://doi.org/10.1088/1361-6633/aadb16} {\bibfield  {journal} {\bibinfo
  {journal} {Rept. Prog. Phys.}\ }\textbf {\bibinfo {volume} {82}},\ \bibinfo
  {pages} {016902} (\bibinfo {year} {2019})},\ \Eprint
  {https://arxiv.org/abs/1808.06011} {arXiv:1808.06011 [gr-qc]} \BibitemShut
  {NoStop}%
\bibitem [{\citenamefont {Ciolfi}(2018)}]{Ciolfi:2018tal}%
  \BibitemOpen
  \bibfield  {author} {\bibinfo {author} {\bibfnamefont {R.}~\bibnamefont
  {Ciolfi}},\ }\bibfield  {title} {\bibinfo {title} {{Short gamma-ray burst
  central engines}},\ }\href {https://doi.org/10.1142/S021827181842004X}
  {\bibfield  {journal} {\bibinfo  {journal} {Int. J. Mod. Phys. D}\ }\textbf
  {\bibinfo {volume} {27}},\ \bibinfo {pages} {1842004} (\bibinfo {year}
  {2018})},\ \Eprint {https://arxiv.org/abs/1804.03684} {arXiv:1804.03684
  [astro-ph.HE]} \BibitemShut {NoStop}%
\bibitem [{\citenamefont {{Ascenzi}}\ \emph {et~al.}(2021)\citenamefont
  {{Ascenzi}}, \citenamefont {{Oganesyan}}, \citenamefont {{Branchesi}},\ and\
  \citenamefont {{Ciolfi}}}]{Ascenzi_2021}%
  \BibitemOpen
  \bibfield  {author} {\bibinfo {author} {\bibfnamefont {S.}~\bibnamefont
  {{Ascenzi}}}, \bibinfo {author} {\bibfnamefont {G.}~\bibnamefont
  {{Oganesyan}}}, \bibinfo {author} {\bibfnamefont {M.}~\bibnamefont
  {{Branchesi}}},\ and\ \bibinfo {author} {\bibfnamefont {R.}~\bibnamefont
  {{Ciolfi}}},\ }\bibfield  {title} {\bibinfo {title} {{Electromagnetic
  counterparts of compact binary mergers}},\ }\href
  {https://doi.org/10.1017/S0022377820001646} {\bibfield  {journal} {\bibinfo
  {journal} {J. Plasma Phys.}\ }\textbf {\bibinfo {volume} {87}},\ \bibinfo
  {eid} {845870102} (\bibinfo {year} {2021})},\ \Eprint
  {https://arxiv.org/abs/2011.04001} {arXiv:2011.04001 [astro-ph.HE]}
  \BibitemShut {NoStop}%
\bibitem [{\citenamefont {Hessels}\ \emph {et~al.}(2006)\citenamefont
  {Hessels}, \citenamefont {Ransom}, \citenamefont {Stairs}, \citenamefont
  {Freire}, \citenamefont {Kaspi},\ and\ \citenamefont
  {Camilo}}]{Hessels:2006ze}%
  \BibitemOpen
  \bibfield  {author} {\bibinfo {author} {\bibfnamefont {J.~W.~T.}\
  \bibnamefont {Hessels}}, \bibinfo {author} {\bibfnamefont {S.~M.}\
  \bibnamefont {Ransom}}, \bibinfo {author} {\bibfnamefont {I.~H.}\
  \bibnamefont {Stairs}}, \bibinfo {author} {\bibfnamefont {P.~C.~C.}\
  \bibnamefont {Freire}}, \bibinfo {author} {\bibfnamefont {V.~M.}\
  \bibnamefont {Kaspi}},\ and\ \bibinfo {author} {\bibfnamefont
  {F.}~\bibnamefont {Camilo}},\ }\bibfield  {title} {\bibinfo {title} {{A radio
  pulsar spinning at 716 Hz}},\ }\href
  {https://doi.org/10.1126/science.1123430} {\bibfield  {journal} {\bibinfo
  {journal} {Science}\ }\textbf {\bibinfo {volume} {311}},\ \bibinfo {pages}
  {1901} (\bibinfo {year} {2006})},\ \Eprint
  {https://arxiv.org/abs/astro-ph/0601337} {arXiv:astro-ph/0601337}
  \BibitemShut {NoStop}%
\bibitem [{\citenamefont {Breu}\ and\ \citenamefont
  {Rezzolla}(2016)}]{Breu:2016ufb}%
  \BibitemOpen
  \bibfield  {author} {\bibinfo {author} {\bibfnamefont {C.}~\bibnamefont
  {Breu}}\ and\ \bibinfo {author} {\bibfnamefont {L.}~\bibnamefont
  {Rezzolla}},\ }\bibfield  {title} {\bibinfo {title} {{Maximum mass, moment of
  inertia and compactness of relativistic stars}},\ }\href
  {https://doi.org/10.1093/mnras/stw575} {\bibfield  {journal} {\bibinfo
  {journal} {Mon. Not. R. Astron. Soc.}\ }\textbf {\bibinfo {volume} {459}},\
  \bibinfo {pages} {646} (\bibinfo {year} {2016})},\ \Eprint
  {https://arxiv.org/abs/1601.06083} {arXiv:1601.06083 [gr-qc]} \BibitemShut
  {NoStop}%
\bibitem [{\citenamefont {Paschalidis}\ and\ \citenamefont
  {Stergioulas}(2017)}]{Paschalidis:2016vmz}%
  \BibitemOpen
  \bibfield  {author} {\bibinfo {author} {\bibfnamefont {V.}~\bibnamefont
  {Paschalidis}}\ and\ \bibinfo {author} {\bibfnamefont {N.}~\bibnamefont
  {Stergioulas}},\ }\bibfield  {title} {\bibinfo {title} {{Rotating stars in
  relativity}},\ }\href {https://doi.org/10.1007/s41114-017-0008-x} {\bibfield
  {journal} {\bibinfo  {journal} {Living Rev. Rel.}\ }\textbf {\bibinfo
  {volume} {20}},\ \bibinfo {pages} {7} (\bibinfo {year} {2017})},\ \Eprint
  {https://arxiv.org/abs/1612.03050} {arXiv:1612.03050 [astro-ph.HE]}
  \BibitemShut {NoStop}%
\bibitem [{\citenamefont {Coulter}\ \emph {et~al.}(2017)\citenamefont {Coulter}
  \emph {et~al.}}]{Coulter:2017wya}%
  \BibitemOpen
  \bibfield  {author} {\bibinfo {author} {\bibfnamefont {D.~A.}\ \bibnamefont
  {Coulter}} \emph {et~al.},\ }\bibfield  {title} {\bibinfo {title} {{Swope
  Supernova Survey 2017a (SSS17a), the optical counterpart to a gravitational
  wave source}},\ }\href {https://doi.org/10.1126/science.aap9811} {\bibfield
  {journal} {\bibinfo  {journal} {Science}\ }\textbf {\bibinfo {volume}
  {358}},\ \bibinfo {pages} {1556} (\bibinfo {year} {2017})},\ \Eprint
  {https://arxiv.org/abs/1710.05452} {arXiv:1710.05452 [astro-ph.HE]}
  \BibitemShut {NoStop}%
\bibitem [{\citenamefont {Drout}\ \emph {et~al.}(2017)\citenamefont {Drout}
  \emph {et~al.}}]{Drout:2017ijr}%
  \BibitemOpen
  \bibfield  {author} {\bibinfo {author} {\bibfnamefont {M.~R.}\ \bibnamefont
  {Drout}} \emph {et~al.},\ }\bibfield  {title} {\bibinfo {title} {{Light
  curves of the neutron star merger GW170817/SSS17a: Implications for r-process
  nucleosynthesis}},\ }\href {https://doi.org/10.1126/science.aaq0049}
  {\bibfield  {journal} {\bibinfo  {journal} {Science}\ }\textbf {\bibinfo
  {volume} {358}},\ \bibinfo {pages} {1570} (\bibinfo {year} {2017})},\ \Eprint
  {https://arxiv.org/abs/1710.05443} {arXiv:1710.05443 [astro-ph.HE]}
  \BibitemShut {NoStop}%
\bibitem [{\citenamefont {Shappee}\ \emph {et~al.}(2017)\citenamefont {Shappee}
  \emph {et~al.}}]{Shappee:2017zly}%
  \BibitemOpen
  \bibfield  {author} {\bibinfo {author} {\bibfnamefont {B.~J.}\ \bibnamefont
  {Shappee}} \emph {et~al.},\ }\bibfield  {title} {\bibinfo {title} {{Early
  spectra of the gravitational wave source GW170817: Evolution of a neutron
  star merger}},\ }\href {https://doi.org/10.1126/science.aaq0186} {\bibfield
  {journal} {\bibinfo  {journal} {Science}\ }\textbf {\bibinfo {volume}
  {358}},\ \bibinfo {pages} {1574} (\bibinfo {year} {2017})},\ \Eprint
  {https://arxiv.org/abs/1710.05432} {arXiv:1710.05432 [astro-ph.HE]}
  \BibitemShut {NoStop}%
\bibitem [{\citenamefont {Kasliwal}\ \emph {et~al.}(2017)\citenamefont
  {Kasliwal} \emph {et~al.}}]{Kasliwal:2017ngb}%
  \BibitemOpen
  \bibfield  {author} {\bibinfo {author} {\bibfnamefont {M.~M.}\ \bibnamefont
  {Kasliwal}} \emph {et~al.},\ }\bibfield  {title} {\bibinfo {title}
  {{Illuminating gravitational waves: A concordant picture of photons from a
  neutron star merger}},\ }\href {https://doi.org/10.1126/science.aap9455}
  {\bibfield  {journal} {\bibinfo  {journal} {Science}\ }\textbf {\bibinfo
  {volume} {358}},\ \bibinfo {pages} {1559} (\bibinfo {year} {2017})},\ \Eprint
  {https://arxiv.org/abs/1710.05436} {arXiv:1710.05436 [astro-ph.HE]}
  \BibitemShut {NoStop}%
\bibitem [{\citenamefont {Tanaka}\ \emph {et~al.}(2017)\citenamefont {Tanaka}
  \emph {et~al.}}]{Tanaka:2017qxj}%
  \BibitemOpen
  \bibfield  {author} {\bibinfo {author} {\bibfnamefont {M.}~\bibnamefont
  {Tanaka}} \emph {et~al.},\ }\bibfield  {title} {\bibinfo {title} {{Kilonova
  from post-merger ejecta as an optical and near-infrared counterpart of
  GW170817}},\ }\href {https://doi.org/10.1093/pasj/psx121} {\bibfield
  {journal} {\bibinfo  {journal} {Publ. Astron. Soc. Jap.}\ }\textbf {\bibinfo
  {volume} {69}},\ \bibinfo {pages} {102} (\bibinfo {year} {2017})},\ \Eprint
  {https://arxiv.org/abs/1710.05850} {arXiv:1710.05850 [astro-ph.HE]}
  \BibitemShut {NoStop}%
\bibitem [{\citenamefont {Arcavi}\ \emph {et~al.}(2017)\citenamefont {Arcavi}
  \emph {et~al.}}]{Arcavi:2017xiz}%
  \BibitemOpen
  \bibfield  {author} {\bibinfo {author} {\bibfnamefont {I.}~\bibnamefont
  {Arcavi}} \emph {et~al.},\ }\bibfield  {title} {\bibinfo {title} {{Optical
  emission from a kilonova following a gravitational-wave-detected neutron-star
  merger}},\ }\href {https://doi.org/10.1038/nature24291} {\bibfield  {journal}
  {\bibinfo  {journal} {Nature}\ }\textbf {\bibinfo {volume} {551}},\ \bibinfo
  {pages} {64} (\bibinfo {year} {2017})},\ \Eprint
  {https://arxiv.org/abs/1710.05843} {arXiv:1710.05843 [astro-ph.HE]}
  \BibitemShut {NoStop}%
\bibitem [{\citenamefont {Pian}\ \emph {et~al.}(2017)\citenamefont {Pian} \emph
  {et~al.}}]{Pian:2017gtc}%
  \BibitemOpen
  \bibfield  {author} {\bibinfo {author} {\bibfnamefont {E.}~\bibnamefont
  {Pian}} \emph {et~al.},\ }\bibfield  {title} {\bibinfo {title}
  {{Spectroscopic identification of r-process nucleosynthesis in a double
  neutron-star merger}},\ }\href {https://doi.org/10.1038/nature24298}
  {\bibfield  {journal} {\bibinfo  {journal} {Nature}\ }\textbf {\bibinfo
  {volume} {551}},\ \bibinfo {pages} {67} (\bibinfo {year} {2017})},\ \Eprint
  {https://arxiv.org/abs/1710.05858} {arXiv:1710.05858 [astro-ph.HE]}
  \BibitemShut {NoStop}%
\bibitem [{\citenamefont {Smartt}\ \emph {et~al.}(2017)\citenamefont {Smartt}
  \emph {et~al.}}]{Smartt:2017fuw}%
  \BibitemOpen
  \bibfield  {author} {\bibinfo {author} {\bibfnamefont {S.~J.}\ \bibnamefont
  {Smartt}} \emph {et~al.},\ }\bibfield  {title} {\bibinfo {title} {{A kilonova
  as the electromagnetic counterpart to a gravitational-wave source}},\ }\href
  {https://doi.org/10.1038/nature24303} {\bibfield  {journal} {\bibinfo
  {journal} {Nature}\ }\textbf {\bibinfo {volume} {551}},\ \bibinfo {pages}
  {75} (\bibinfo {year} {2017})},\ \Eprint {https://arxiv.org/abs/1710.05841}
  {arXiv:1710.05841 [astro-ph.HE]} \BibitemShut {NoStop}%
\bibitem [{\citenamefont {Soares-Santos}\ \emph {et~al.}(2017)\citenamefont
  {Soares-Santos} \emph {et~al.}}]{Soares-Santos:2017lru}%
  \BibitemOpen
  \bibfield  {author} {\bibinfo {author} {\bibfnamefont {M.}~\bibnamefont
  {Soares-Santos}} \emph {et~al.},\ }\bibfield  {title} {\bibinfo {title} {{The
  electromagnetic counterpart of the binary neutron star merger LIGO/Virgo
  GW170817. I. Discovery of the optical counterpart using the dark energy
  camera}},\ }\href {https://doi.org/10.3847/2041-8213/aa9059} {\bibfield
  {journal} {\bibinfo  {journal} {Astrophys. J. Lett.}\ }\textbf {\bibinfo
  {volume} {848}},\ \bibinfo {pages} {L16} (\bibinfo {year} {2017})},\ \Eprint
  {https://arxiv.org/abs/1710.05459} {arXiv:1710.05459 [astro-ph.HE]}
  \BibitemShut {NoStop}%
\bibitem [{\citenamefont {Nicholl}\ \emph {et~al.}(2017)\citenamefont {Nicholl}
  \emph {et~al.}}]{Nicholl:2017ahq}%
  \BibitemOpen
  \bibfield  {author} {\bibinfo {author} {\bibfnamefont {M.}~\bibnamefont
  {Nicholl}} \emph {et~al.},\ }\bibfield  {title} {\bibinfo {title} {{The
  electromagnetic counterpart of the binary neutron star merger LIGO/VIRGO
  GW170817. III. Optical and UV spectra of a blue kilonova from fast polar
  ejecta}},\ }\href {https://doi.org/10.3847/2041-8213/aa9029} {\bibfield
  {journal} {\bibinfo  {journal} {Astrophys. J. Lett.}\ }\textbf {\bibinfo
  {volume} {848}},\ \bibinfo {pages} {L18} (\bibinfo {year} {2017})},\ \Eprint
  {https://arxiv.org/abs/1710.05456} {arXiv:1710.05456 [astro-ph.HE]}
  \BibitemShut {NoStop}%
\bibitem [{\citenamefont {Cowperthwaite}\ \emph {et~al.}(2017)\citenamefont
  {Cowperthwaite} \emph {et~al.}}]{Cowperthwaite:2017dyu}%
  \BibitemOpen
  \bibfield  {author} {\bibinfo {author} {\bibfnamefont {P.~S.}\ \bibnamefont
  {Cowperthwaite}} \emph {et~al.},\ }\bibfield  {title} {\bibinfo {title} {{The
  electromagnetic counterpart of the binary neutron star merger LIGO/Virgo
  GW170817. II. UV, optical, and near-infrared light curves and comparison to
  kilonova models}},\ }\href {https://doi.org/10.3847/2041-8213/aa8fc7}
  {\bibfield  {journal} {\bibinfo  {journal} {Astrophys. J. Lett.}\ }\textbf
  {\bibinfo {volume} {848}},\ \bibinfo {pages} {L17} (\bibinfo {year}
  {2017})},\ \Eprint {https://arxiv.org/abs/1710.05840} {arXiv:1710.05840
  [astro-ph.HE]} \BibitemShut {NoStop}%
\bibitem [{Note6()}]{Note6}%
  \BibitemOpen
  \bibinfo {note} {Using a higher value for the mass of PSR J0740+6620 would
  result in a more constrained EoS ensemble, mainly because some EoSs do not
  support NSs of such a higher mass. For this reason, a more conservative
  choice is using $M = 2M_{\odot }$ as the mass parameter.}\BibitemShut {Stop}%
\bibitem [{Note7()}]{Note7}%
  \BibitemOpen
  \bibinfo {note} {As shown in \cite {Riley:2021pdl,Miller:2021qha}, modeling
  PSR J0740+6620 as non-rotating induces at most an error of 0.2~km in radius
  for the stiffest EoSs, with softer EoSs showing smaller errors of 0.05~km.
  Since the softer EoSs are those which are constrained by a lower bound on the
  radius, this is not a large effect with our implementation.}\BibitemShut
  {Stop}%
\bibitem [{Note8()}]{Note8}%
  \BibitemOpen
  \bibinfo {note} {In \cite {Annala:2019puf}, the onset of QM was defined to
  take place at the lowest density from which $\gamma $ remains below 1.75 all
  the way to asymptotically larger densities, where it approaches
  1.}\BibitemShut {Stop}%
\bibitem [{\citenamefont {Weih}\ \emph {et~al.}(2019)\citenamefont {Weih},
  \citenamefont {Most},\ and\ \citenamefont {Rezzolla}}]{Weih:2019rzo}%
  \BibitemOpen
  \bibfield  {author} {\bibinfo {author} {\bibfnamefont {L.~R.}\ \bibnamefont
  {Weih}}, \bibinfo {author} {\bibfnamefont {E.~R.}\ \bibnamefont {Most}},\
  and\ \bibinfo {author} {\bibfnamefont {L.}~\bibnamefont {Rezzolla}},\
  }\bibfield  {title} {\bibinfo {title} {{Optimal neutron-star mass ranges to
  constrain the equation of state of nuclear matter with electromagnetic and
  gravitational-wave observations}},\ }\href
  {https://doi.org/10.3847/1538-4357/ab2edd} {\bibfield  {journal} {\bibinfo
  {journal} {Astrophys. J.}\ }\textbf {\bibinfo {volume} {881}},\ \bibinfo
  {pages} {73} (\bibinfo {year} {2019})},\ \Eprint
  {https://arxiv.org/abs/1905.04900} {arXiv:1905.04900 [astro-ph.HE]}
  \BibitemShut {NoStop}%
\bibitem [{\citenamefont {Coughlin}\ \emph {et~al.}(2018)\citenamefont
  {Coughlin} \emph {et~al.}}]{Coughlin:2018miv}%
  \BibitemOpen
  \bibfield  {author} {\bibinfo {author} {\bibfnamefont {M.~W.}\ \bibnamefont
  {Coughlin}} \emph {et~al.},\ }\bibfield  {title} {\bibinfo {title}
  {{Constraints on the neutron star equation of state from AT2017gfo using
  radiative transfer simulations}},\ }\href
  {https://doi.org/10.1093/mnras/sty2174} {\bibfield  {journal} {\bibinfo
  {journal} {Mon. Not. R. Astron. Soc.}\ }\textbf {\bibinfo {volume} {480}},\
  \bibinfo {pages} {3871} (\bibinfo {year} {2018})},\ \Eprint
  {https://arxiv.org/abs/1805.09371} {arXiv:1805.09371 [astro-ph.HE]}
  \BibitemShut {NoStop}%
\bibitem [{\citenamefont {Wang}\ \emph {et~al.}(2019)\citenamefont {Wang},
  \citenamefont {Shao}, \citenamefont {Jiang}, \citenamefont {Tang},
  \citenamefont {Ren}, \citenamefont {Zhang}, \citenamefont {Jin},
  \citenamefont {Fan},\ and\ \citenamefont {Wei}}]{Wang:2018nye}%
  \BibitemOpen
  \bibfield  {author} {\bibinfo {author} {\bibfnamefont {Y.-Z.}\ \bibnamefont
  {Wang}}, \bibinfo {author} {\bibfnamefont {D.-S.}\ \bibnamefont {Shao}},
  \bibinfo {author} {\bibfnamefont {J.-L.}\ \bibnamefont {Jiang}}, \bibinfo
  {author} {\bibfnamefont {S.-P.}\ \bibnamefont {Tang}}, \bibinfo {author}
  {\bibfnamefont {X.-X.}\ \bibnamefont {Ren}}, \bibinfo {author} {\bibfnamefont
  {F.-W.}\ \bibnamefont {Zhang}}, \bibinfo {author} {\bibfnamefont {Z.-P.}\
  \bibnamefont {Jin}}, \bibinfo {author} {\bibfnamefont {Y.-Z.}\ \bibnamefont
  {Fan}},\ and\ \bibinfo {author} {\bibfnamefont {D.-M.}\ \bibnamefont {Wei}},\
  }\bibfield  {title} {\bibinfo {title} {{GW170817: The energy extraction
  process of the off-axis relativistic outflow and the constraint on the
  equation of state of neutron stars}},\ }\href
  {https://doi.org/10.3847/1538-4357/ab1914} {\bibfield  {journal} {\bibinfo
  {journal} {Astrophys. J.}\ }\textbf {\bibinfo {volume} {877}},\ \bibinfo
  {pages} {2} (\bibinfo {year} {2019})},\ \Eprint
  {https://arxiv.org/abs/1811.02558} {arXiv:1811.02558 [astro-ph.HE]}
  \BibitemShut {NoStop}%
\bibitem [{\citenamefont {Radice}\ and\ \citenamefont
  {Dai}(2019)}]{Radice:2018ozg}%
  \BibitemOpen
  \bibfield  {author} {\bibinfo {author} {\bibfnamefont {D.}~\bibnamefont
  {Radice}}\ and\ \bibinfo {author} {\bibfnamefont {L.}~\bibnamefont {Dai}},\
  }\bibfield  {title} {\bibinfo {title} {{Multimessenger parameter estimation
  of GW170817}},\ }\href {https://doi.org/10.1140/epja/i2019-12716-4}
  {\bibfield  {journal} {\bibinfo  {journal} {Eur. Phys. J. A}\ }\textbf
  {\bibinfo {volume} {55}},\ \bibinfo {pages} {50} (\bibinfo {year} {2019})},\
  \Eprint {https://arxiv.org/abs/1810.12917} {arXiv:1810.12917 [astro-ph.HE]}
  \BibitemShut {NoStop}%
\bibitem [{\citenamefont {Tsokaros}\ \emph {et~al.}(2020)\citenamefont
  {Tsokaros}, \citenamefont {Ruiz},\ and\ \citenamefont
  {Shapiro}}]{Tsokaros:2020hli}%
  \BibitemOpen
  \bibfield  {author} {\bibinfo {author} {\bibfnamefont {A.}~\bibnamefont
  {Tsokaros}}, \bibinfo {author} {\bibfnamefont {M.}~\bibnamefont {Ruiz}},\
  and\ \bibinfo {author} {\bibfnamefont {S.~L.}\ \bibnamefont {Shapiro}},\
  }\bibfield  {title} {\bibinfo {title} {{GW190814: Spin and equation of state
  of a neutron star companion}},\ }\href
  {https://doi.org/10.3847/1538-4357/abc421} {\bibfield  {journal} {\bibinfo
  {journal} {Astrophys. J.}\ }\textbf {\bibinfo {volume} {905}},\ \bibinfo
  {pages} {48} (\bibinfo {year} {2020})},\ \Eprint
  {https://arxiv.org/abs/2007.05526} {arXiv:2007.05526 [astro-ph.HE]}
  \BibitemShut {NoStop}%
\bibitem [{\citenamefont {Godzieba}\ \emph {et~al.}(2021)\citenamefont
  {Godzieba}, \citenamefont {Radice},\ and\ \citenamefont
  {Bernuzzi}}]{Godzieba:2020tjn}%
  \BibitemOpen
  \bibfield  {author} {\bibinfo {author} {\bibfnamefont {D.~A.}\ \bibnamefont
  {Godzieba}}, \bibinfo {author} {\bibfnamefont {D.}~\bibnamefont {Radice}},\
  and\ \bibinfo {author} {\bibfnamefont {S.}~\bibnamefont {Bernuzzi}},\
  }\bibfield  {title} {\bibinfo {title} {{On the maximum mass of neutron stars
  and GW190814}},\ }\href {https://doi.org/10.3847/1538-4357/abd4dd} {\bibfield
   {journal} {\bibinfo  {journal} {Astrophys. J.}\ }\textbf {\bibinfo {volume}
  {908}},\ \bibinfo {pages} {122} (\bibinfo {year} {2021})},\ \Eprint
  {https://arxiv.org/abs/2007.10999} {arXiv:2007.10999 [astro-ph.HE]}
  \BibitemShut {NoStop}%
\bibitem [{\citenamefont {Demircik}\ \emph {et~al.}(2021)\citenamefont
  {Demircik}, \citenamefont {Ecker},\ and\ \citenamefont
  {J\"arvinen}}]{Demircik:2020jkc}%
  \BibitemOpen
  \bibfield  {author} {\bibinfo {author} {\bibfnamefont {T.}~\bibnamefont
  {Demircik}}, \bibinfo {author} {\bibfnamefont {C.}~\bibnamefont {Ecker}},\
  and\ \bibinfo {author} {\bibfnamefont {M.}~\bibnamefont {J\"arvinen}},\
  }\bibfield  {title} {\bibinfo {title} {{Rapidly spinning compact stars with
  deconfinement phase transition}},\ }\href
  {https://doi.org/10.3847/2041-8213/abd853} {\bibfield  {journal} {\bibinfo
  {journal} {Astrophys. J. Lett.}\ }\textbf {\bibinfo {volume} {907}},\
  \bibinfo {pages} {L37} (\bibinfo {year} {2021})},\ \Eprint
  {https://arxiv.org/abs/2009.10731} {arXiv:2009.10731 [astro-ph.HE]}
  \BibitemShut {NoStop}%
\bibitem [{\citenamefont {Kanakis-Pegios}\ \emph {et~al.}(2021)\citenamefont
  {Kanakis-Pegios}, \citenamefont {Koliogiannis},\ and\ \citenamefont
  {Moustakidis}}]{Kanakis-Pegios:2020kzp}%
  \BibitemOpen
  \bibfield  {author} {\bibinfo {author} {\bibfnamefont {A.}~\bibnamefont
  {Kanakis-Pegios}}, \bibinfo {author} {\bibfnamefont {P.~S.}\ \bibnamefont
  {Koliogiannis}},\ and\ \bibinfo {author} {\bibfnamefont {C.~C.}\ \bibnamefont
  {Moustakidis}},\ }\bibfield  {title} {\bibinfo {title} {Probing the nuclear
  equation of state from the existence of a {$\sim 2.6~M_{\odot}$} neutron
  star: The {GW190814} puzzle},\ }\href {https://doi.org/10.3390/sym13020183}
  {\bibfield  {journal} {\bibinfo  {journal} {Symmetry}\ }\textbf {\bibinfo
  {volume} {13}},\ \bibinfo {pages} {183} (\bibinfo {year} {2021})},\ \Eprint
  {https://arxiv.org/abs/2012.09580} {arXiv:2012.09580 [astro-ph.HE]}
  \BibitemShut {NoStop}%
\bibitem [{\citenamefont {Most}\ \emph {et~al.}(2020)\citenamefont {Most},
  \citenamefont {Papenfort}, \citenamefont {Weih},\ and\ \citenamefont
  {Rezzolla}}]{Most:2020bba}%
  \BibitemOpen
  \bibfield  {author} {\bibinfo {author} {\bibfnamefont {E.~R.}\ \bibnamefont
  {Most}}, \bibinfo {author} {\bibfnamefont {L.~J.}\ \bibnamefont {Papenfort}},
  \bibinfo {author} {\bibfnamefont {L.~R.}\ \bibnamefont {Weih}},\ and\
  \bibinfo {author} {\bibfnamefont {L.}~\bibnamefont {Rezzolla}},\ }\bibfield
  {title} {\bibinfo {title} {{A lower bound on the maximum mass if the
  secondary in GW190814 was once a rapidly spinning neutron star}},\ }\href
  {https://doi.org/10.1093/mnrasl/slaa168} {\bibfield  {journal} {\bibinfo
  {journal} {Mon. Not. R. Astron. Soc. Lett.}\ }\textbf {\bibinfo {volume}
  {499}},\ \bibinfo {pages} {L82} (\bibinfo {year} {2020})},\ \Eprint
  {https://arxiv.org/abs/2006.14601} {arXiv:2006.14601 [astro-ph.HE]}
  \BibitemShut {NoStop}%
\bibitem [{Note9()}]{Note9}%
  \BibitemOpen
  \bibinfo {note} {Note that due to the tight correlation between the NS radius
  and its tidal deformability, upper limits for $\protect \tilde {\Lambda }$
  effectively act as upper limits for NS radii. Enforcing e.g.~$\protect \tilde
  {\Lambda }_\protect \text {GW170817}<200$ would correspond to $1.4M_\odot $
  NSs having radii well below 10~km.}\BibitemShut {Stop}%
\bibitem [{\citenamefont {Kiuchi}\ \emph {et~al.}(2019)\citenamefont {Kiuchi},
  \citenamefont {Kyutoku}, \citenamefont {Shibata},\ and\ \citenamefont
  {Taniguchi}}]{Kiuchi:2019lls}%
  \BibitemOpen
  \bibfield  {author} {\bibinfo {author} {\bibfnamefont {K.}~\bibnamefont
  {Kiuchi}}, \bibinfo {author} {\bibfnamefont {K.}~\bibnamefont {Kyutoku}},
  \bibinfo {author} {\bibfnamefont {M.}~\bibnamefont {Shibata}},\ and\ \bibinfo
  {author} {\bibfnamefont {K.}~\bibnamefont {Taniguchi}},\ }\bibfield  {title}
  {\bibinfo {title} {{Revisiting the lower bound on tidal deformability derived
  by AT 2017gfo}},\ }\href {https://doi.org/10.3847/2041-8213/ab1e45}
  {\bibfield  {journal} {\bibinfo  {journal} {Astrophys. J. Lett.}\ }\textbf
  {\bibinfo {volume} {876}},\ \bibinfo {pages} {L31} (\bibinfo {year}
  {2019})},\ \Eprint {https://arxiv.org/abs/1903.01466} {arXiv:1903.01466
  [astro-ph.HE]} \BibitemShut {NoStop}%
\bibitem [{\citenamefont {Pordes}\ \emph {et~al.}(2007)\citenamefont {Pordes}
  \emph {et~al.}}]{Pordes:2007zzb}%
  \BibitemOpen
  \bibfield  {author} {\bibinfo {author} {\bibfnamefont {R.}~\bibnamefont
  {Pordes}} \emph {et~al.},\ }\bibfield  {title} {\bibinfo {title} {{The open
  science grid}},\ }\href {https://doi.org/10.1088/1742-6596/78/1/012057}
  {\bibfield  {journal} {\bibinfo  {journal} {J. Phys. Conf. Ser.}\ }\textbf
  {\bibinfo {volume} {78}},\ \bibinfo {pages} {012057} (\bibinfo {year}
  {2007})}\BibitemShut {NoStop}%
\bibitem [{\citenamefont {Sfiligoi}\ \emph {et~al.}(2009)\citenamefont
  {Sfiligoi}, \citenamefont {Bradley}, \citenamefont {Holzman}, \citenamefont
  {Mhashilkar}, \citenamefont {Padhi},\ and\ \citenamefont
  {Wurthwein}}]{osg09}%
  \BibitemOpen
  \bibfield  {author} {\bibinfo {author} {\bibfnamefont {I.}~\bibnamefont
  {Sfiligoi}}, \bibinfo {author} {\bibfnamefont {D.~C.}\ \bibnamefont
  {Bradley}}, \bibinfo {author} {\bibfnamefont {B.}~\bibnamefont {Holzman}},
  \bibinfo {author} {\bibfnamefont {P.}~\bibnamefont {Mhashilkar}}, \bibinfo
  {author} {\bibfnamefont {S.}~\bibnamefont {Padhi}},\ and\ \bibinfo {author}
  {\bibfnamefont {F.}~\bibnamefont {Wurthwein}},\ }\bibfield  {title} {\bibinfo
  {title} {The pilot way to grid resources using glideinwms},\ }in\ \href
  {https://doi.org/10.1109/CSIE.2009.950} {\emph {\bibinfo {booktitle} {2009
  WRI World Congress on Computer Science and Information Engineering}}},\
  \bibinfo {series} {2}, Vol.~\bibinfo {volume} {2}\ (\bibinfo {year} {2009})\
  pp.\ \bibinfo {pages} {428--432}\BibitemShut {NoStop}%
\bibitem [{\citenamefont {Read}\ \emph {et~al.}(2009)\citenamefont {Read},
  \citenamefont {Lackey}, \citenamefont {Owen},\ and\ \citenamefont
  {Friedman}}]{Read:2008iy}%
  \BibitemOpen
  \bibfield  {author} {\bibinfo {author} {\bibfnamefont {J.~S.}\ \bibnamefont
  {Read}}, \bibinfo {author} {\bibfnamefont {B.~D.}\ \bibnamefont {Lackey}},
  \bibinfo {author} {\bibfnamefont {B.~J.}\ \bibnamefont {Owen}},\ and\
  \bibinfo {author} {\bibfnamefont {J.~L.}\ \bibnamefont {Friedman}},\
  }\bibfield  {title} {\bibinfo {title} {{Constraints on a phenomenologically
  parametrized neutron-star equation of state}},\ }\href
  {https://doi.org/10.1103/PhysRevD.79.124032} {\bibfield  {journal} {\bibinfo
  {journal} {Phys. Rev. D}\ }\textbf {\bibinfo {volume} {79}},\ \bibinfo
  {pages} {124032} (\bibinfo {year} {2009})},\ \Eprint
  {https://arxiv.org/abs/0812.2163} {arXiv:0812.2163 [astro-ph]} \BibitemShut
  {NoStop}%
\bibitem [{\citenamefont {{\"O}zel}\ and\ \citenamefont
  {Psaltis}(2009)}]{Ozel:2009da}%
  \BibitemOpen
  \bibfield  {author} {\bibinfo {author} {\bibfnamefont {F.}~\bibnamefont
  {{\"O}zel}}\ and\ \bibinfo {author} {\bibfnamefont {D.}~\bibnamefont
  {Psaltis}},\ }\bibfield  {title} {\bibinfo {title} {{Reconstructing the
  neutron-star equation of state from astrophysical measurements}},\ }\href
  {https://doi.org/10.1103/PhysRevD.80.103003} {\bibfield  {journal} {\bibinfo
  {journal} {Phys. Rev. D}\ }\textbf {\bibinfo {volume} {80}},\ \bibinfo
  {pages} {103003} (\bibinfo {year} {2009})},\ \Eprint
  {https://arxiv.org/abs/0905.1959} {arXiv:0905.1959 [astro-ph.HE]}
  \BibitemShut {NoStop}%
\bibitem [{\citenamefont {Lindblom}(2010)}]{Lindblom:2010bb}%
  \BibitemOpen
  \bibfield  {author} {\bibinfo {author} {\bibfnamefont {L.}~\bibnamefont
  {Lindblom}},\ }\bibfield  {title} {\bibinfo {title} {{Spectral
  representations of neutron-star equations of state}},\ }\href
  {https://doi.org/10.1103/PhysRevD.82.103011} {\bibfield  {journal} {\bibinfo
  {journal} {Phys. Rev. D}\ }\textbf {\bibinfo {volume} {82}},\ \bibinfo
  {pages} {103011} (\bibinfo {year} {2010})},\ \Eprint
  {https://arxiv.org/abs/1009.0738} {arXiv:1009.0738 [astro-ph.HE]}
  \BibitemShut {NoStop}%
\bibitem [{\citenamefont {Lindblom}(2018)}]{Lindblom:2018rfr}%
  \BibitemOpen
  \bibfield  {author} {\bibinfo {author} {\bibfnamefont {L.}~\bibnamefont
  {Lindblom}},\ }\bibfield  {title} {\bibinfo {title} {{Causal representations
  of neutron-star equations of state}},\ }\href
  {https://doi.org/10.1103/PhysRevD.97.123019} {\bibfield  {journal} {\bibinfo
  {journal} {Phys. Rev. D}\ }\textbf {\bibinfo {volume} {97}},\ \bibinfo
  {pages} {123019} (\bibinfo {year} {2018})},\ \Eprint
  {https://arxiv.org/abs/1804.04072} {arXiv:1804.04072 [astro-ph.HE]}
  \BibitemShut {NoStop}%
\bibitem [{\citenamefont {Weinberg}(1990)}]{Weinberg:1990rz}%
  \BibitemOpen
  \bibfield  {author} {\bibinfo {author} {\bibfnamefont {S.}~\bibnamefont
  {Weinberg}},\ }\bibfield  {title} {\bibinfo {title} {{Nuclear forces from
  chiral Lagrangians}},\ }\href {https://doi.org/10.1016/0370-2693(90)90938-3}
  {\bibfield  {journal} {\bibinfo  {journal} {Phys. Lett. B}\ }\textbf
  {\bibinfo {volume} {251}},\ \bibinfo {pages} {288} (\bibinfo {year}
  {1990})}\BibitemShut {NoStop}%
\bibitem [{\citenamefont {Epelbaum}\ \emph {et~al.}(2009)\citenamefont
  {Epelbaum}, \citenamefont {Hammer},\ and\ \citenamefont
  {Mei{\ss}ner}}]{Epelbaum:2008ga}%
  \BibitemOpen
  \bibfield  {author} {\bibinfo {author} {\bibfnamefont {E.}~\bibnamefont
  {Epelbaum}}, \bibinfo {author} {\bibfnamefont {H.-W.}\ \bibnamefont
  {Hammer}},\ and\ \bibinfo {author} {\bibfnamefont {U.-G.}\ \bibnamefont
  {Mei{\ss}ner}},\ }\bibfield  {title} {\bibinfo {title} {{Modern theory of
  nuclear forces}},\ }\href {https://doi.org/10.1103/RevModPhys.81.1773}
  {\bibfield  {journal} {\bibinfo  {journal} {Rev. Mod. Phys.}\ }\textbf
  {\bibinfo {volume} {81}},\ \bibinfo {pages} {1773} (\bibinfo {year}
  {2009})},\ \Eprint {https://arxiv.org/abs/0811.1338} {arXiv:0811.1338
  [nucl-th]} \BibitemShut {NoStop}%
\bibitem [{\citenamefont {Machleidt}\ and\ \citenamefont
  {Entem}(2011)}]{Machleidt:2011zz}%
  \BibitemOpen
  \bibfield  {author} {\bibinfo {author} {\bibfnamefont {R.}~\bibnamefont
  {Machleidt}}\ and\ \bibinfo {author} {\bibfnamefont {D.~R.}\ \bibnamefont
  {Entem}},\ }\bibfield  {title} {\bibinfo {title} {{Chiral effective field
  theory and nuclear forces}},\ }\href
  {https://doi.org/10.1016/j.physrep.2011.02.001} {\bibfield  {journal}
  {\bibinfo  {journal} {Phys. Rept.}\ }\textbf {\bibinfo {volume} {503}},\
  \bibinfo {pages} {1} (\bibinfo {year} {2011})},\ \Eprint
  {https://arxiv.org/abs/1105.2919} {arXiv:1105.2919 [nucl-th]} \BibitemShut
  {NoStop}%
\bibitem [{\citenamefont {Hebeler}\ and\ \citenamefont
  {Schwenk}(2010)}]{Hebeler:2009iv}%
  \BibitemOpen
  \bibfield  {author} {\bibinfo {author} {\bibfnamefont {K.}~\bibnamefont
  {Hebeler}}\ and\ \bibinfo {author} {\bibfnamefont {A.}~\bibnamefont
  {Schwenk}},\ }\bibfield  {title} {\bibinfo {title} {{Chiral three-nucleon
  forces and neutron matter}},\ }\href
  {https://doi.org/10.1103/PhysRevC.82.014314} {\bibfield  {journal} {\bibinfo
  {journal} {Phys. Rev. C}\ }\textbf {\bibinfo {volume} {82}},\ \bibinfo
  {pages} {014314} (\bibinfo {year} {2010})},\ \Eprint
  {https://arxiv.org/abs/0911.0483} {arXiv:0911.0483 [nucl-th]} \BibitemShut
  {NoStop}%
\bibitem [{\citenamefont {Hebeler}\ \emph {et~al.}(2011)\citenamefont
  {Hebeler}, \citenamefont {Bogner}, \citenamefont {Furnstahl}, \citenamefont
  {Nogga},\ and\ \citenamefont {Schwenk}}]{Hebeler:2010xb}%
  \BibitemOpen
  \bibfield  {author} {\bibinfo {author} {\bibfnamefont {K.}~\bibnamefont
  {Hebeler}}, \bibinfo {author} {\bibfnamefont {S.~K.}\ \bibnamefont {Bogner}},
  \bibinfo {author} {\bibfnamefont {R.~J.}\ \bibnamefont {Furnstahl}}, \bibinfo
  {author} {\bibfnamefont {A.}~\bibnamefont {Nogga}},\ and\ \bibinfo {author}
  {\bibfnamefont {A.}~\bibnamefont {Schwenk}},\ }\bibfield  {title} {\bibinfo
  {title} {{Improved nuclear matter calculations from chiral low-momentum
  interactions}},\ }\href {https://doi.org/10.1103/PhysRevC.83.031301}
  {\bibfield  {journal} {\bibinfo  {journal} {Phys. Rev. C}\ }\textbf {\bibinfo
  {volume} {83}},\ \bibinfo {pages} {031301(R)} (\bibinfo {year} {2011})},\
  \Eprint {https://arxiv.org/abs/1012.3381} {arXiv:1012.3381 [nucl-th]}
  \BibitemShut {NoStop}%
\bibitem [{\citenamefont {Freedman}\ and\ \citenamefont
  {McLerran}(1977)}]{Freedman:1976ub}%
  \BibitemOpen
  \bibfield  {author} {\bibinfo {author} {\bibfnamefont {B.~A.}\ \bibnamefont
  {Freedman}}\ and\ \bibinfo {author} {\bibfnamefont {L.~D.}\ \bibnamefont
  {McLerran}},\ }\bibfield  {title} {\bibinfo {title} {{Fermions and gauge
  vector mesons at finite temperature and density. III. The ground-state energy
  of a relativistic quark gas}},\ }\href
  {https://doi.org/10.1103/PhysRevD.16.1169} {\bibfield  {journal} {\bibinfo
  {journal} {Phys. Rev. D}\ }\textbf {\bibinfo {volume} {16}},\ \bibinfo
  {pages} {1169} (\bibinfo {year} {1977})}\BibitemShut {NoStop}%
\bibitem [{\citenamefont {Shuryak}(1980)}]{Shuryak:1980tp}%
  \BibitemOpen
  \bibfield  {author} {\bibinfo {author} {\bibfnamefont {E.~V.}\ \bibnamefont
  {Shuryak}},\ }\bibfield  {title} {\bibinfo {title} {{Quantum chromodynamics
  and the theory of superdense matter}},\ }\href
  {https://doi.org/10.1016/0370-1573(80)90105-2} {\bibfield  {journal}
  {\bibinfo  {journal} {Phys. Rept.}\ }\textbf {\bibinfo {volume} {61}},\
  \bibinfo {pages} {71} (\bibinfo {year} {1980})}\BibitemShut {NoStop}%
\bibitem [{\citenamefont {Vuorinen}(2003)}]{Vuorinen:2003fs}%
  \BibitemOpen
  \bibfield  {author} {\bibinfo {author} {\bibfnamefont {A.}~\bibnamefont
  {Vuorinen}},\ }\bibfield  {title} {\bibinfo {title} {{The Pressure of QCD at
  finite temperatures and chemical potentials}},\ }\href
  {https://doi.org/10.1103/PhysRevD.68.054017} {\bibfield  {journal} {\bibinfo
  {journal} {Phys. Rev. D}\ }\textbf {\bibinfo {volume} {68}},\ \bibinfo
  {pages} {054017} (\bibinfo {year} {2003})},\ \Eprint
  {https://arxiv.org/abs/hep-ph/0305183} {arXiv:hep-ph/0305183} \BibitemShut
  {NoStop}%
\end{thebibliography}%

\end{document}